\documentclass{aa}
\usepackage[utf8]{inputenc}
\usepackage[T1]{fontenc}
\usepackage[english]{babel}

\usepackage{graphicx}
\usepackage{amsmath}
\usepackage{amssymb}

\usepackage[tight]{subfigure}

\usepackage[table,svgnames]{xcolor}

\usepackage{natbib}

\title{Protostellar disk formation and transport of angular momentum during magnetized core collapse}
\titlerunning{Protostellar disk formation}
\author{Marc \textsc{Joos}\inst{\ref{full}} \and Patrick \textsc{Hennebelle}\inst{\ref{ens}} \and Andrea \textsc{Ciardi}\inst{\ref{full}}}
\authorrunning{Marc \textsc{Joos} et al.}
\institute{Laboratoire de radioastronomie, LERMA, Observatoire de Paris, \'Ecole Normale Supérieure, Université Pierre et Marie Curie (UMR 8112 CNRS), 24 rue Lhomond, 75231 \textsc{Paris} Cedex 05, \textsc{France}\label{full} \and Laboratoire de radioastronomie, LERMA, Observatoire de Paris, \'Ecole Normale Supérieure (UMR 8112 CNRS), 24 rue Lhomond, 75231 \textsc{Paris} Cedex 05, \textsc{France}\label{ens}}

\abstract{Theoretical studies of collapsing clouds have found that even a relatively weak magnetic field may prevent the formation of disks and their fragmentation. However, most previous studies have been limited to cases where the magnetic field and the rotation axis of the cloud are aligned.}
{We study the transport of angular momentum, and its effects on disk formation, for non-aligned initial configurations and a range magnetic intensities.}
{We perform three-dimensional, adaptive mesh, numerical simulations of magnetically supercritical collapsing dense cores using the magneto-hydrodynamic code \textsc{Ramses}. We compute the contributions of all the relevant processes transporting angular momentum, in both the envelope and the region of the disk. We clearly define what could be defined as centrifugally supported disks and thoroughly study their properties.}
{At variance with earlier analyses, we show that the transport of angular momentum acts less efficiently in collapsing cores with non-aligned rotation and magnetic field. Analytically, this result can be understood by taking into account the bending of field lines occurring during the gravitational collapse. For the transport of angular momentum, we conclude that magnetic braking in the mean direction of the magnetic field tends to dominate over both the gravitational and outflow transport of angular momentum. We find that massive disks, containing at least 10\% of the initial core mass, can form during the earliest stages of star formation even for mass-to-flux ratios as small as three to five times the critical value. At higher field intensities, the early formation of massive disks is prevented.}
{ Given the ubiquity of Class I disks, and because the early formation of massive disks can take place at moderate magnetic intensities, we speculate that for stronger fields, disks will form later, when most of the envelope will have been accreted. In addition, we speculate that some observed early massive disks may actually be outflow cavities, mistaken for disks by projection effects.}


\begin{document}

\maketitle

\section{Introduction}

The formation of protostellar disks plays a central role in the context of star and planet formation. Protostars probably grow by accreting material from protostellar accretion disks \citep{Larson03}, and these disks, at later stages, are the natural progenitors of planets \citep{Lissauer93}. While observations of circumstellar disks around late young stellar objects (YSO), from Class I to T Tauri stars, are well-established \citep{Watson07}, it is still unclear when circumstellar disks form during the early collapse of prestellar dense cores and the early Class 0 phase, and what their initial properties are (mass, radius, magnetic flux, and temperature). For these embedded sources, direct observations are indeed more difficult than for YSOs, since disk emission is difficult to distinguish from the envelope signature \citep{Belloche02}, even with a relatively high spatial resolution \citep[50~AU,][]{Maury10}. However, other studies observing at lower resolution (about 250 AU) and without resolving the disks, infer from detailed emission modeling the presence of disks as massive as one solar mass, corresponding to about 12\% of the envelope mass. Although as stressed by these authors, these estimates depend on the assumptions made regarding the envelope \citep{Jorgensen07, Jorgensen09, Enoch09, Enoch11}.

From a theoretical point of view, it has been shown that, in the absence of a magnetic field, disks grow from small radii and masses by angular momentum conservation during the collapse of prestellar dense cores \citep[e.g.][]{Terebey84}.

However, observations infer that cores are magnetized and typically slightly super-critical \citep{Crutcher99}, that is to say the mass-to-flux ratio, $M/\Phi$, is comparable to a few times larger than its critical value $\simeq1/(2\pi G^{1/2})$. Theoretically, the presence of a magnetic field of such intensities has been found to modify substantially the collapse \citep{Allen03, Machida05, Fromang06} and in particular the formation of disks. Multidimensional simulations using different numerical techniques (\emph{e.g.} grid-based in 2D or 3D (including adaptive mesh refinement, AMR), smooth particle hydrodynamics, SPH, codes) have shown that the efficient transport of angular momentum through magnetic braking may suppress the formation of a centrifugally supported disk, even at relatively low magnetic intensities \citep[$\mu\lesssim5-10$,][]{Mellon08, Price07, Hennebelle08a}. Similar conclusions were reached by \cite{Galli06}, who performed analytical studies of magnetized collapsing cores.
\newline

To illustrate the problem, we can estimate the strength of the magnetic field, parametrized by $\mu=(M/\Phi)/(1/(2\pi G^{1/2}))$, for which efficient magnetic braking is expected. We consider the simple magnetic braking time, $\tau_{\rm br}$, defined as that taken by a torsional Alfv\'en wave to redistribute angular momentum from the inner to the outer regions of a cloud. This can be most naturally expressed as 
\begin{equation}
\tau_{\rm br}\sim\frac{Z_{d}}{v_{A}},
\end{equation}
where $Z_{d}$ is a characteristic scale-height, $v_{A}$ is the Alfv\'en speed, $v_{A}=B_{z}/\sqrt{4\pi\rho}$, $\rho$ is the characteristic density, and $B_{z}$ the vertical component of the magnetic field. This magnetic braking time should be compared to the characteristic rotation time of the central region of the cloud, where a disk would form if the braking were not strong enough. This can be written as
\begin{equation}
\tau_{\rm rot}\sim\frac{2\pi r_{d}}{v_{\phi}},
\end{equation}
 where $r_{d}$ is the disk radius, $v_{\phi}$ the Keplerian rotational velocity, $v_{\phi}=\sqrt{GM/r_{d}}$, and the mass $M=2\pi r_{d}^{2}Z_{d}\rho$.

The ratio of these two timescales then gives
\begin{eqnarray}
\frac{\tau_{\rm br}}{\tau_{\rm rot}} & \sim & \sqrt{2}\left(\frac{Z_{d}}{r_{d}}\right)^{1/2}\left(\frac{Z\rho}{B_{z}}G^{1/2}\right)\nonumber \\
 & \sim & \frac{1}{\sqrt{2}}\left(\frac{Z_{d}}{r_{d}}\right)^{1/2}\frac{\mu_{\rm eff}}{2\pi},\label{eq:brrot}
\end{eqnarray}
Where $\mu_{\rm eff}$ is an ``effective'' $\mu$ for the dynamically collapsing structure. In general, $\mu_{\rm eff}$ is smaller than the initial $\mu$ because only a fraction of the mass contracts along the field line and should be used in the estimate. We recall that in a disk $Z_d < r_d$; the ratio of times can then be approximated as
\begin{equation}
\frac{\tau_{\rm br}}{\tau_{\rm rot}} \lesssim \frac{\mu_{\rm eff}}{10}.
\end{equation}
This estimate shows that for $\mu_{\rm eff} \lesssim 10$ magnetic braking should be sufficiently efficient to remove a significant fraction of angular momentum from the inner region of the cloud, and thus greatly affect disk formation there.

Recent studies have attempted to avoid this magnetic braking catastrophe by invoking non-ideal magnetohydrodynamics (MHD) to effectively remove magnetic flux from the collapsing core. Ambipolar diffusion appears to be inefficient: by diffusing the magnetic field out of the central regions, it allows instead the build up of a strong magnetic field over a small circumstellar region. In this so-called ambipolar diffusion-induced accretion shock \citep{Mellon09,Li11}, the magnetic braking is greatly enhanced and can efficiently prevent the formation of rotationally supported disks. The effects of Ohmic dissipation remain uncertain. \cite{Krasnopolsky10} claim that an enhanced resistivity of about two to three orders of magnitude larger than the classical value is required although the physical origin of such a resistivity remains unclear. However, with a classical resistivity, \cite{Dapp10} and \cite{Machida11b} find tiny disks of radius the order of a few tens of AU, which can grow larger at later times. The discrepancies between these different studies of Ohmic dissipation might be caused by their different initial conditions. In addition, \cite{Santos-Lima11} investigated the effect of turbulence: they argued that an effective turbulent diffusivity \citep[of the same order of magnitude as the enhanced resistivity of][]{Krasnopolsky10} is sufficient to remove the magnetic flux excess and decrease the magnetic braking efficiency.

Most previous simulations have been performed in a somewhat idealized configuration, where the magnetic field and the rotation axis are initially aligned. As emphasized in \cite{Hennebelle09} \citep[see also][]{Price07}, the results of the collapse depend critically on the initial angle $\alpha$ between the magnetic field ${\bf B}$ and the rotation axis (which is the direction of the angular momentum ${\bf J}$). In particular, magnetic braking has been found to be more efficient when the magnetic field is initially aligned with the rotation axis, rather than when it is not. This is somewhat at odds with the theoretical conclusions of \cite{Mouschovias79}, as we discuss in section 3.

Following the previous studies of \cite{Hennebelle09} \citep[see also][]{Ciardi10}, we investigate in detail the transport of angular momentum, and the effects of magnetic braking in collapsing prestellar cores with aligned and misaligned configurations ($\alpha$ between $0^{\circ}$ and $90^{\circ}$). 

The plan of the paper is as follows. Section 2 provides a general description of a collapsing core. Analytical results describing magnetic braking are discussed in section 3, where we show that in a collapsing core where the field lines are strongly bent, magnetic braking is more efficient when the rotation axis is parallel to the direction of the magnetic field, than when it is perpendicular. In section 4, we present our numerical setup and initial conditions. The numerical results are presented in the section 5 and 6, where we focus first on the physical processes transporting angular momentum, and then on the physics and properties of the disk. Section 7 concludes the paper.

\section{Collapse, magnetized pseudo-disks, and centrifugally supported disks} \label{sec:collapse}

\begin{figure*}
  \includegraphics[width = 1. \textwidth]{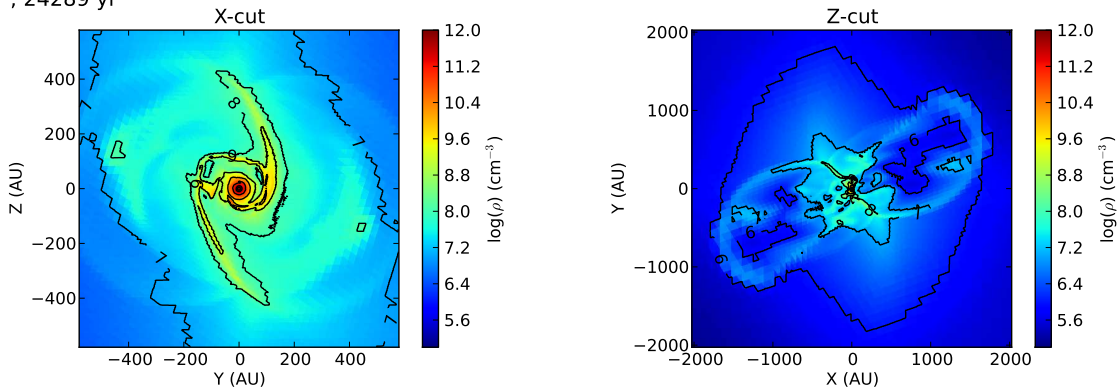}
  \caption{Slice in density in the equatorial plane and a plane aligned with the rotation axis for $\mu = 5, \alpha = 45^{\circ}$, at $t = 24290$~yr. The contours show levels of density on a logarithmic scale ($n > 10^6,\, 10^7,\, 10^8,\, 10^9$ and $10^{10}$~cm$^{-3}$).}
  \label{img:intro}
\end{figure*}

The gravitationally driven collapse of a magnetized core proceeds from an initially spherical cloud that tends to flatten along the magnetic field lines, leading to the formation of an oblate overdensity, the pseudo-disk. Pseudo-disks are magnetized, disk-like structures \citep{Galli93, Li96} that are \emph{not} centrifugally supported. Unlike centrifugally supported disks, they have no characteristic scale and are in a sense self-similar. In this paper, we emphasize the role of pseudo-disks as the place within the prestellar core where magnetic braking takes place. Figure \ref{img:intro} shows a slice in the equatorial plane (left panel) and along the rotation axis (right panel) of a dense-core collapse calculation, for $\mu = 5, \alpha = 45^{\circ}$. In the following, we loosely define a pseudo-disk as the structure with a density $n > 10^7$~cm$^{-3}$, as can be seen in the right panel in Fig. \ref{img:intro}. A more precise definition is given later.

After the isothermal phase of the protostellar core collapse, an adiabatic core (the first Larson's core) with a density $\gtrsim 10^{10}$~cm$^{-3}$ and a radius of about 10-20~AU forms in the center of the pseudo-disk. This is the central object in Fig. \ref{img:intro}. We do not treat the formation of the protostar itself.

The subsequent build-up of a centrifugally supported disk critically depends on the transport of angular momentum in the cloud. In contrast to pseudo-disks, which again are only geometrical overdensities, disks are rotationally supported structures formed around collapsing adiabatic cores. Unlike pseudo-disks, centrifugally supported disks possess a characteristic scale, namely the centrifugal radius. We discuss extensively their formation and properties in the next few sections. In Fig. \ref{img:intro}, the disk corresponds roughly to the gas with a density $\gtrsim10^{9}$~cm$^{-3}$, and a typical radius of 100-200~AU (on the left panel).

At the same time, outflows are launched in the direction of the rotation axis for all angles $\alpha \lesssim 80^{\circ}$ \citep{Ciardi10}. In Fig. \ref{img:intro}, the bipolar outflows are shown with an extent of about 2000~AU, and also their associated magnetic cavity (with a density below $10^{6}$~cm$^{-3}$, inside the outflows).

The adiabatic core, the disk, and the pseudo-disk are embedded in an envelope, with a density of between $10^{6}$~cm$^{-3}$ and $10^{8}$~cm$^{-3}$.

\section{Analytical study} \label{sec:anal}

Before presenting our numerical results, we develop analytical estimates of the braking timescales for the two extreme cases of an \emph{aligned rotator}, where ${\bf B}$ and ${\bf J}$ are initially parallel, and a \emph{perpendicular rotator}, where ${\bf B}$ and ${\bf J}$ are initially perpendicular. The main result is that, in a collapsing core, magnetic braking is more efficient for an aligned rotator, and that disks should then form more easily in perpendicular rotators.

This result contrasts somewhat with the classical analyses of \cite{Mouschovias79, Mouschovias80}, who showed that in a simple geometry with straight-parallel field lines, magnetic braking is more efficient for a perpendicular rotator. However, in a collapsing core, the gravitational pull strongly bends the magnetic field lines, which are frozen in the gas, toward the center of the cloud. As suggested by \cite{Mouschovias91}, the braking efficiency should then increase. As we show below, using more realistic assumptions that are appropriate to a collapsing prestellar core, the braking time for a perpendicular rotator is longer than for an aligned one.

Here we focus on cores, in contrast to the classical work of \cite{Mouschovias79}, which was applied to clouds, although it does not modify the analysis.

\subsection{Magnetic braking timescales}
\subsubsection{Aligned rotator ($\alpha=0$)}

Before investigating the more complex case of a core with strongly bent field lines, we first recall the classical analysis performed by \cite{Mouschovias79}. The important point here is that in their analysis the field lines are straight and parallel. We first consider an aligned rotator consisting of a core of mass $M$, density $\rho_c$, radius $R_c$, and half-height $Z$, surrounded by an external medium -- the envelope -- of density $\rho_{\rm ext}$. The core has an initial angular velocity $\Omega$, and the magnetic field ${\bf B}$ is uniform and parallel to the rotation axis. A magnetic braking timescale, $\tau_{\parallel}$, can be defined as the time needed for a torsional Alfv\'en wave to transfer the initial angular momentum of the core to the external medium
\begin{equation}
\rho_{\rm ext}v_{\rm A,ext}\tau_{\parallel} \sim \rho_{c}Z. \label{eq:parallel}
\end{equation}
Using the expression for the Alfv\'en speed in the external medium, $v_{\rm A,ext} = B/\sqrt{4\pi\rho_{\rm ext}}$ , together with expressions for the mass of the core, $M \sim 2\pi\rho_cR_c^2Z$, and the magnetic flux through it, $\Phi_B \sim \pi R_c^2B$, one obtains \citep{Mouschovias85}
\begin{equation}
\tau_{\parallel} \sim \left(\frac{\pi}{\rho_{\rm ext}}\right)^{1/2}\frac{M}{\Phi_B}. \label{eq:tau_classic}
\end{equation}
This demonstrates in particular that the magnetic braking timescale depends only on the initial conditions, namely the density of the external medium, $\rho_{\rm ext}$, and $M/\Phi_{B}$, the mass-to-flux ratio of the core. 

\subsubsection{Aligned rotator ($\alpha=0$) with fanning-out}

\begin{figure}[h!]
\centering
\includegraphics[width = .28\textwidth]{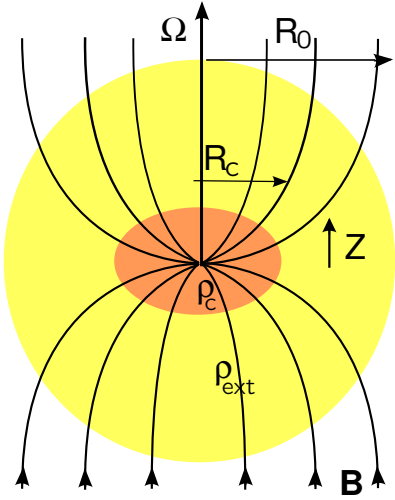}
\caption{Schematic view of a collapsing core in aligned configuration with field lines that fan-out. $B$ is the magnetic field, $R_c$ the radius of the core, $R_0$ the initial radius, $Z$ the half-height of the transition region, $\rho_c$ the density of the core, and $\rho_{\rm ext}$ the density of the external medium.}
\label{img:timescale}
\end{figure} 

We now consider an aligned rotator that is contracting and whose inner part -- the core -- has a density $\rho_c$ and a radius $R_c$. It is embedded in a medium of density $\rho_{\rm ext}$. The magnetic field is initially uniform. However, upon contraction the field strength increases, through a \emph{transition region} (\emph{i.e.} its envelope), from its original value $B_{\rm ext}$ in the external medium, to the compressed value in the core, $B_c$. Magnetic braking efficiently slows down the core rotation if this braking sets into co-rotation an amount of matter in the envelope with a moment of inertia equal to that of the core. Since the angular momentum in the envelope must be approximately equal to the momentum of the core, and assuming a co-rotation of the field lines, we find that
\begin{equation}
\Omega R_0^2\ \pi R_0^2\ \rho_{\rm ext}v_{\rm A,ext}\tau_{\parallel,fo} \sim \Omega R_c^2\ \pi R_c^2\ \rho_cZ,
\end{equation}
where $R_0$ is the initial radius of the flattened core, $Z$ its half-height, $v_{\rm A,ext}$ the Alfv\'en speed in the external medium ($v_{\rm A,ext} = B_{\rm ext}/\sqrt{4\pi\rho_{\rm ext}}$), and $\tau_{\parallel}$ the magnetic braking timescale for the aligned configuration. We thus obtain the magnetic braking time for the case of fan-out, $\tau_{\parallel,fo}$, which is given by
\begin{equation}
\tau_{\parallel,fo} = \frac{\rho_c}{\rho_{\rm ext}}\frac{Z}{v_{\rm A,ext}}\left(\frac{R_c}{R_0}\right)^4. \label{eq:tau_alfo}
\end{equation}
This expression is identical to Eq.~\ref{eq:parallel}, apart from the coefficient $(R_c/R_0)^4$, whose origin is twofold. First, because the field lines fan-out, the volume of the external medium swept by the Alfv\'en waves increases more rapidly than when the field lines are parallel. This accounts for a factor $(R_c/R_0)^2$. Second, as co-rotation of the field lines is assumed, fluid elements of the external medium lying along diverging field lines have higher specific angular momentum than if they were on straight field lines. This accounts for another factor $(R_c/R_0)^2$.

Using again the mass and the magnetic flux of the core, $M \sim 2\pi\rho_cR_c^2Z$ and $\Phi_B \sim \pi R_0^2B_{\rm ext} = \pi R_c^2B_c$, and the expression for $v_{\rm A,ext}$, we can rewrite Eq.~(\ref{eq:tau_alfo}) as \citep{Mouschovias85}
\begin{equation}
\tau_{\parallel,fo} = \left(\frac{\pi}{\rho_{\rm ext}}\right)^{1/2}\frac{M}{\Phi_B}\left(\frac{R_c}{R_0}\right)^2. \label{eq:tau_alfomf}
\end{equation}
Therefore, when the field lines fan-out, the magnetic braking timescale depends not only on $\rho_{\rm ext}$ and $M/\Phi_B$, but also on the contraction factor of the core $R_c/R_0$. In a collapsing core $R_c \ll R_0$, and this geometrical factor can significantly reduce the characteristic magnetic braking timescale in an aligned rotator, the ratio of timescale being $\tau_{\parallel}/\tau_{\parallel,fo} = (R_0/R_c)^2$.

Although this analysis gives a more realistic estimate of the braking time in a collapsing prestellar core, it is still greatly simplified. In particular, it is assumed that the field lines are immediately set into co-rotation. Because of the collapse, as the waves propagate outwards, the radius $R_0$ and the density $\rho_{\rm ext}$ vary continuously, before reaching approximately constant values.

\subsubsection{Perpendicular rotator ($\alpha=90^{\circ}$) with radially decreasing Alfv\'en speed}

\begin{figure}[h!]
  \centering
  \includegraphics[width = .28\textwidth]{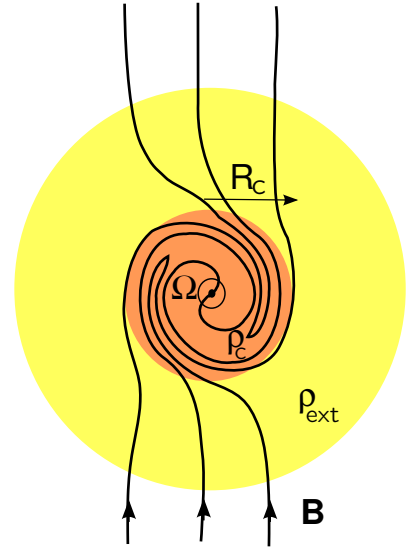}
  \caption{Schematic view of a collapsing core in perpendicular configuration. $B$ is the magnetic field, $R_c$ the radius of the core, $\rho_c$ the density of the core, and $\rho_{\rm ext}$ the density of the external medium.}
  \label{img:perpwofo}
\end{figure} 

In the case of a perpendicular rotator, the analysis of \cite{Mouschovias79} considers the braking timescale corresponding to the time needed for Alfv\'en waves to reach $R_{\perp}$, the radius for which the angular momentum of the external medium is equal to the initial angular momentum of the core. In this case, Alfv\'en waves propagate in the equatorial plane, rather than along the rotation axis, and sweep a cylinder of half-height $Z$ and radius $R_{\perp}$, thus
\begin{equation}
\rho_{\rm ext}(R_{\perp}^4 - R_c^4) \sim \rho_c R_c^4. \label{radius}
\end{equation}
Assuming further that the magnetic field has a radial dependence, $B(r) \propto r^{-1}$, so that $v_{\rm A}(r) = v_{\rm A}(R_c) \times R_c/r$, the perpendicular rotator magnetic braking time is then given by
\begin{equation}
\tau_{\perp} = \int_{R_c}^{R_{\perp}}\frac{{\rm d}r}{v_{\rm A}(r)} = \frac{1}{2}\frac{R_c}{v_{\rm A}(R_c)} \left[\left(1 + \frac{\rho_c}{\rho_{\rm ext}}\right)^{1/2} - 1 \right]. \label{eq:tau_perpmf1}
\end{equation}
Using the expressions for the mass, $M = 2\pi\rho_c R_c^2 Z$, magnetic flux, $\Phi_B = 4\pi R_c Z B(R_c)$, and Alfv\'en speed $v_{\rm A}(R_c)$, together with the approximation $\rho_c \gg \rho_{\rm ext}$, Eq.~\ref{eq:tau_perpmf1} becomes
\begin{equation}
\tau_{\perp} \sim 2\left(\frac{\pi}{\rho_c}\right)^{1/2}\frac{M}{\Phi_B}. \label{eq:tau_perpmf}
\end{equation}

\subsubsection{Perpendicular rotator ($\alpha=90^{\circ}$) with constant Alfv\'en speed}

The expression in Eq.~(\ref{eq:tau_perpmf}) was derived assuming a magnetic field $B(r) \propto r^{-1}$, and consequently an Alfv\'en speed that decreases with radius. However, the field lines are twisted because of the rotation of the core and are not purely radial. It can easily be inferred from the divergence-free constraint on the magnetic field, and from the simulations, that the structure of the magnetic field is more complex, and is not radial, as shown schematically in Fig.~\ref{img:perpwofo}. It is therefore unlikely that the Alfv\'en speed drops with radius as $1/r$, and collapse calculations indeed show that the Alfv\'en speed remains roughly constant in the dense cores \citep[e.g.][]{Hennebelle11}. In this case, the braking time is given by
\begin{eqnarray}
\tau_{\perp,cv_{\rm A}} = \int_{R_c}^{R_{\perp}}\frac{{\rm d}r}{v_{\rm A}} & = & \frac{R_{\perp} - R_c}{v_{\rm A}} \nonumber \\
 & = & \frac{R_c}{v_{\rm A}}\left[\left(1 + \frac{\rho_c}{\rho_{\rm ext}}\right)^{1/4} - 1\right].
\end{eqnarray}
Since $\rho_c/\rho_{\rm ext} \gg 1$, the braking time then becomes
\begin{equation}
\tau_{\perp,cv_{\rm A}}\sim 4 \sqrt{\pi}\frac{\rho_{\rm ext}^{1/4}}{\rho_c^{3/4}}\frac{M}{\Phi_B}. \label{eq:tau_perpcavmf}
\end{equation}
This braking time is shorter than that obtained for a radially decreasing Alfv\'en speed (cf. Eq.~\ref{eq:tau_perpmf}), their ratio being
\begin{equation}
\frac{\tau_{\perp}}{\tau_{\perp,cv_{\rm A}}} \sim \frac{1}{2}\left(\frac{\rho_c}{\rho_{\rm ext}}\right)^{1/4}.
\end{equation}
  
\subsubsection{Comparison of timescales}

in the preceding sections, we have derived four characteristic magnetic-braking timescales: two for aligned rotators consisting of one with straight field lines, $\tau_{\parallel},$ and one with field lines that are fanning-out, $\tau_{\parallel,fo}$, and two for perpendicular rotators, namely one for a radially decreasing Alfv\'en speed, $\tau_{\perp}$, and one for a constant Alfv\'en speed, $\tau_{\perp,cv_{\rm A}}$. Comparing the braking timescales given in Eq.~\ref{eq:tau_classic} and \ref{eq:tau_perpmf} gives 
\begin{equation}
\frac{\tau_{\parallel}}{\tau_{\perp}} = \frac{1}{2}\left(\frac{\rho_c}{\rho_{\rm ext}}\right)^{1/2} \label{comp1}
\end{equation}
and since $\rho_c \gg \rho_{\rm ext}$, this leads to the conclusion that the magnetic braking is more efficient in the perpendicular case than the aligned one \citep{Mouschovias79}\footnote{Although $R$ and $Z$ do not appear explicitly in this expression, we recall that the definitions of the magnetic flux are not the same in both cases. A more correct expression should include a factor $\Phi_{\parallel}/\Phi_{\perp}$.}. This conclusion may apply to prestellar cores whose density is not centrally condensed, and for which the fanning-out is weak. It probably applies to the outer part of the cores, making it quite possible that during the prestellar phase, before the collapse occurs, magnetic field and angular momentum may become, to some extent, aligned. Owing to the turbulent motions in the ISM, it is not unlikely however that the cores have a misalignment between the rotation axis and the magnetic field. But this conclusion that the magnetic braking is more efficient in the perpendicular configuration does not apply to the internal part of the collapsing cores where magnetic field lines are strongly squeezed toward the center.

To be more quantitative, we consider the external medium to be the core's envelope, and estimate the average $\rho_{\rm ext}$ to be a few times the density of the singular isothermal sphere, $\rho_{\rm ext}(R_0) \propto R_0^{-2}$, and similarly for the core density $\rho_c(R_c) \propto R_c^{-2}$. Using Eqs.~(\ref{eq:tau_alfomf}) and (\ref{eq:tau_perpmf}), gives for the ratio of timescales 
\begin{equation}
\frac{\tau_{\parallel,fo}}{\tau_{\perp}} = \frac{R_c}{R_0}.
\end{equation}
Since $R_c/R_0 \ll 1$, the angular momentum is more efficiently transferred to the envelope in an aligned rotator than in a perpendicular one. This is still so when considering the more realistic case of a perpendicular rotator with a constant Alfv\'en speed. Using Eq.~(\ref{eq:tau_alfomf}) and (\ref{eq:tau_perpcavmf}), the ratio of the timescales is
\begin{equation}
\frac{\tau_{\parallel,fo}}{\tau_{\perp,cv_{\rm A}}} = \left(\frac{R_c}{R_0}\right)^{1/2}.
\end{equation}
Evidently, the previous conclusion still holds in this case; magnetic braking is more efficient in an aligned configuration than in a perpendicular one, although the difference is smaller. As both configurations somehow represent extreme cases, we expect our simulations to have properties that are in-between.

To conclude, there are four magnetic-braking timescale of interest, which have the following ordering $\tau_{\parallel} > \tau_{\perp} > \tau_{\perp,cv_{\rm A}} > \tau_{\parallel,fo}$. The first two inequalities hold in (non-collapsing) prestellar cores, whose density is not centrally condensed and fanning-out is weak. However, for conditions that are more appropriate to collapsing prestellar cores, it is the last inequality that is appropriate, and aligned rotators are more efficiently braked than perpendicular ones.

\section{Numerical setup and initial conditions} 

\subsection{Numerical setup} 

We perform three-dimensional (3D) numerical simulations with the AMR code \textsc{Ramses}~\citep{Teyssier02, Fromang06}. \textsc{Ramses} can treat ideal MHD problems with self-gravity and cooling. The magnetic field evolves using the constrained transport method, preserving the nullity of the divergence of the magnetic field. The high resolution needed to investigate the problem is provided by the AMR scheme. Our simulations are performed using the HLLD solver \citep{Miyoshi05}. 

The calculations start with 128$^{3}$ grid cells. As the collapse proceeds, new cells are introduced to ensure the Jeans length with at least ten cells. Altogether, we typically use 8 AMR levels during the calculation, providing a maximum spatial resolution of $\sim$ 0.5~AU. 

\subsection{Initial conditions} 

We consider simple initial conditions consisting of a spherical cloud of 1 $M_{\odot}$. The density profile of the initial cloud
\begin{equation*} 
\rho = \frac{\rho_0}{1 + (r/r_0)^2}, 
\end{equation*} 
where $\rho_0$ is the central density and $r_0$ the initial radius of the spherical cloud is in accordance with observations \citep{Andre00,Belloche02}. The ratio of the thermal to gravitational energy is about 0.25, whereas the ratio of the rotational to gravitational energy $\beta$ is about 0.03. We run 17 simulations. Various magnetization cases are studied: $\mu =$ 2, 3, and 5 (magnetized super-critical cloud, in agreement with observations, as pointed out in the introduction) and 17 (very super-critical cloud). The angle between the initial magnetic field and the initial rotation axis $\alpha$ is taken to be between 0 and 90$^\circ$. Table \ref{tab:simu} lists all the simulation parameters. 

To avoid the formation of a singularity and mimic that at high density the gas becomes opaque i.e. nearly adiabatic, we use a barotropic equation of state
\begin{equation*} 
\frac{P}{\rho} = c_s^2 = c_{s,0}^2\left[1 + \left(\frac{\rho}{\rho_{\rm ad}}\right)^{2/3}\right],
\end{equation*} 
where $\rho_{\rm ad}$ is the critical density over which the gas becomes adiabatic; we assume that $\rho_{\rm ad} = 10^{-13}$ g cm$^{-3}$. When $\rho > \rho_{\rm ad}$, the adiabatic index $\gamma$ is therefore equal to 5/3, which corresponds to an adiabatic mono-atomic gas. At lower density, when the gas is isothermal, $P/\rho$ is constant, with $c_{s,0} \sim 0.2$ km s$^{-1}$. The corresponding free-fall time is $t_{\rm ff} \sim 12$~kyr (for an initial density peak of about $3 \times 10^{-17}$ g cm$^{-3}$).

\begin{table} 
 \centering  
 \begin{tabular}{c c || c c}   
  \hline   
  \hline   
  $\mu$ & $\alpha$ & $\mu$ & $\alpha$ \\   
  \hline    
  17 & 0  &  3 & 0 \\
     & 45 &    & 20 \\
     & 90 &    & 45 \\ 
     &    &    & 90 \\ 
  \hline    
  5 & 0  &  2 & 0 \\
    & 20 &    & 20 \\ 
    & 45 &    & 45 \\ 
    & 70 &    & 80 \\
    & 80 &    & 90 \\ 
    & 90 &    &  \\   
  \hline   
  \hline  
 \end{tabular}  
 \caption{List of performed calculations}  
 \label{tab:simu} 
\end{table}

\section{Transport of angular momentum}

We analyze in detail the transport of angular momentum in our numerical simulations.

\subsection{Temporal evolution}

\begin{figure*}
\subfigure[\label{img:01am}]{\includegraphics[width = 1\textwidth]{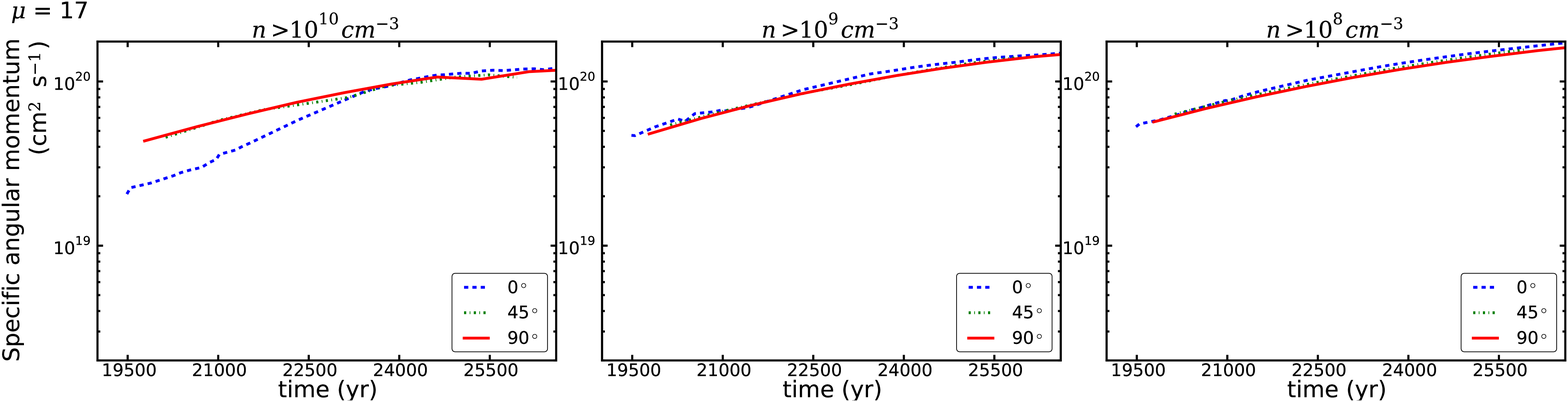}}
\subfigure[\label{img:03am}]{\includegraphics[width = 1\textwidth]{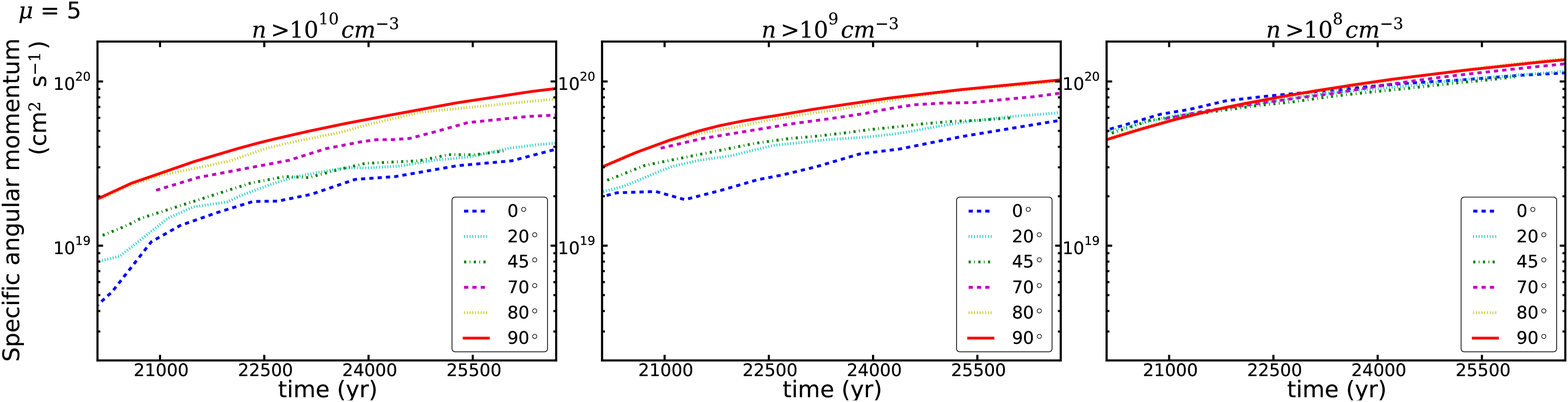}}
\subfigure[\label{img:05am}]{\includegraphics[width = 1\textwidth]{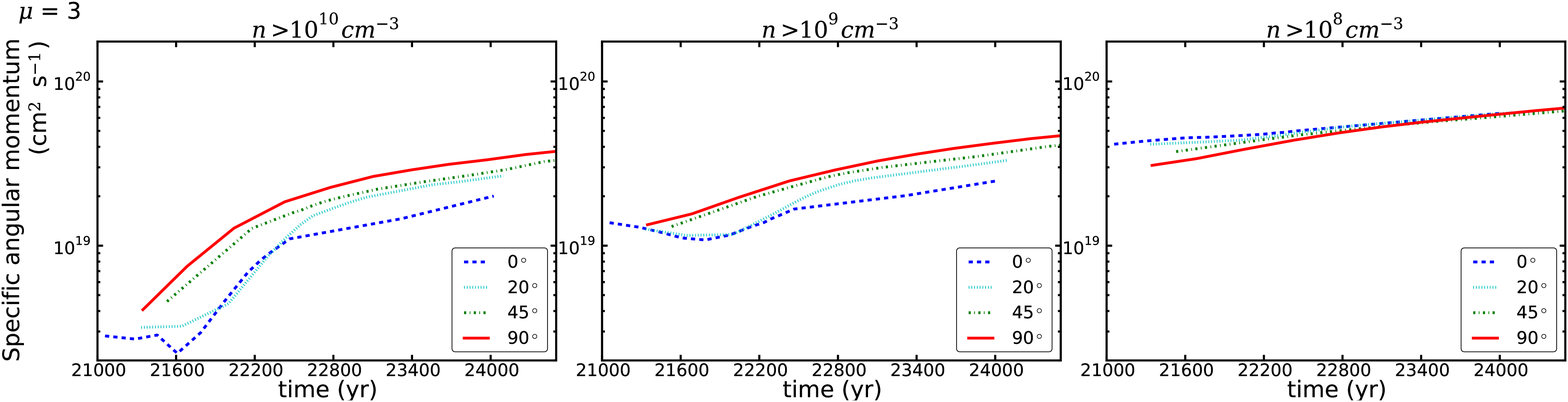}}
\subfigure[\label{img:07am}]{\includegraphics[width = 1\textwidth]{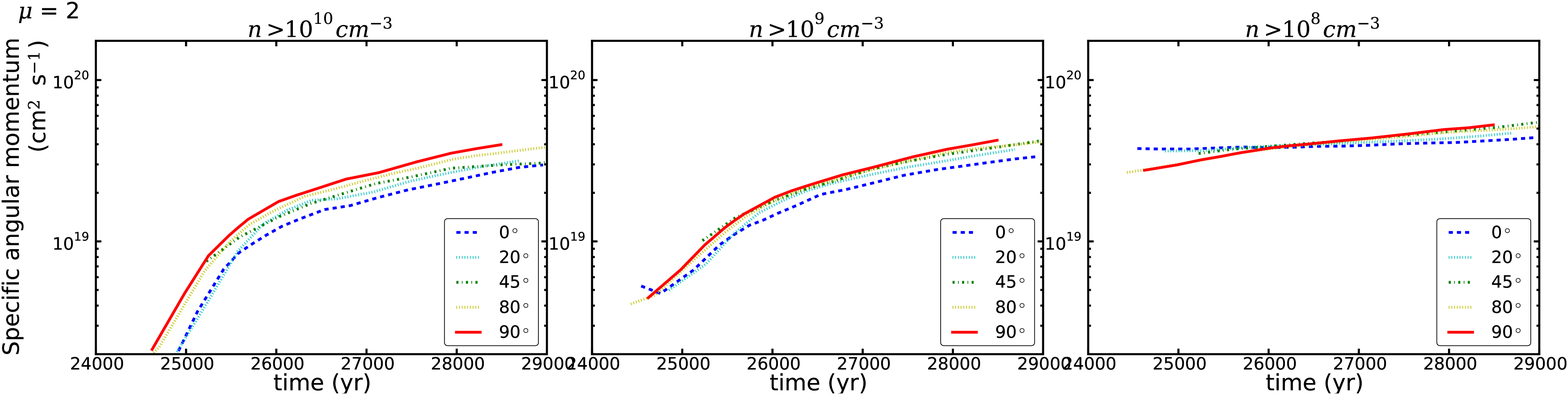}}
\caption{Evolution of the specific angular momentum $\frac{1}{M}\left(\int_{\rho > \rho_{\rm c}} {\bf r} \times \rho{\bf v} \ {\rm d}V\right)$ for $\mu = 17$ (Fig. \ref{img:01am}), $\mu = 5$ (Fig. \ref{img:03am}), $\mu = 3$ (Fig. \ref{img:05am}), and $\mu = 2$ (Fig. \ref{img:07am}), for three different density thresholds $\rho_{\rm cr}$ that correspond to: $n > 10^{10}$ cm$^{-3}$, $n > 10^{9}$ cm$^{-3}$, and $n > 10^{8}$ cm$^{-3}$.}
\label{img:am}
\end{figure*}

\begin{figure*}
\includegraphics[width = 1\textwidth]{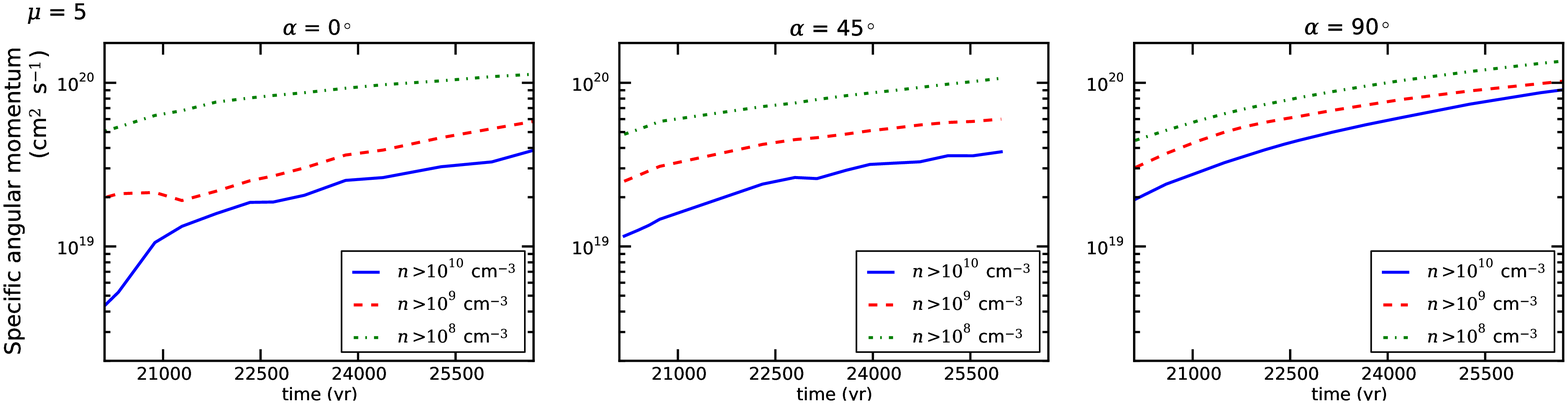}
\caption{Evolution of the specific angular momentum $\frac{1}{M}\left(\int_{\rho > \rho_{\rm cr}} {\bf r} \times \rho{\bf v} \ {\rm d}V\right)$ for $\mu = 5$, three orientations ($\alpha = 0, 45$ and 90$^{\circ}$; respectively left, central and right panel), and three density thresholds $\rho_{\rm cr}$ corresponding to $n > 10^{10}$ cm$^{-3}$, $n > 10^{9}$ cm$^{-3}$, and $n > 10^{8}$ cm$^{-3}$.}
\label{img:angmom5}
\end{figure*}

We begin by describing the temporal dependence of the angular momentum, in particular the specific angular momentum, which is defined by 
\begin{equation}
{\bf J}=\frac{1}{M}\int_{V}{\bf r}\times\rho{\bf v}\ {\rm d}V,
\end{equation}
where $M$ is the mass contained within the volume $V$, ${\bf r}$ the position (with respect to the center of mass), $\rho$ the mass density, and ${\bf v}$ the velocity. In general, we compute three values of ${\bf J}$ using three density thresholds corresponding roughly to the adiabatic core ($n > 10^{10}$~cm$^{-3}$), the disk ($n > 10^{9}$~cm$^{-3}$), and the densest parts of the envelope ($n > 10^{8}$~cm$^{-3}$). These are nested structures: the adiabatic core is embedded in the structure described by $n > 10^{8}$ cm$^{-3}$. Figure \ref{img:am} displays the norm of the specific angular momentum $|{\bf J}|$ for all considered orientations, magnetizations ($\mu=17,\,5,\,3$) and density thresholds ($n > 10^8,\,10^9,\,10^{10}$ cm$^{-3}$). Figure \ref{img:angmom5} displays $|{\bf J}|$ for all the density thresholds for $\mu = 5$ and three different orientations ($\alpha = 0,\,45,\,90^{\circ}$).

Figure \ref{img:angmom5} illustrates that, as expected, the angular momentum increases with decreasing densities, which correspond to the outer regions of the collapsing prestellar core. 

Figure \ref{img:am} shows that, as matter is continuously accreted, the angular momentum increases with time, and it is smaller for larger magnetizations, an indication that magnetic braking is more efficient in transporting angular momentum from the inner to the outer parts of the prestellar core.

As for the dependence on the angle $\alpha$, there are several interesting aspects worth discussing. First of all, for the disk ($n > 10^{9}$~cm$^{-3}$) and adiabatic core ($n > 10^{10}$~cm$^{-3}$) the angular momentum increases with $\alpha$ (see Fig. \ref{img:am}). Conversely, magnetic braking rapidly decreases with $\alpha$, which is consistent with the prediction that the magnetic braking timescale for a perpendicular rotator is longer than for an aligned one (see section \ref{sec:anal}). For the disk region, the angular momentum in the perpendicular case is indeed almost three times larger than in the aligned one ($\mu = 5$). This proportion decreases with the magnetization: it is a factor of two when $\mu = 3$, and for $\mu = 2$ the values are comparable, albeit the angular momentum still increases slightly with the angle $\alpha$. Therefore, in misaligned rotators and for intermediate magnetizations, more angular momentum will be available to ``build'' centrifugally supported disks.

Interestingly for $\mu=5$ and 3, the angular momentum below a density of $10^{8}$ cm$^{-3}$ is independent of $\alpha$, which suggests that its efficient transport only occurs in the highest density regions ($n \gtrsim 10^{8}$ cm$^{-3}$). This corresponds therefore to a ``braking region''. This is no longer true for $\mu = 2$: magnetic braking occurs earlier, simply because the magnetic field is stronger.

\subsection{General considerations}\label{subsec:transGC}

The azimuthal component of the conservation of angular momentum, in cylindrical coordinates, is the starting point of our analysis. In conservative form, it is given by  
\begin{eqnarray}
\partial_{t}\left(\rho rv_{\phi}\right) & + & \nabla \cdot r\left[\rho v_{\phi}{\bf v} + \left(P + \frac{B^2}{8\pi}-\frac{g^2}{8 \pi G}\right){\bf e_{\phi}}\right. \nonumber \\
 & - & \left.\frac{B_{\phi}}{4\pi}{\bf B} + \frac{g_{\phi}}{4\pi G}{\bf g}\right] = 0, \label{euler}
\end{eqnarray}
where $\rho$ is the density, ${\bf v}$ the velocity, $P$ the gas pressure, ${\bf B}$ the magnetic field, and the gravitational acceleration ${\bf g} = \nabla \Phi$, where $\Phi$ is the gravitational potential. For the sake of completeness, its derivation is detailed in the appendix. 

The corresponding fluxes of angular momentum in this equation are $r\rho v_{\phi}{\bf v}$ for the mass flow, $rB_{\phi}{\bf B}/4\pi$ for the magnetic field, and $rg_{\phi}{\bf g}/4\pi$ for the gravitational field; they represent the contribution of each of those processes to the transport of angular momentum. In general, magnetic braking can be efficient in both the vertical and radial directions. However, the outflows remove angular momentum mainly in the vertical direction (mass accretion occurs instead in the radial direction, carrying in angular momentum). The gravitational transport of angular momentum is most efficient in the radial direction, because of the spiral arms that develop around the first core.

The pressure terms $(P + B^2/8\pi - g^2/8\pi G)$ do not contribute significantly to the transport of angular momentum. In the following, to quantify the contribution of each of these processes, these fluxes are considered and compared one by one. One should obviously add all these terms to obtain the total angular momentum transport.
\newline

To calculate the fluxes defined above, one needs to define the main axis of a cylindrical frame of reference. Two different choices have been made below: we use either the inertia matrix or simply the rotation axis of the system.

The pseudo-disk is defined as all matter with a particle number density $n>10^{7}$ cm$^{-3}$. We then calculate the inertia matrix over the volume $V$ of the pseudo-disk as 
\begin{equation}
I_{ij}=\int_{V}\rho r_{i}r_{j}\ dV,
\end{equation}
where $\rho$ is the density, and $r_{i}$ the coordinates (with $i,j\in\{1,2,3\}$, so $\{r_{1},r_{2},r_{3}\}=\{x,y,z\}$) of a fluid element with respect to the center of mass. The eigenvectors of this matrix are the axis of the frame of the pseudo-disk, which we call $\mathcal{R}_{p}$ in what follows. The z-axis of the frame $\mathcal{R}_{p}$ is defined as the eigenvector associated with the minimum eigenvalue of the matrix. It corresponds to the z-axis of the cylindrical coordinates in the following analysis. As the pseudo-disk is essentially perpendicular to the magnetic field, the z-axis of this frame is close to the direction of the actual ${\bf B}$.

The fluxes defined above are then computed on surfaces at the edge of the pseudo-disk. Annuli are then defined to fit its surface: we consider a set of one hundred annuli of radius $r$, between 0 and $R_{0}$, height $h(r)=r/4$, and thickness $R_{0}/100$, where $R_0$ is about 1000~AU. The fluxes are computed on the cells belonging to these annuli.

We note that we also tried to estimate the actual value of $h(r)$, as a function of radius, although we found that the surface integrated fluxes did not change significantly. In addition, we also considered a frame whose main axis is the rotation axis of the pseudo-disk instead of the main axis of the inertia matrix, and our conclusions remain again qualitatively unchanged.

For the disk, we simply choose for the cylinder axis, the rotation axis of gas denser than $10^{10}$~cm$^{-3}$, which corresponds to the rotation axis of the core itself. We refer later to this frame as $\mathcal{R}_{d}$. We define annuli as previously, but restrict the analysis to a maximum radius of 400~AU since disks are not larger.

The integrated vertical and radial fluxes of angular momentum transported by the magnetic field are then defined as 
\begin{eqnarray*}
F_{v}^{B}(R)=\\
\left|\int_{0}^{2\pi}\int_{0}^{R}\right. &  & \left.r\frac{B_{\phi}(r,\phi,\pm h(r)/2)B_{z}(r,\phi,\pm h(r)/2)}{4\pi}r{\rm d}r{\rm d}\phi\right|,
\end{eqnarray*}
 and 
\begin{eqnarray}
F_{r}^{B}(R)=\\
\left|\int_{0}^{2\pi}\int_{-h(R)/2}^{h(R)/2}\right. &  & \left.R\frac{B_{\phi}(R,\phi,z)B_{r}(R,\phi,z)}{4\pi}R{\rm d}z{\rm d}\phi\right|,\nonumber 
\end{eqnarray}
where $B_i(R,\phi,z) \equiv B_i(r=R,\phi,z)$. We note that $F_v^B$ is the sum of the fluxes through the faces defined by $h(r)/2$ and $-h(r)/2$. To compare the various cases and time-steps, it seems appropriate to consider specific quantities. In the following, we study $F_{r}^{B}/M$ and $F_{v}^{B}/M$, where $M$ is the mass enclosed in the volume of interest. We do the same for the outflows and the gravity terms.

\subsection{Transport of angular momentum in the envelope}

We first investigate the transport of angular momentum in the envelope (see section \ref{sec:collapse}), focusing on the magnetic braking. As we show below, this is the most efficient means of transporting angular momentum, in particular in the strongly magnetized clouds. We work here in the frame $\mathcal{R}_p$.

\begin{figure*}
\subfigure[\label{img:0300mag}]{\includegraphics[width = 1\textwidth]{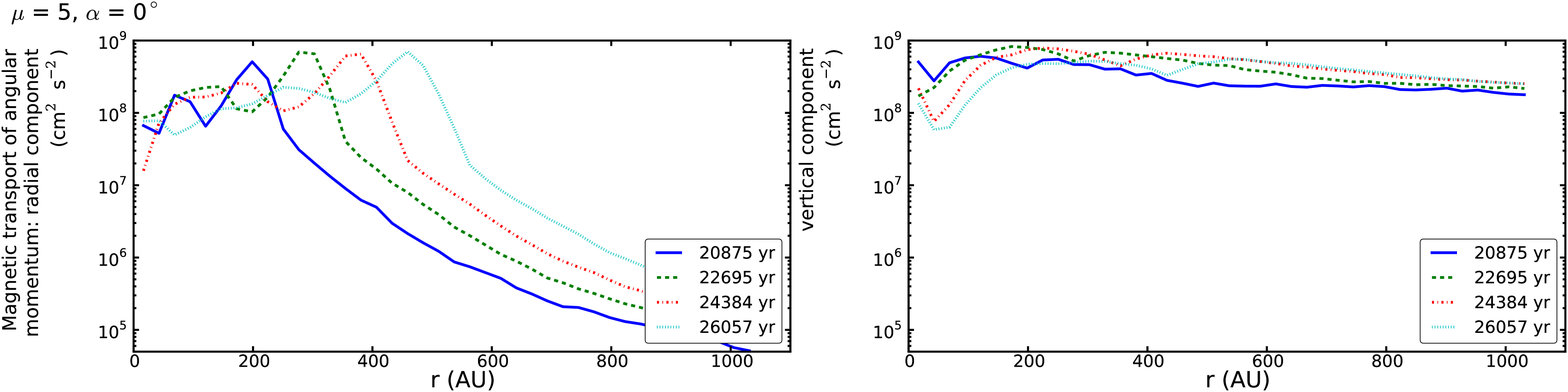}}
\subfigure[\label{img:0308mag}]{\includegraphics[width = 1\textwidth]{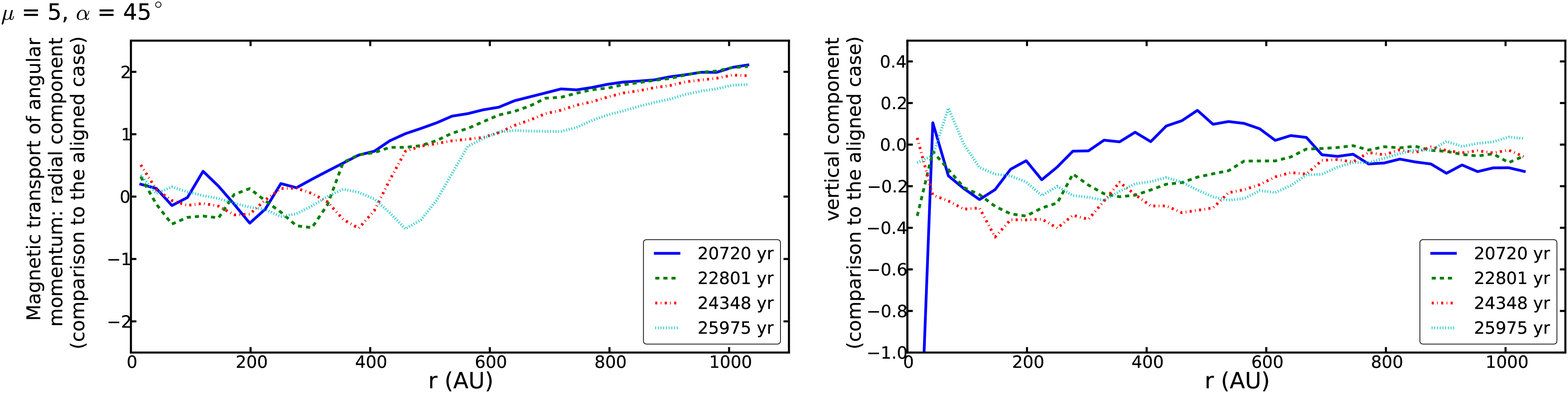}}
\subfigure[\label{img:0316mag}]{\includegraphics[width = 1\textwidth]{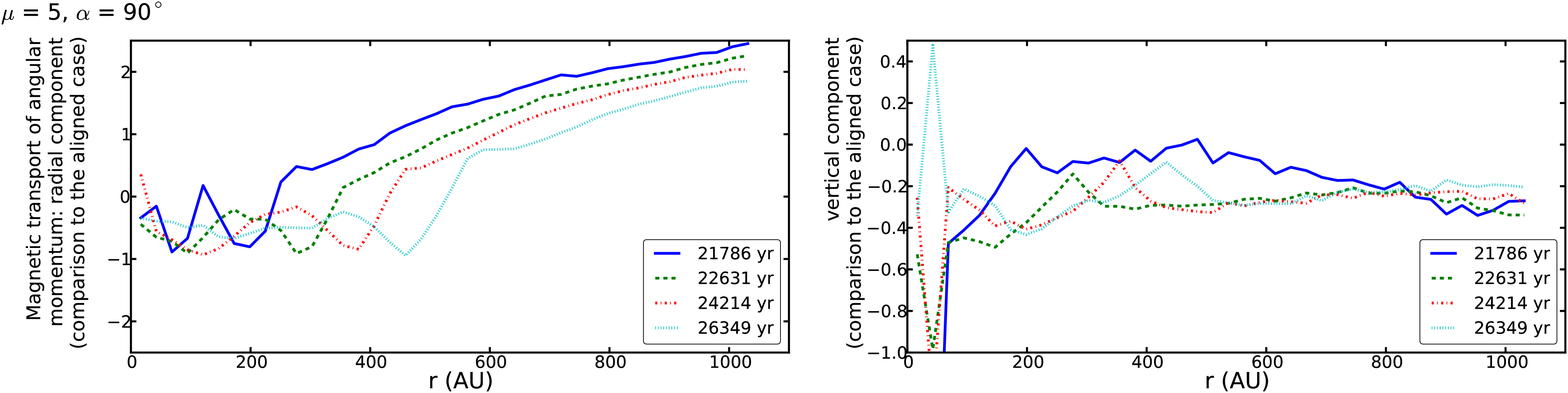}}
\caption{Magnetic transport of angular momentum in logarithmic scale, for $\mu = 5$, $\alpha = 0^{\circ}$ (Fig. \ref{img:0300mag}), $\alpha = 45^{\circ}$ (Fig. \ref{img:0308mag}), and $\alpha = 90^{\circ}$ (Fig. \ref{img:0316mag}). Figure \ref{img:0300mag} represents $\log(F_r^B/M)$ (left panel) and $\log(F_v^B/M)$ (right panel) for four time-steps. Figures \ref{img:0308mag} and \ref{img:0316mag} display $\log((F_v^B/M)/(F_{v,0}^B/M_0))$ and $\log((F_r^B/M)/(F_{r,0}^B/M_0))$ where $F_{v,0}^B/M_0$ and $F_{r,0}^B/M_0$ correspond to the aligned case, at corresponding timesteps.}
\label{img:mag}
\end{figure*}

Figure \ref{img:0300mag} shows the evolution of specific radial flux ($F_r^B/M$) in the left panel, and specific vertical flux ($F_v^B/M$) in the right panel, at four different time-steps, for $\mu = 5, \alpha = 0^{\circ}$.

The two panels of Fig.~\ref{img:0300mag} show that the magnetic braking depends on the radius: it is more efficient in the inner region of the envelope than in the outer region. In the outer part of the cloud, both components decrease with the radius $r$. In particular, $B_r$ and $B_{\phi}$ drastically decrease outside the cavity of the outflows (see section \ref{sec:collapse}) since the twisting of the magnetic field lines -- which generates both the radial and azimuthal components of the magnetic field -- occurs essentially inside the cavity of the outflows.

Figure~\ref{img:0300mag} shows that the vertical component of the magnetic braking is larger than the radial component (by about one order of magnitude in the inner part of the envelope). The ratio of those two components increases with the radius, since the radial component of the magnetic field is almost zero at large radius.

Figures \ref{img:0308mag} and \ref{img:0316mag} display a comparison of the radial (left panel) and vertical (right panel) components of the magnetic braking for $\alpha = 45^{\circ}$ and $90^{\circ}$ with the components in the aligned case. The left panels of Fig. \ref{img:0308mag} and \ref{img:0316mag} show that the magnetic braking in the radial direction is less efficient in the tilted cases than in the aligned case in the inner part of the envelope by a factor $\sim 2 - 10$ (especially for $\alpha = 90^{\circ}$). The limit of the cavity is shown by the sharp increase in the ratio (around $\simeq$ 200~AU for the first time-step, and $\simeq$ 500~AU for the later time-step). Outside the cavity, the ratio increases because $B_r$ decreases more dramatically in the aligned case than in the misaligned cases; this is because the radial component vanishes initially in the aligned case whereas there is an initial $B_r$ in the misaligned cases.

The right panels of Fig. \ref{img:0308mag} and \ref{img:0316mag} show that magnetic braking is less efficient in the vertical direction in the tilted cases than in the aligned case. This is true everywhere by a factor of $\sim2 - 5$. As we see later, the vertical component dominates the transport of angular momentum by the magnetic field, it clearly shows that the magnetic braking is more efficient in the aligned case than in the misaligned cases; this is consistent with our previous analytical analysis.

\subsection{Transport of angular momentum in the disk}

In the region of the disk, magnetic braking, outflows, and gravitational torques all contribute to the transport of angular momentum. We here work in the frame $\mathcal{R}_d$.

\subsubsection{Comparison between magnetic braking and extraction by the outflows}

Outflows are one of the most important tracers of star formation. From the very beginning of protostar evolution to the T Tauri stage, they are thought to be launched by magneto-centrifugal means (\citealp{Blandford82,Pudritz83}, \citealp{Uchida85}), and may play an important role in the efficient transport of angular momentum (\citealp{Bacciotti02}). The early formation of outflows during the collapse of dense cores was investigated recently by 2D and 3D MHD simulations \citep{Mellon08,Hennebelle08a}, and where the second collapse was included (e.g. \citealp{Banerjee06, Machida08}).

\begin{figure*}
\subfigure[\label{img:tamd03}]{\includegraphics[width = 1\textwidth]{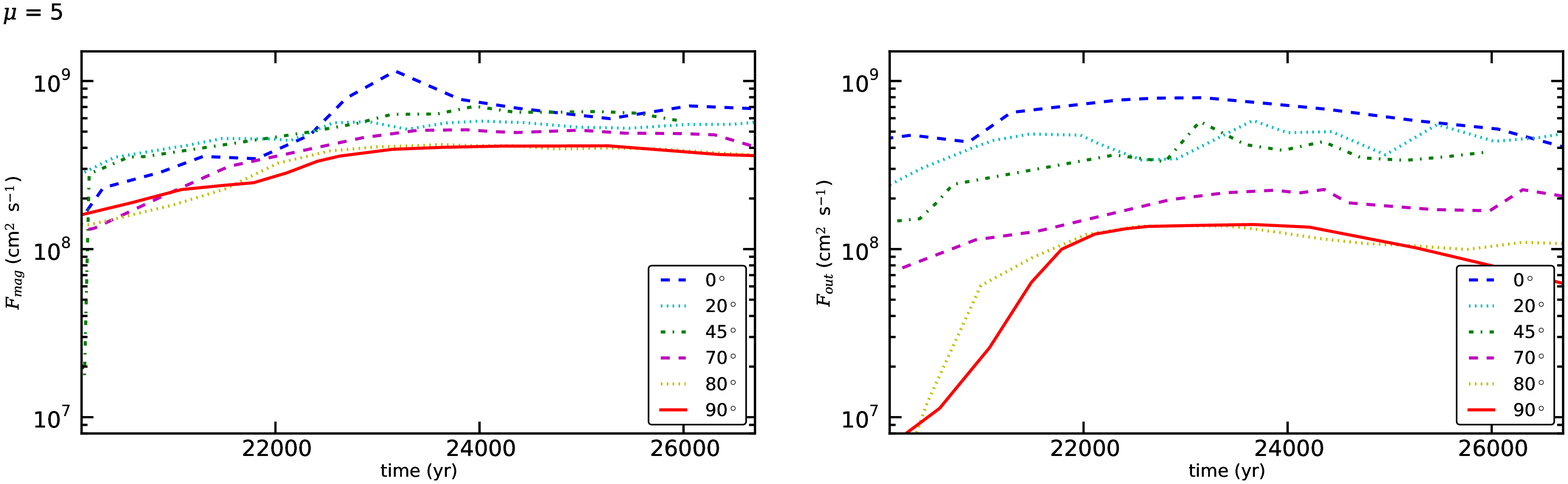}}
\subfigure[\label{img:tamd05}]{\includegraphics[width = 1\textwidth]{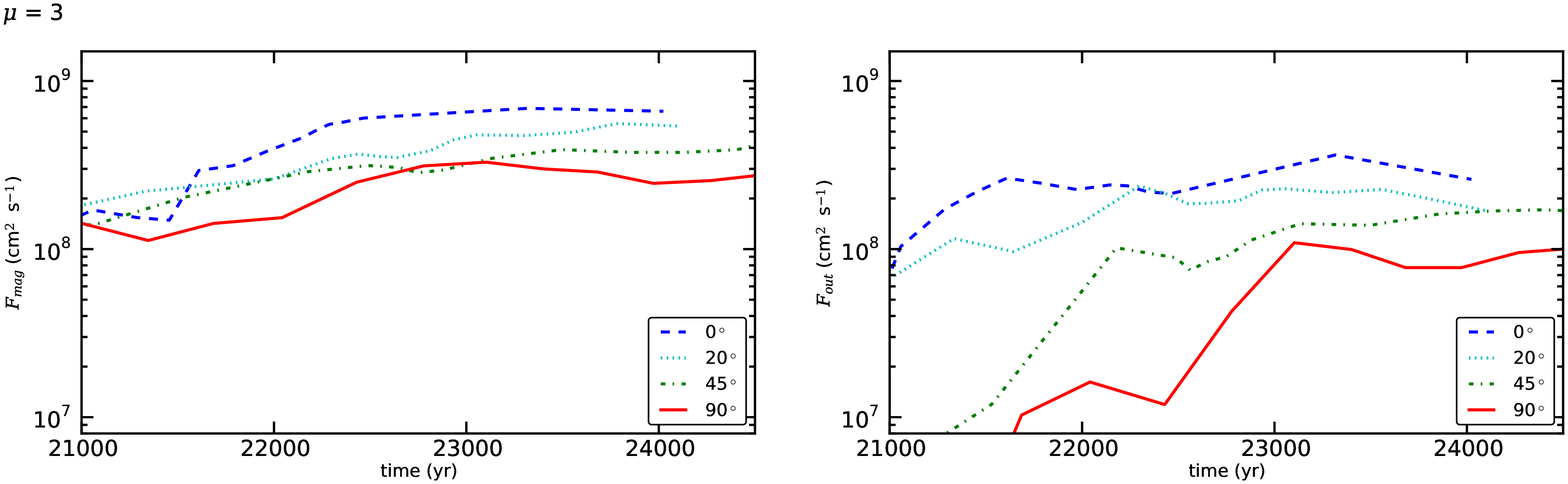}}
\subfigure[\label{img:tamd07}]{\includegraphics[width = 1\textwidth]{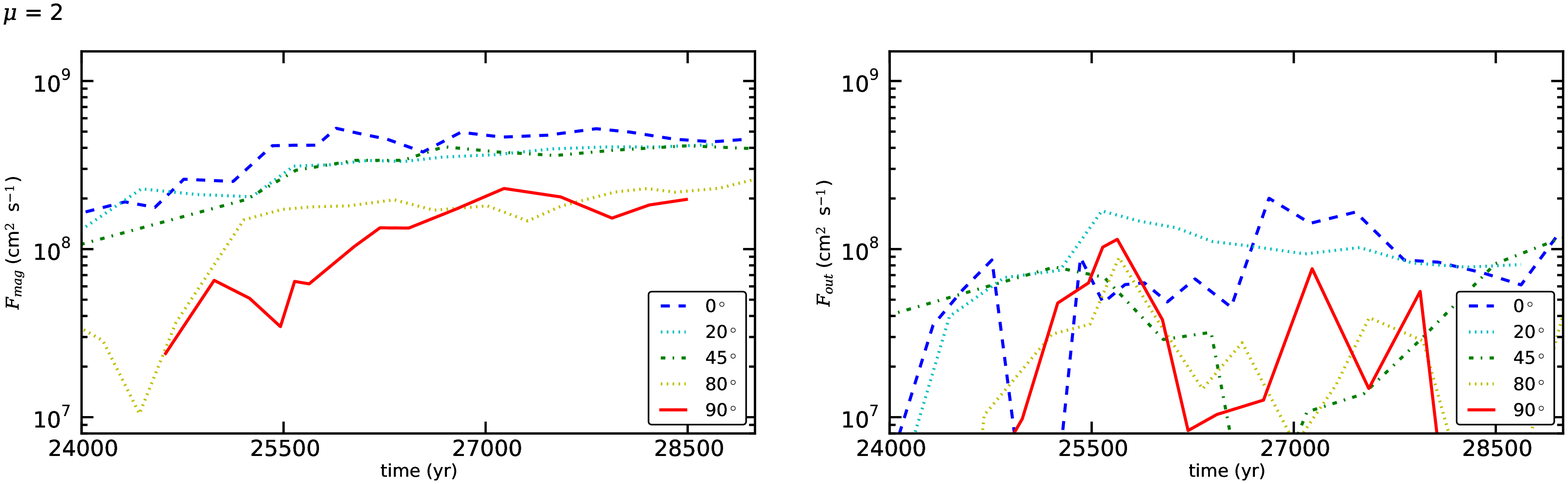}}
\caption{Evolution of angular momentum transported by the magnetic field and by the outflows within a cylinder of radius 300~AU and height 150~AU, for $\mu = 5$ (Fig. \ref{img:tamd03}), $\mu = 3$ (Fig. \ref{img:tamd05}), and $\mu = 2$ (Fig. \ref{img:tamd07}).}
\label{img:tamd}
\end{figure*}

To study the impact of the outflows on the transport of angular momentum, we begin by comparing the integrated flux of angular momentum transported by the magnetic field, $F_{\rm mag}$, with that of the outflows, $F_{\rm out}$, as a function of time. The respective integrals are given by
\begin{equation}
F_{\rm mag} = \left|\int_S r\frac{B_{\phi}}{4\pi}{\bf B}\cdot{\rm d}{\bf S}\right|,
\end{equation}
for the magnetic braking,
\begin{equation}
F_{\rm out} = \left|\int_S \rho rv_{\phi}{\bf v}\cdot{\rm d}{\bf S}\right|, {\bf v}\cdot{\rm d}{\bf S} > 0
\end{equation}
for the outflows, and
\begin{equation}
F_{\rm in} = \left|\int_S \rho rv_{\phi}{\bf v}\cdot{\rm d}{\bf S}\right|, {\bf v}\cdot{\rm d}{\bf S} < 0
\end{equation}
for the accretion flow.

The integrals are taken over the surface $S$ of a cylinder, corresponding approximately to the disk, of radius $R \simeq$ 300~AU and height $h \simeq$ 150~AU, whose axis is taken to be that of rotation. As previously, we study the specific quantities $F_{mag}/M, \, F_{out}/M$, and $F_{in}/M$ in the following. 

Figure \ref{img:tamd} displays the evolution of angular momentum carried away by magnetic braking (left panel), and the outflows (right panel) for $\mu = 5, 3$, and 2, respectively. More precisely, the ratio of this quantity to the total mass enclosed in $S$ is computed. The left panel of Fig.~\ref{img:tamd} shows that more angular momentum is carried away from the central region of the collapsing core for relatively small $\alpha$ (below 70$^{\circ}$) than for larger $\alpha$ (above 70$^{\circ}$). This is the case for all the magnetizations but is particularly evident for lower $\mu$ (Fig. \ref{img:tamd05} and \ref{img:tamd07}). In the right panel of Fig.~\ref{img:tamd03}, we can see that the angular momentum carried away by the outflows is comparable to that transported by the magnetic braking in the aligned case. When the angle $\alpha$ increases, the amount of angular momentum carried away decreases and in the perpendicular case, the total angular momentum transported by the flow is about ten times smaller than in the aligned case. The suppression of the outflows with increasing $\alpha$ is clearly responsible for this decrease. Figures~\ref{img:tamd05} and \ref{img:tamd07} show that this effect is even stronger, since the increasing magnetic intensity (\emph{i.e.} decreasing $\mu$) reduces the strength of the outflows. Less and less momentum is therefore carried away by the outflows with increasing $\alpha$ and decreasing $\mu$.

\begin{figure}[!h]
  \includegraphics[width=.5\textwidth]{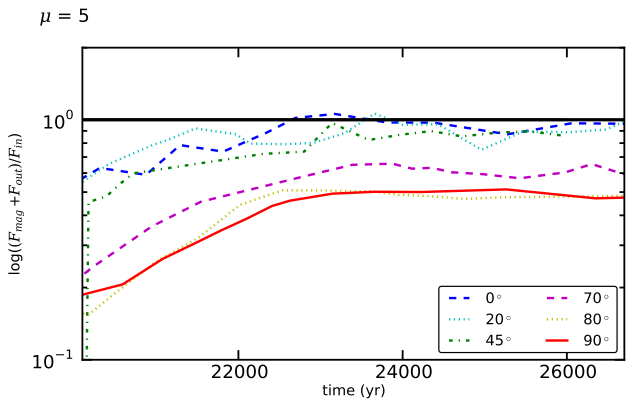}
  \caption{Ratio of the flux of angular momentum transported outward ($F_{mag} + F_{out}$) to the flux of angular momentum transported inward ($F_{in}$), within a cylinder of radius 300~AU and height 150~AU, for $\mu = 5$. The straight line corresponds to $(F_{mag}+F_{out})/F_{in} = 1$.}
  \label{img:am5ratio}
\end{figure}

Figure \ref{img:am5ratio} displays the ratio of the flux of angular momentum transported outward (by the magnetic field, $F_{mag}$, and the ouflows, $F_{out}$) to that transported inward (by the accretion flow, $F_{in}$) for $\mu = 5$. While accretion dominates for $\alpha > 45^{\circ}$, the angular momentum both accreted and expelled are comparable in the other cases. This can be understood by recalling that, in steady-state, Eq.~(\ref{euler}) reduces to
\begin{equation}
\nabla \cdot r\left[\rho v_{\phi}{\bf v} - \frac{B_{\phi}}{4\pi}{\bf B}\right] = 0,
\end{equation}
where the other terms, in particular the gravitional torques, have been neglected (see discussion in section \ref{subsec:gravtrans}). The steady-state condition therefore reduces to $F_{mag} + F_{out} \sim F_{in}$, which is approximately the case for $\alpha \leq 45^{\circ}$.

\begin{figure*}
\subfigure[\label{img:0300mass}]{\includegraphics[width = 1\textwidth]{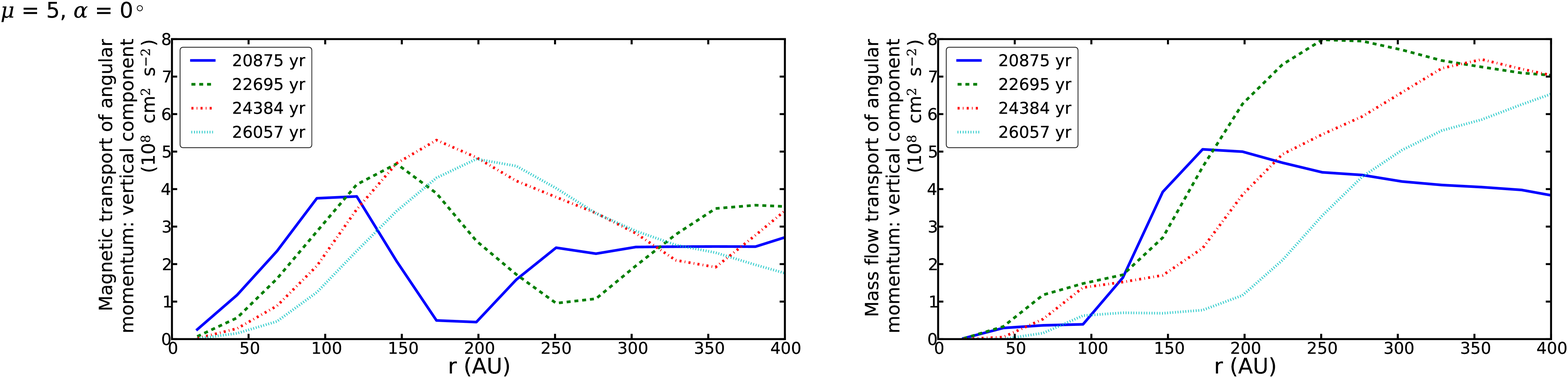}}
\subfigure[\label{img:0308mass}]{\includegraphics[width = 1\textwidth]{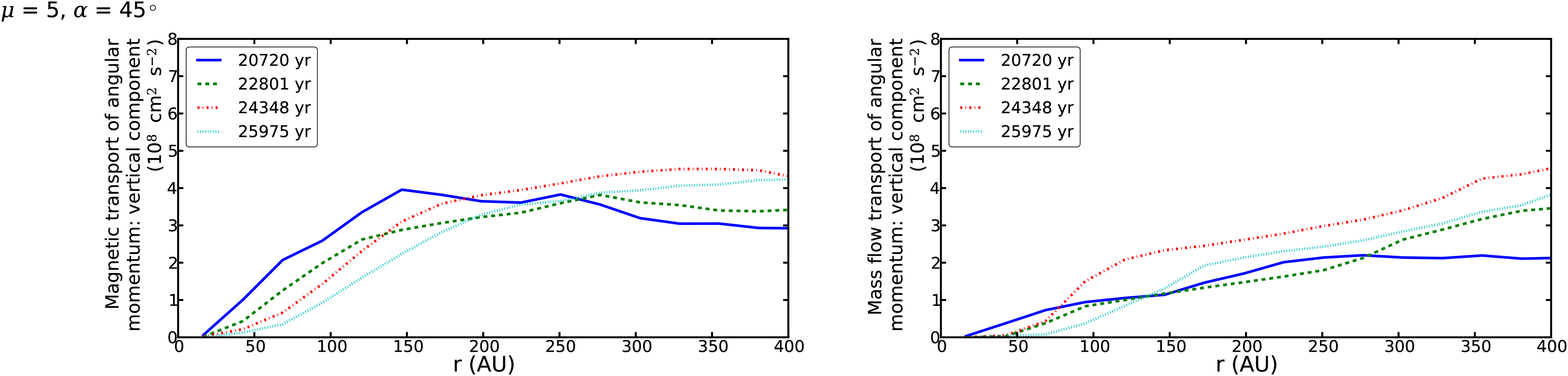}}
\subfigure[\label{img:0316mass}]{\includegraphics[width = 1\textwidth]{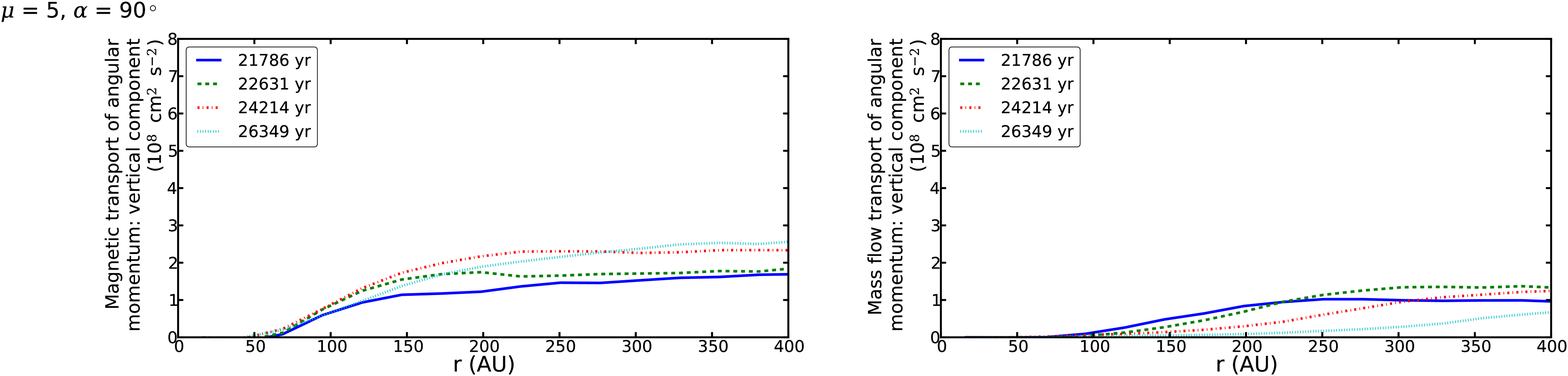}}
\caption{Magnetic transport $\frac{1}{M}\int \ rB_{\phi}{\bf B}/4\pi \cdot d{\bf S}$ and transport by the outflows $\frac{1}{M}\int \ r \rho v_{\phi}{\bf v} \cdot d{\bf S}$, $\mu = 5$, $\alpha = 0^{\circ}$ (Fig. \ref{img:0300mass}), $\alpha = 45^{\circ}$ (Fig. \ref{img:0308mass}), and $\alpha = 90^{\circ}$ (Fig. \ref{img:0316mass}).}
\label{img:03mass}
\end{figure*}

To get a deeper understanding of the impact of the flows on the transport of angular momentum, we look at the spatial distribution of the fluxes and compute the total flux of angular momentum transported by outflows and magnetic braking in concentric cylinders of constant height $H$. Those cylinders are oriented along the rotation axis of the core, since the outflows are approximately aligned with it. We consider only the mass expelled from the core (\emph{i.e.} with a positive vertical velocity) hence only the vertical component of the flux, since mass is accreted mostly along the radial direction. Figure \ref{img:03mass} displays the vertical flux of angular momentum transported by the magnetic field (left panel) and the outflows (right panel), for $\mu = 5$ and three different angles ($0^{\circ}, 45^{\circ}$, and $90^{\circ}$). The integrated flux of the angular momentum carried by the magnetic field (left panel of Fig. \ref{img:0300mass}, \ref{img:0308mass}, and \ref{img:0316mass}) increases with the radius, until the limit of the cavity of the outflows (its radius is from about 150 to 200~AU). There, a local reversal of the magnetic field usually happens, which provokes a local variation in the integrated flux. Outside the cavity, $B_{\phi}$ is close to 0, which means that braking no longer occurs anymore and the integrated flux remains almost constant (the differential flux $\delta F_{\rm mag}$ being close to zero).
 
The integrated flux of angular momentum carried by the outflows (right panel of Fig. \ref{img:0300mass}, \ref{img:0308mass}, and \ref{img:0316mass}) similarly increases with the radius inside the cavity. Outside the cavity, no outflow occurs and the integrated flux therefore remains almost constant. As pointed out in \cite{Ciardi10}, there is almost no outflow in the perpendicular case (Fig.~\ref{img:0316mass}). Thus, almost no angular momentum (or more precisely one order of magnitude less than in the aligned case, Fig.~\ref{img:0300mass}) is transported by the outflows in this configuration. A comparison of the two previous integrated fluxes confirms that magnetic braking is globally more efficient than the outflows in removing angular momentum from the central part of the cloud (within a radius of 100-150~AU, where the ratio is meaningful). 

In the misaligned cases (Fig.~\ref{img:0308mass} and \ref{img:0316mass}), it is also clear that magnetic braking dominates in the central part of the cloud.

\subsubsection{Gravitational transport of angular momentum} \label{subsec:gravtrans}

\begin{figure}[!h]
\subfigure[\label{img:0100grav}]{\includegraphics[width = .5\textwidth]{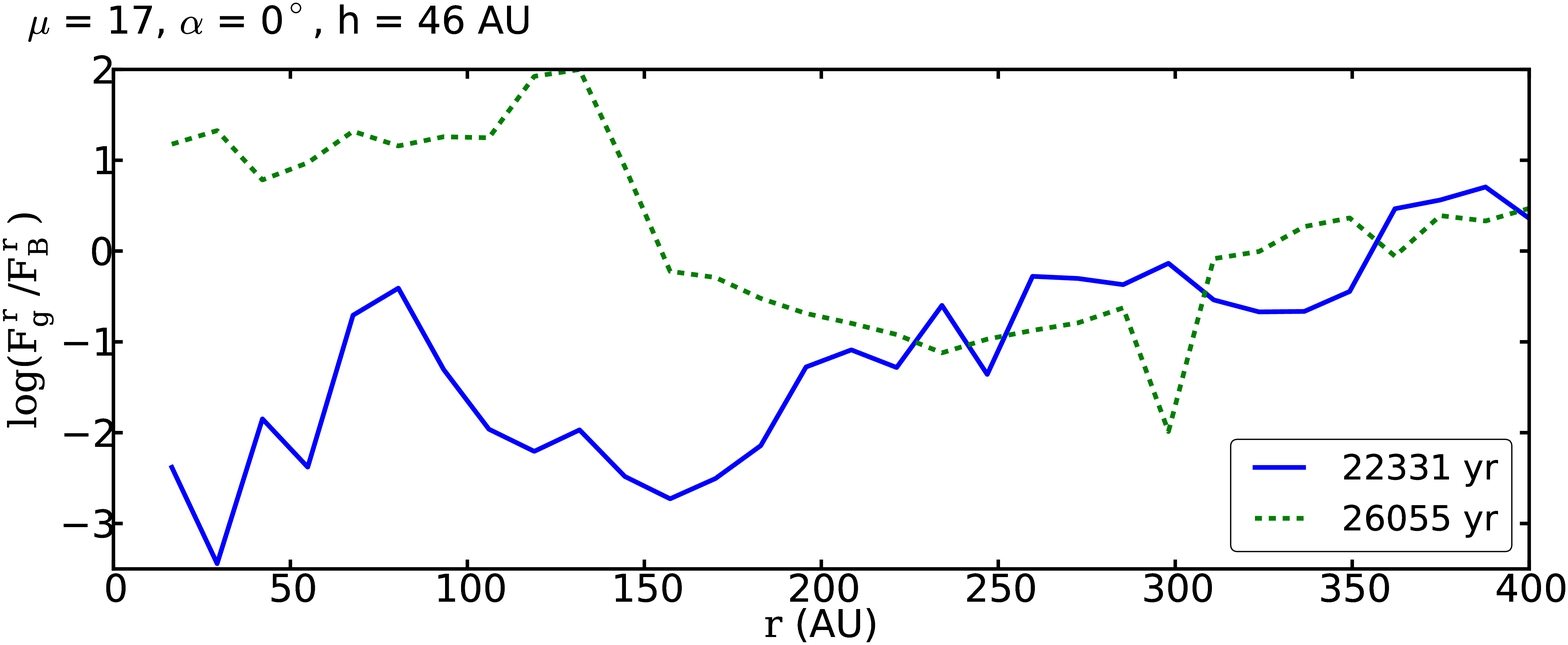}}
\subfigure[\label{img:0108grav}]{\includegraphics[width = .5\textwidth]{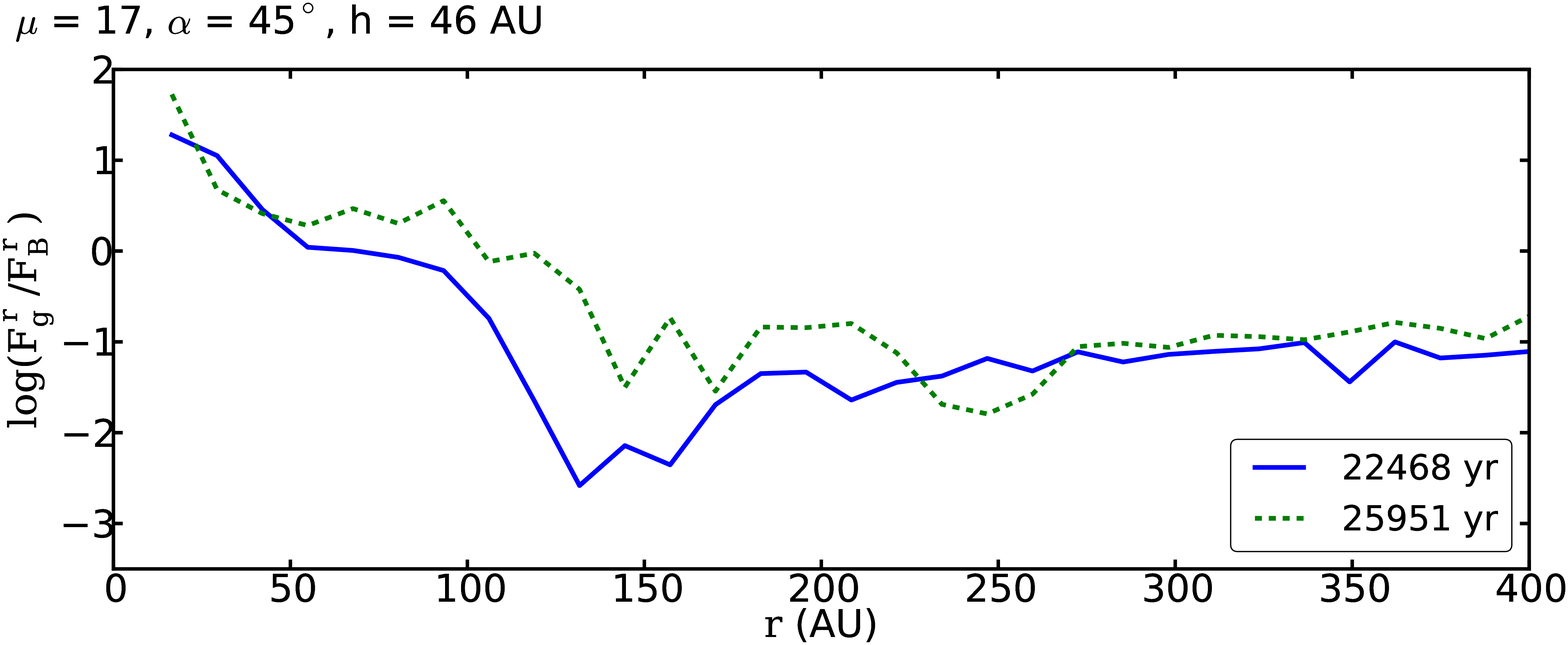}}
\subfigure[\label{img:0116grav}]{\includegraphics[width = .5\textwidth]{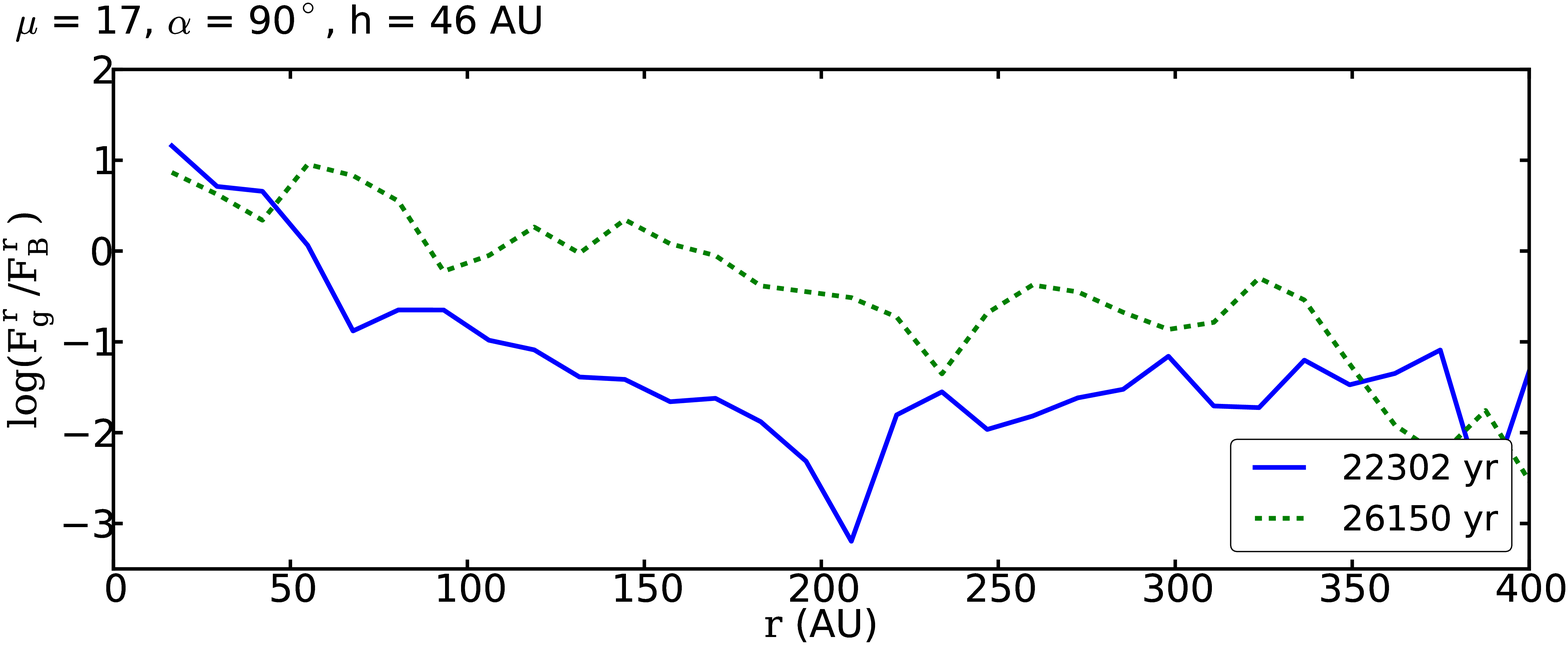}}
\caption{Gravitational transport of angular momentum for $\mu = 17$ and $\alpha = 0, 45$, and $90^{\circ}$, at $h = 46$~AU. It represents the ratio (in logarithmic scale) of the radial component of gravitational ($F_g^r/M$) to magnetic ($F_B^r/M$) transports of angular momentum.}
\label{img:01grav}
\end{figure}

\begin{figure}[!h]
\subfigure[\label{img:0300grav}]{\includegraphics[width = .5\textwidth]{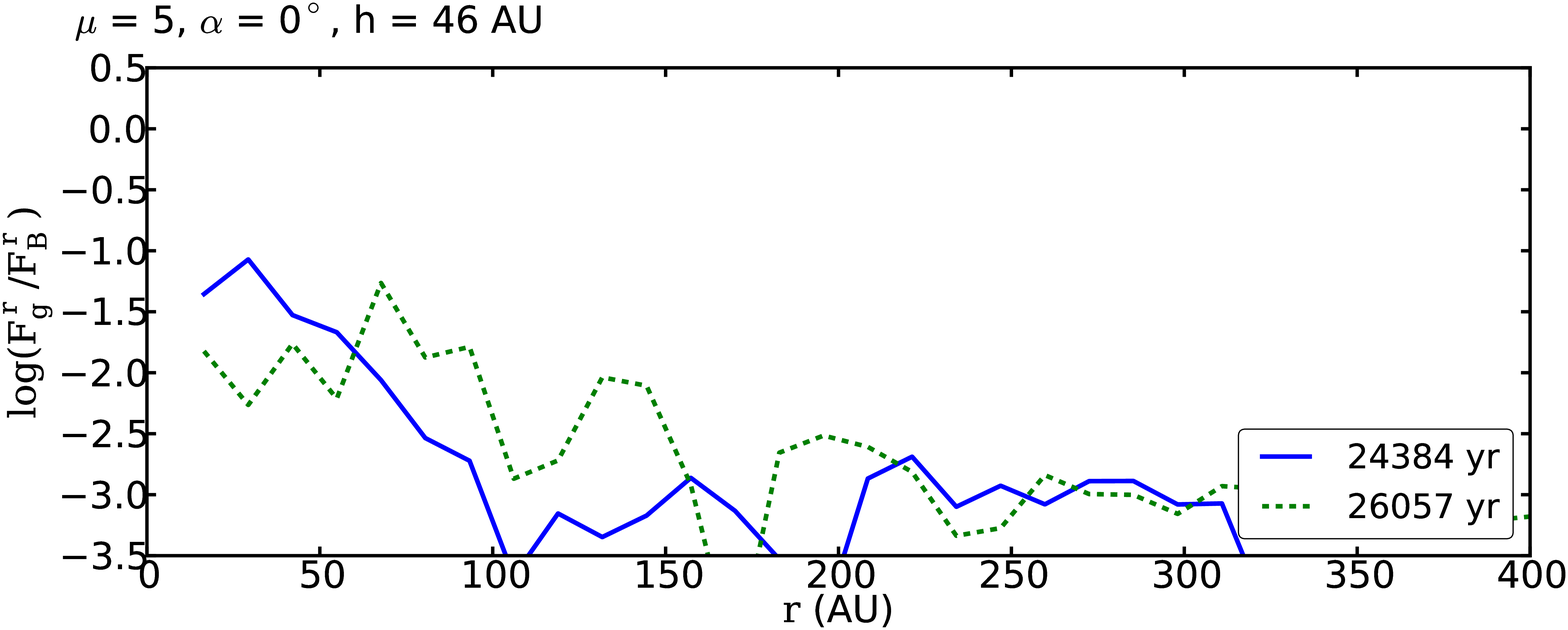}}
\subfigure[\label{img:0308grav}]{\includegraphics[width = .5\textwidth]{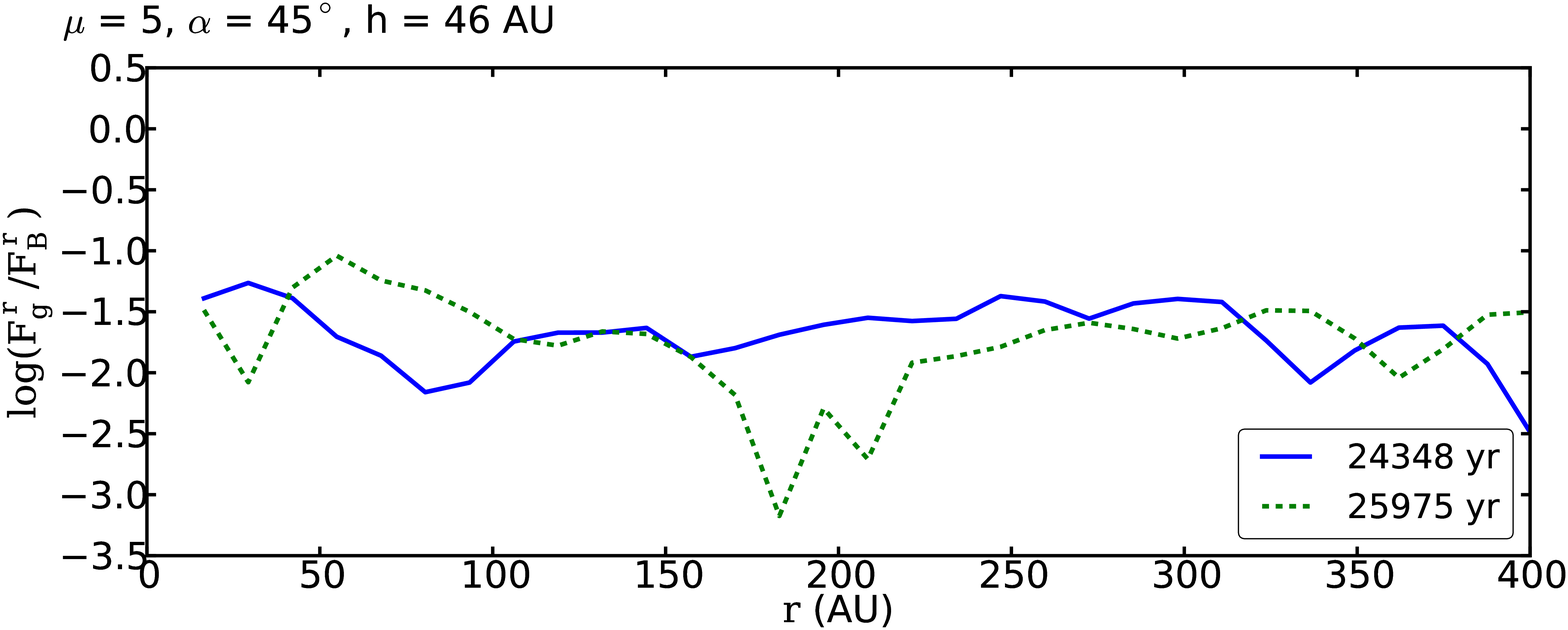}}
\subfigure[\label{img:0316grav}]{\includegraphics[width = .5\textwidth]{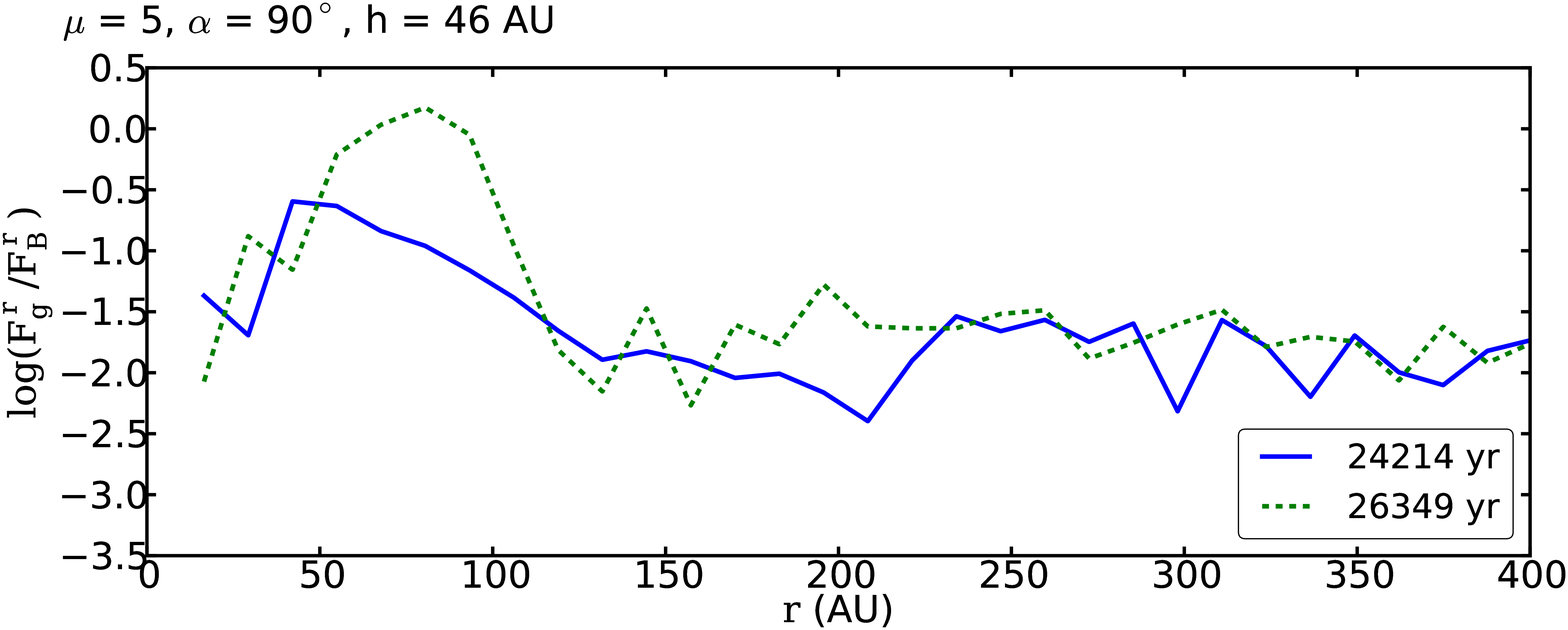}}
\caption{Same as Fig. \ref{img:01grav} but for $\mu = 5$.}
\label{img:03grav}
\end{figure}

The integrated gravitational fluxes of angular momentum are given by
\begin{equation}
F_g^r = \left| \int_{-h/2}^{h/2} \int _0 ^{2 \pi} R \frac{g_{\phi}(r = R, z)g_r(r = R, z)}{4\pi G}2\pi R {\rm d}\phi {\rm d}z \right|.
\end{equation}
As for the other fluxes, we compute the specific quantity $F_g^r/M$. We focus on the radial component of these angular momentum transport processes because the gravitational transport acts mostly in the radial direction.

Figures \ref{img:01grav} and \ref{img:03grav} show a comparison of the magnetic and gravitational braking in the region of the disk for $\mu = 2$ and $\mu = 5$. It represents the evolution of the logarithm of the ratio of their radial components on the surface of cylinders of fixed height ($h = 46$~AU) and a radius between 0 and 400~AU. The height of the cylinder is defined to take into account the whole gravitational transport in the disk. There is no (or negligible) gravitational transport above and below the disk and the typical height of an hydrostatic disk is about 30~AU (see section \ref{subsec:diskshape}); integrating over these cylinders gives the whole gravitational transport in the disk.

The $\mu = 17$ case (Fig. \ref{img:01grav}) corresponds to a quasi-hydrodynamic case, where the magnetic field strength is low. The magnetic braking is less efficient than in higher magnetization cases, and in every configuration, a centrifugally supported disk forms. Therefore, the gravitational transport of angular momentum is always larger (up to ten times larger) than the magnetic transport of angular momentum in the disk (within a radius of 150~AU). Outside the disk, the gravitational transport is weaker (10 to 100 times weaker than the magnetic transport).

Since there is no disk in the aligned case for $\mu = 5$, (Fig. \ref{img:0300grav}), the radial component of the gravitational transport of angular momentum is about 10 to 30 times weaker than the radial component of the magnetic braking under 150~AU and around 500 times weaker for larger radii. In the misaligned cases, for $\mu = 5$ (Fig. \ref{img:0308grav} \& \ref{img:0316grav}), the gravitational contribution to the transport of angular momentum gradually increases: it is 10 times weaker than the magnetic transport for $\alpha = 45^{\circ}$ and of the same order of magnitude to 3 times weaker in the perpendicular case, for radius $r < 100$~AU. For $r > 100$~AU, for all the misaligned cases the gravitational contribution is between 10 and 100 weaker than the magnetic one. On the one hand, the magnetic transport becomes weaker as $\alpha$ increases and on the other hand, gravity transports momentum more efficiently in the presence of a disk, owing to density waves (the spiral arms) that propagate in the radial direction. The gravitational transport is nonetheless less efficient because of the symmetry of the disks, which is stabilized by the magnetic field, as emphasized in \cite{Hennebelle08b}; less symmetric disks would transport more momentum by means of gravitational torques. The gravitational transport is stronger in the perpendicular case than in the other misaligned cases because disks are more massive.

For other magnetizations, these conclusions hold, since without a disk the magnetic braking is the most efficient process of angular momentum transport, whereas in the presence of a disk, a significant -- although not predominant -- fraction of the momentum can be transported by gravity. In the low magnetized cases ($\mu = 17$), gravity can transport even more momentum in the radial direction than the magnetic field.

\section{Disk properties}

When enough angular momentum is left in the envelope, a disk can form around the adiabatic core. Here, we first discuss in detail how to define disks, then study some of their properties.

\subsection{Disk formation} \label{subsec:diskform}

We work in the frame of the disk $\mathcal{R}_d$ (see section \ref{subsec:transGC}), where the main axis of the frame is defined by the angular momentum. Several criteria must be used to define disks. As we show below, a simple rotation criterion is insufficient to define a disk because several parts of the envelope are rotating but do not belong to the disk. For example, defining the disk as all material whose rotation velocity is larger than a few times the infall velocity would also pick up the walls of the cavity of the outflows, which are rapidly rotating. A single geometric criterion is also insufficient; disks are not well-approximated by cylinders. We define disks by employing a combination of five different criteria. As disks are expected to be reasonably axisymmetric, these criteria are defined for concentric and superposed rings in which density, velocity, pressure, and magnetic field are averaged.

\begin{enumerate}
\item As disks are expected to be Keplerian, we first use a velocity criterion. A ring of matter should not collapse too rapidly along the radial direction, which implies that the azimuthal velocity must be larger than the radial velocity ($v_{\phi} > f_{\rm thres} v_r$).
\item As disks are expected to be near the hydrostatic equilibrium, the azimuthal velocity is larger than the vertical velocity ($v_{\phi} > f_{\rm thres} v_z$).
\item The central adiabatic core is also rotating but is not in the disk; another criterion is therefore added, to take into account only areas of the simulation that are rotationally supported. We thus check whether the rotational support (the rotational energy, $\rho v_{\phi}^2/2$) is larger than the thermal support (the thermal pressure, $P_{\rm th}$) by a factor $f_{\rm thres}$.
\item A connectivity criterion is also used: a ring area belongs to the disk if it is linked to the equatorial plane.
\item As discussed in section \ref{subsec:diskshape}, we add a density criterion ($n > 10^9$~cm$^{-3}$) to avoid the large spiral arms and obtain more realistic estimates of the shape of the disk.
\end{enumerate}
For the three first criteria, a value $f_{\rm thres} = 2$ is chosen below.

\begin{figure}
\subfigure{\includegraphics[width = .5\textwidth]{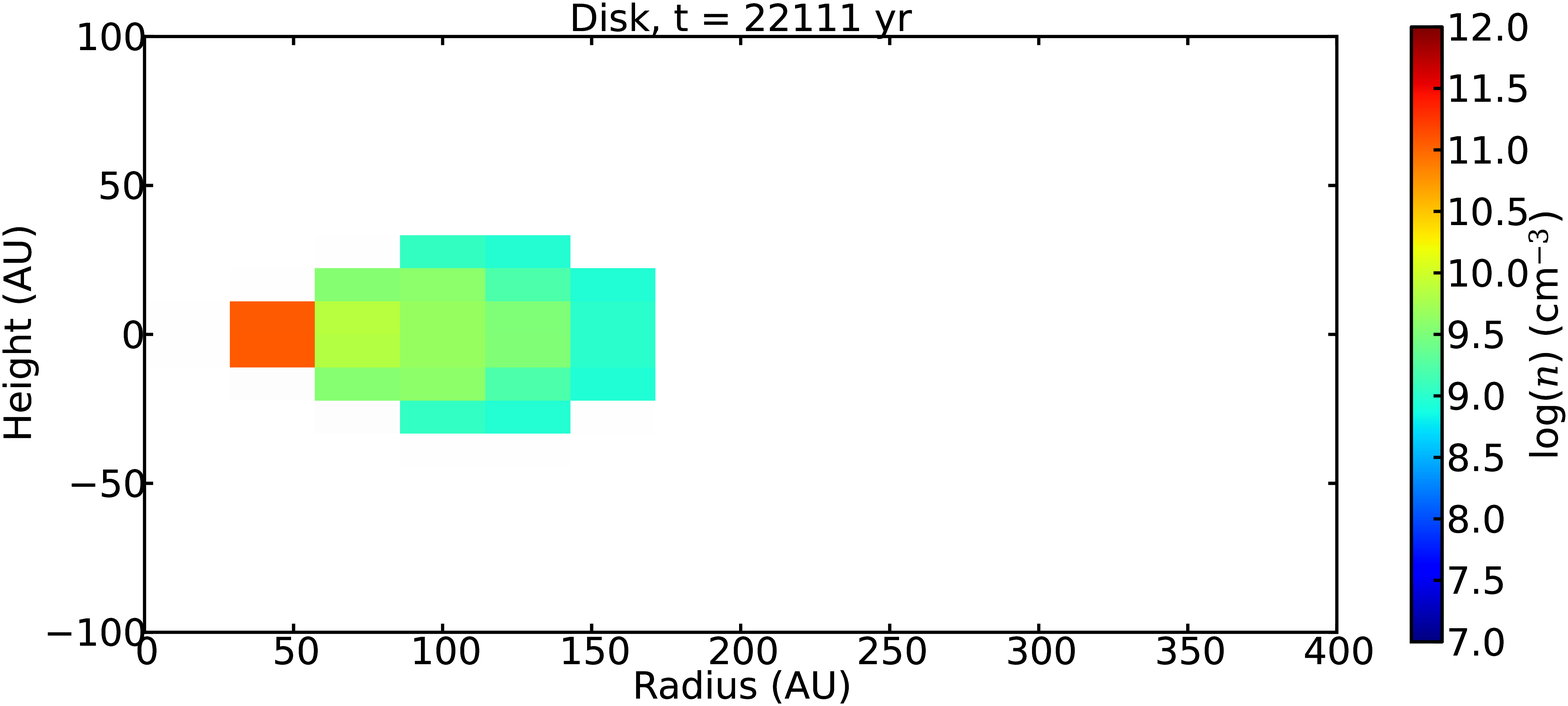}}
\subfigure{\includegraphics[width = .5\textwidth]{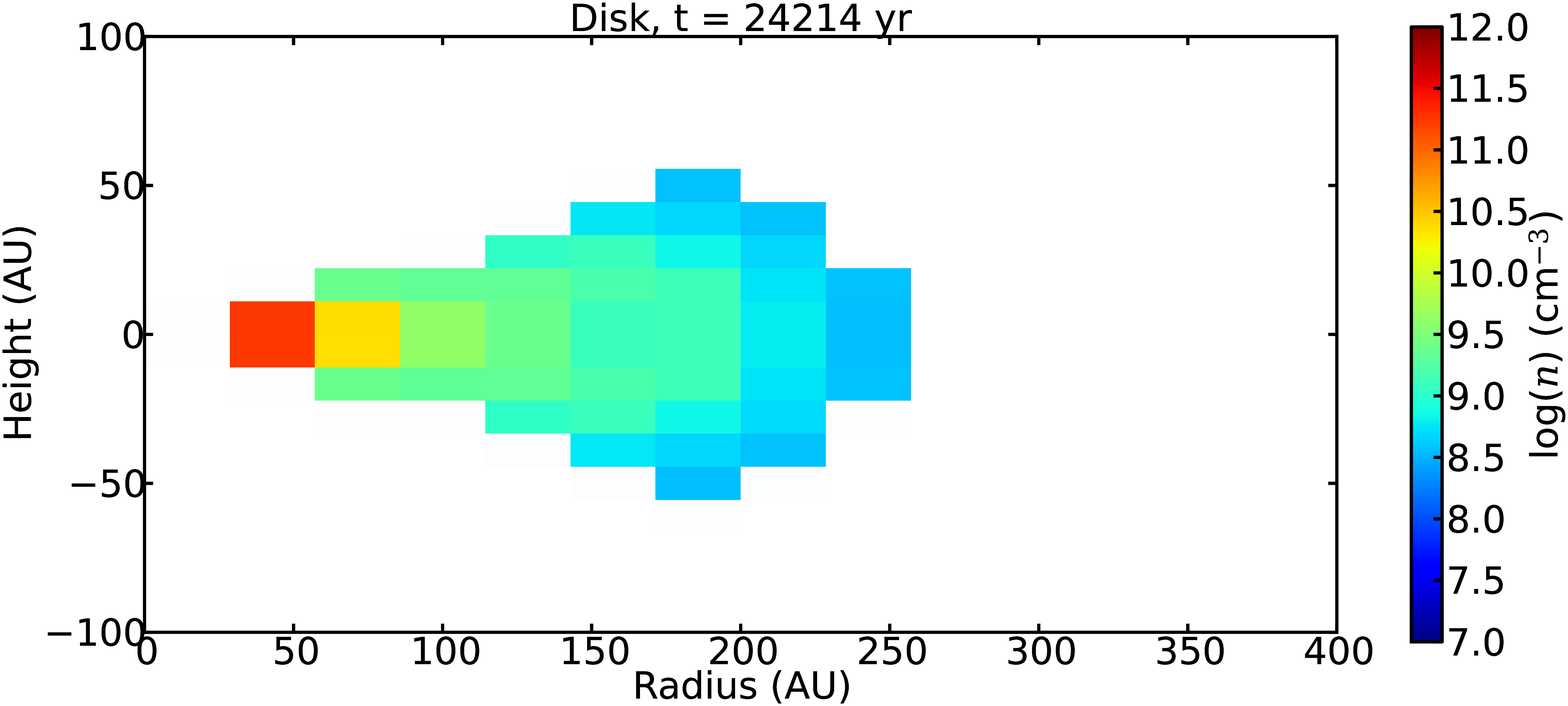}}
\subfigure{\includegraphics[width = .5\textwidth]{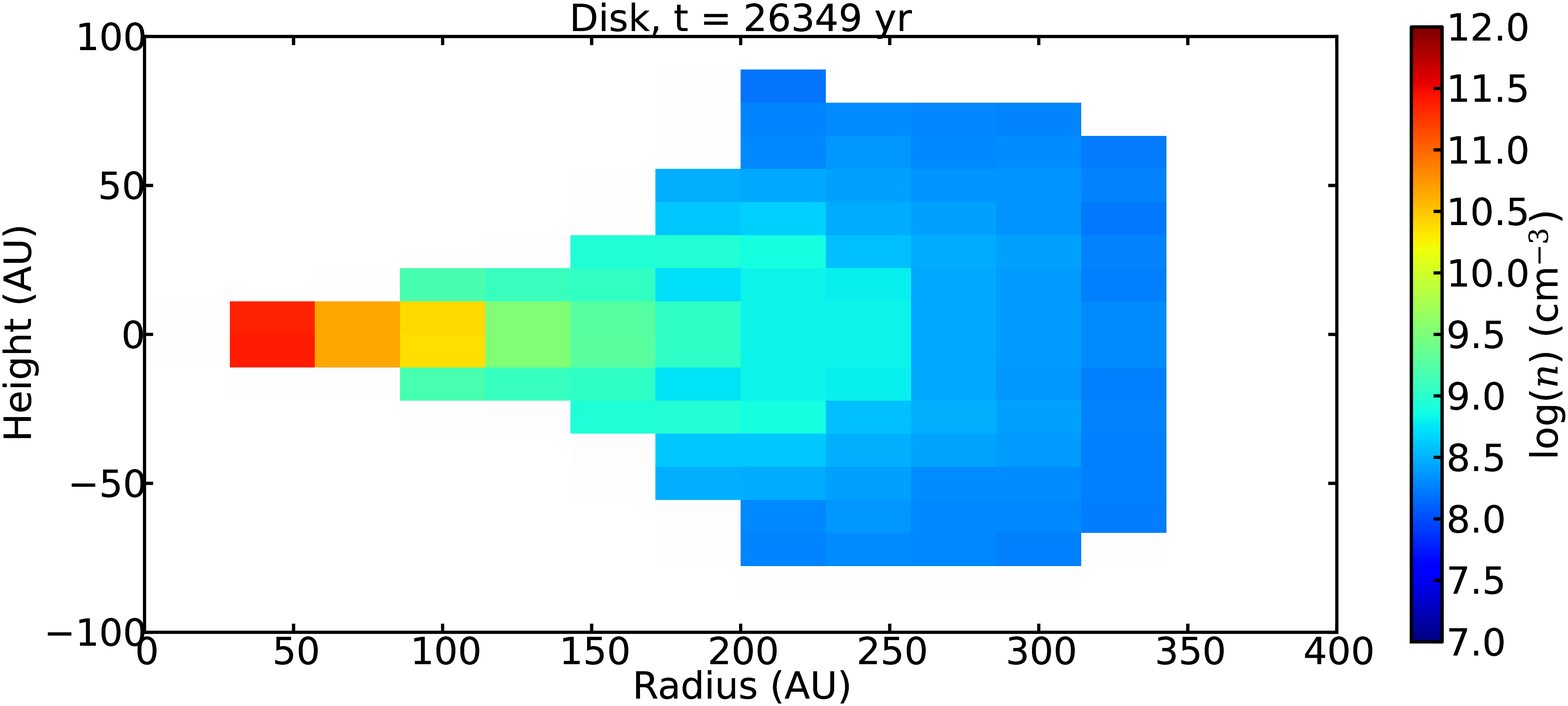}}
\caption{Density map on a logarithmic scale in the disk for $\mu = 5, \alpha = 90^{\circ}$, for three different time-steps. The disk grows with time as the central part becomes denser; the maximum radius corresponds to the edge of the spiral arms of the disk structure.}
\label{img:shape0316}
\end{figure}

Figure \ref{img:shape0316} shows the azimuthally averaged shape of the disk using these criteria, for $\mu = 5, \alpha = 90^{\circ}$, and three different time-steps.

\subsection{Mass}

\begin{figure}
 \subfigure[\label{img:mass01}]{\includegraphics[width = 0.5\textwidth]{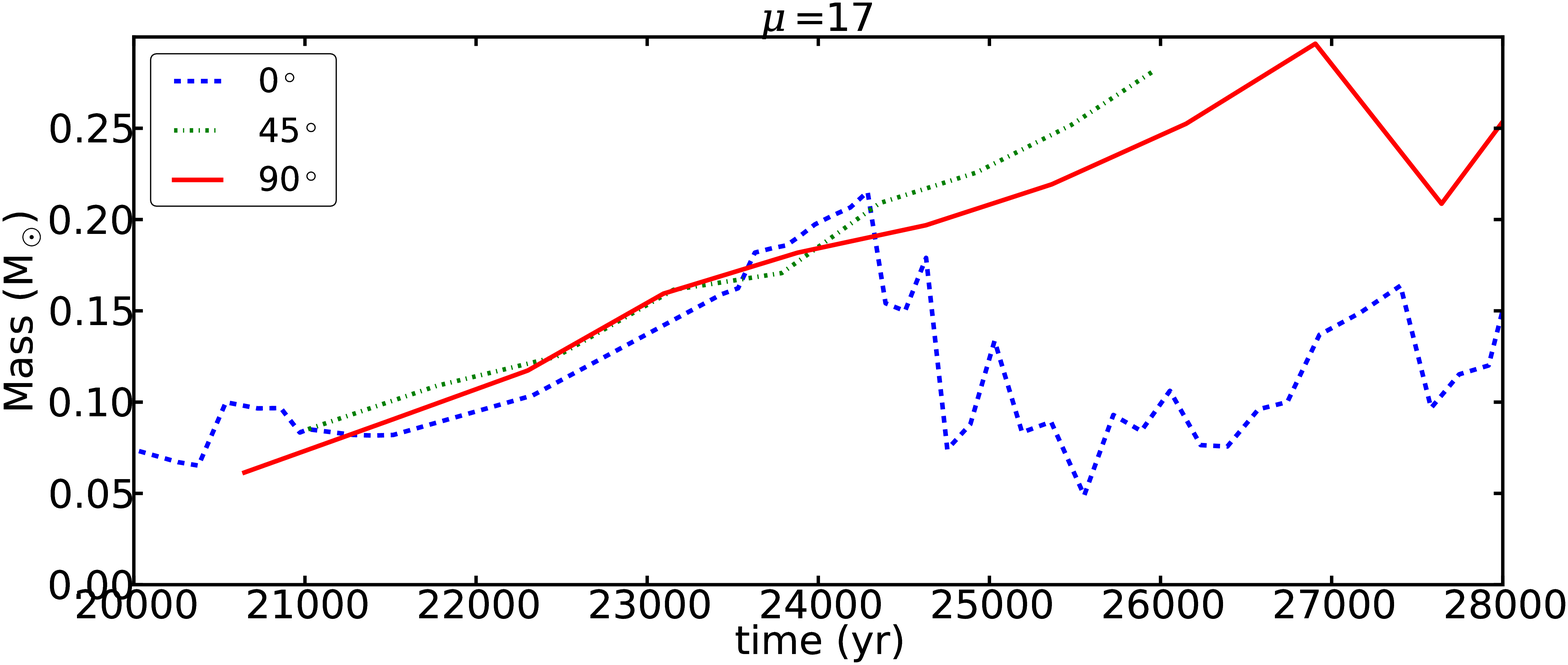}}
 \subfigure[\label{img:mass03}]{\includegraphics[width = 0.5\textwidth]{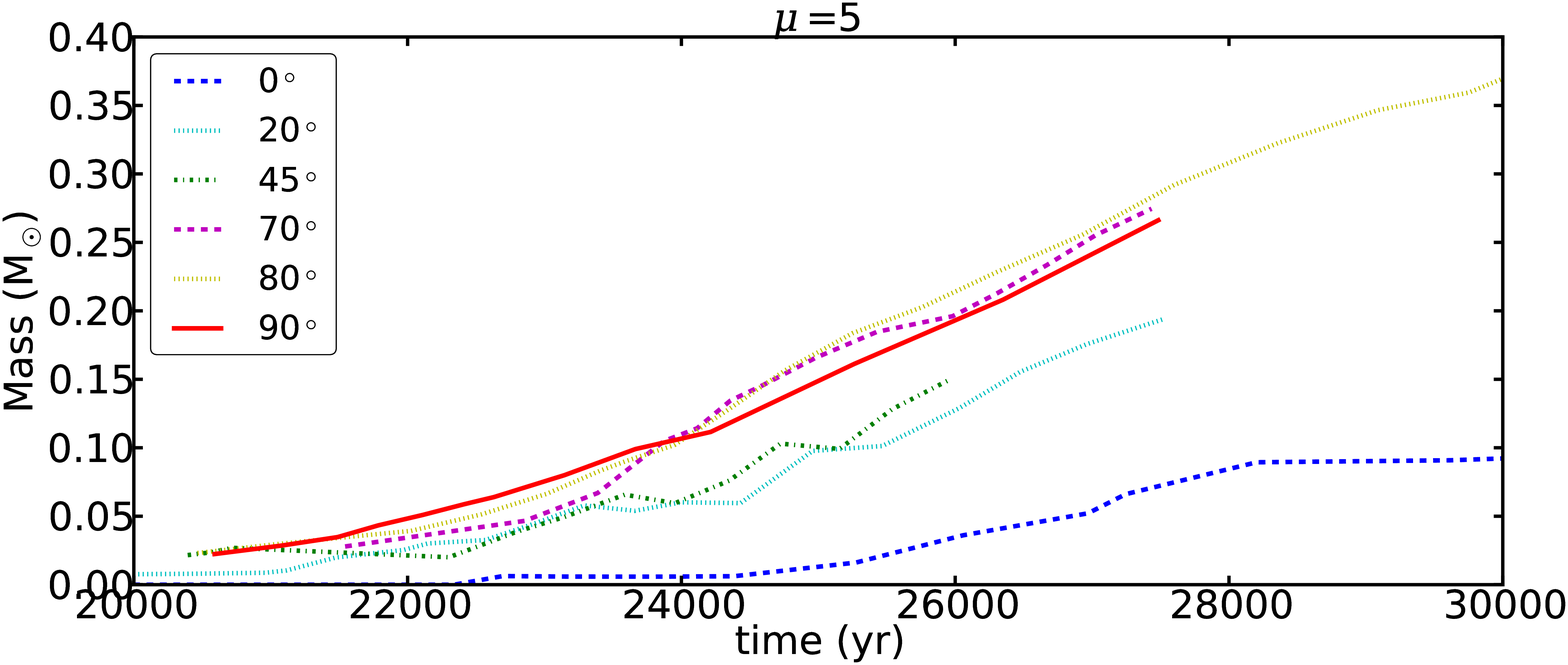}}
 \subfigure[\label{img:mass05}]{\includegraphics[width = 0.5\textwidth]{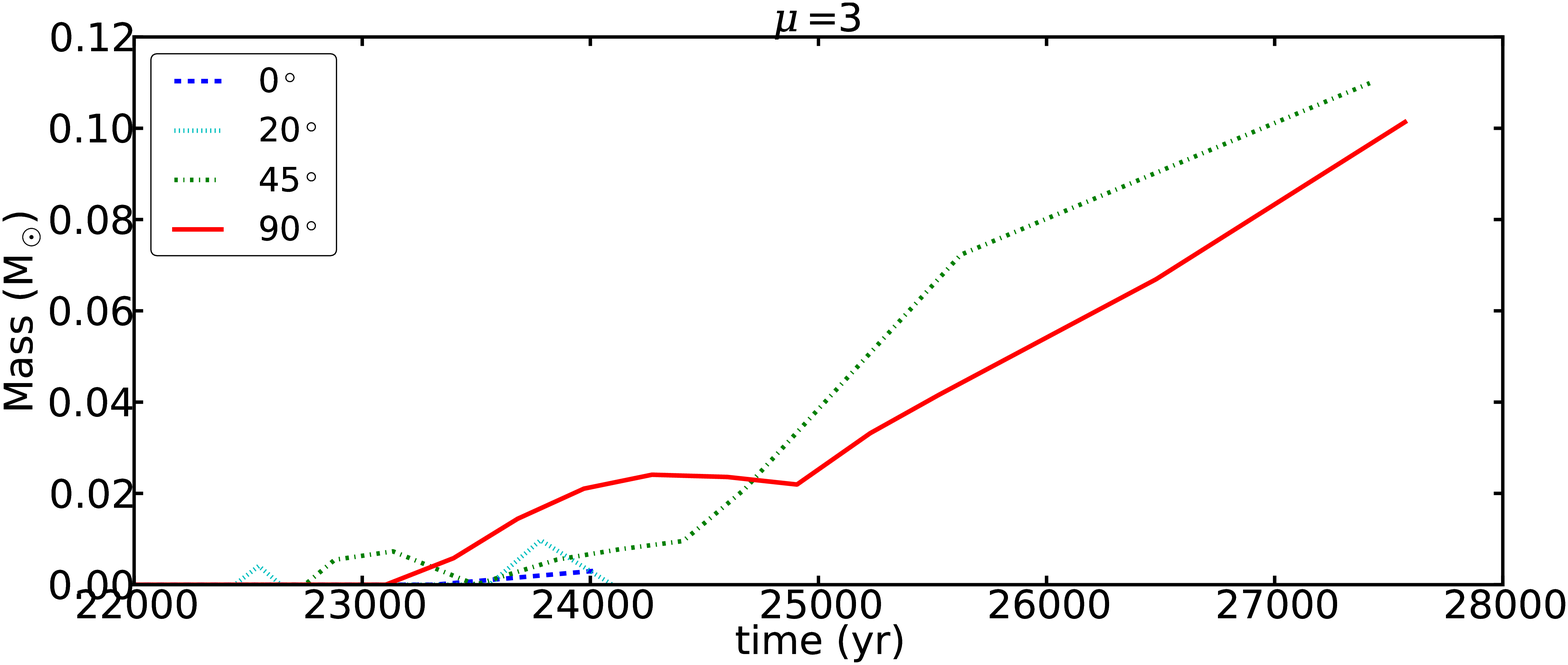}}
 \subfigure[\label{img:mass07}]{\includegraphics[width = 0.5\textwidth]{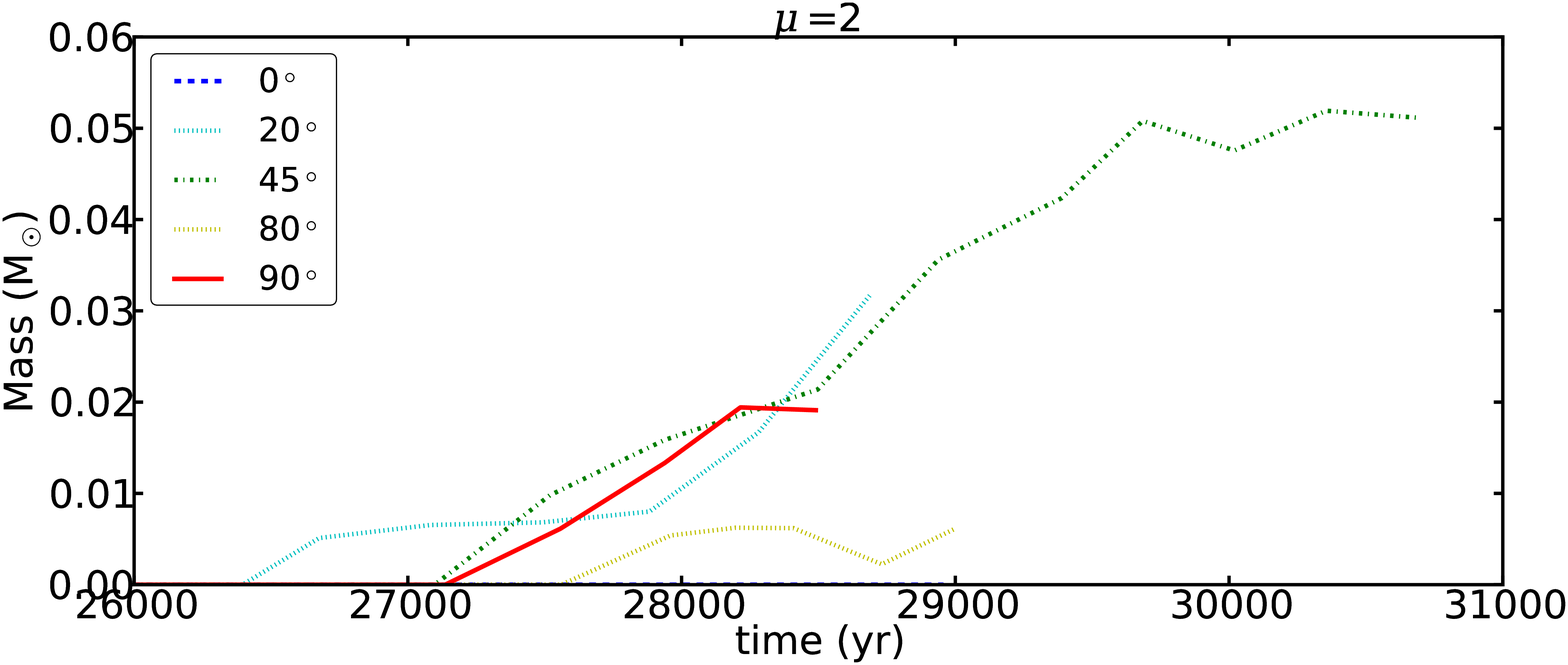}}
 \caption{Mass of the disk as a function of time for $\mu = 17$ (Fig. \ref{img:mass01}), $\mu = 5$ (Fig. \ref{img:mass03}), $\mu = 3$ (Fig. \ref{img:mass05}), and $\mu = 2$ (Fig. \ref{img:mass07}).}
 \label{img:massdisk}
\end{figure}

The mass of the disk as a function of time, for different magnetizations and angles, is presented in Fig.~\ref{img:massdisk}. The general trend shows an increase in the disk mass with the angle $\alpha$. This agrees with our previous discussions, which indicated that the braking time in the parallel case is shorter, leading to a more rapid removal of angular momentum in the infalling envelope, thus limiting the effective mass of disks. At the same time, it is clear that for increasing magnetic field strength, thus increasing magnetic braking, disks with masses greater than $0.05 M_{\odot}$ are only found in misaligned configurations. The limiting case corresponds to a magnetization of $\mu = 2$, where the removal of angular momentum by the magnetic field is so efficient that the mass of rotating gas does not exceed $0.05 M_{\odot}$, even in the perpendicular case. We note that higher resolution simulations tend to indicate masses that are even smaller than this value (see fig. \ref{img:massConv}\subref{img:massConv2}).

At the other end for low magnetization ($\mu = 17$), disks always form with masses that increase to about $0.3 M_{\odot}$. We note that in the aligned case, the disk fragments, leading to a decrease in its mass after a time $t \simeq 24$ kyrs.

In the intermediate regime of magnetizations ($\mu = 5$), the mass of the disk starts to increase significantly even for small $\alpha$. For the intermediate angles (20 and 45$^{\circ}$), disk masses increase within the range $0.15 - 0.2 M_{\odot}$. For the more tilted cases ($70$, $80^{\circ}$, and the perpendicular case), disks grow to $0.25 - 0.4 M_{\odot}$.

For $\mu = 3$, disks do not form for angles $\lesssim 20^{\circ}$. In the $45^{\circ}$ case and the perpendicular case, the masses of the disk increase to $0.1 M_{\odot}$.

\begin{figure}
  \includegraphics[width = 0.47\textwidth]{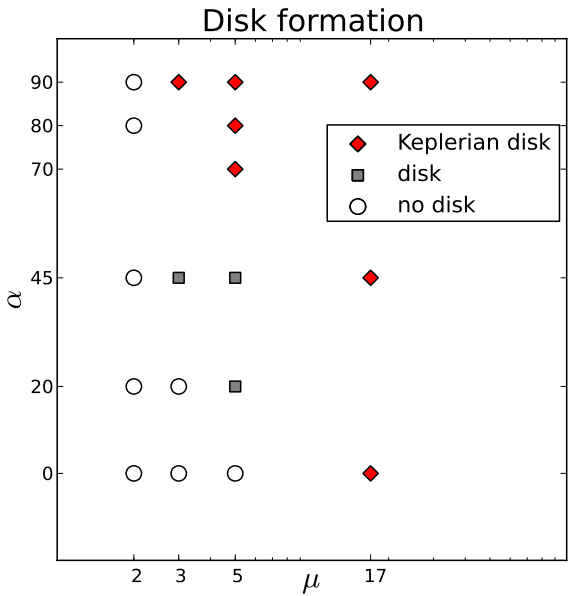}
  \caption{Disk formation in the parameter space investigated by the simulations (inclination angle $\alpha$ versus magnetization $\mu$). The red diamonds are configurations in which approximately Keplerian disks form, the grey squares denote configurations in which disks form with flat rotation curve (\emph{cf.} Fig. \ref{img:vkep} and the related discussion for more details); the white circles are configurations with no significant disk ($M_{\rm disk} < 5.10^{-2} M_{\odot}$).}
  \label{img:muvsal}
\end{figure}

Figure~\ref{img:muvsal} summarizes our results and shows the parameters (orientation and magnetization) for which disks can form.

\subsection{Velocity}

\begin{figure}
\subfigure[\label{img:vkep0308}]{\includegraphics[width = 0.47\textwidth]{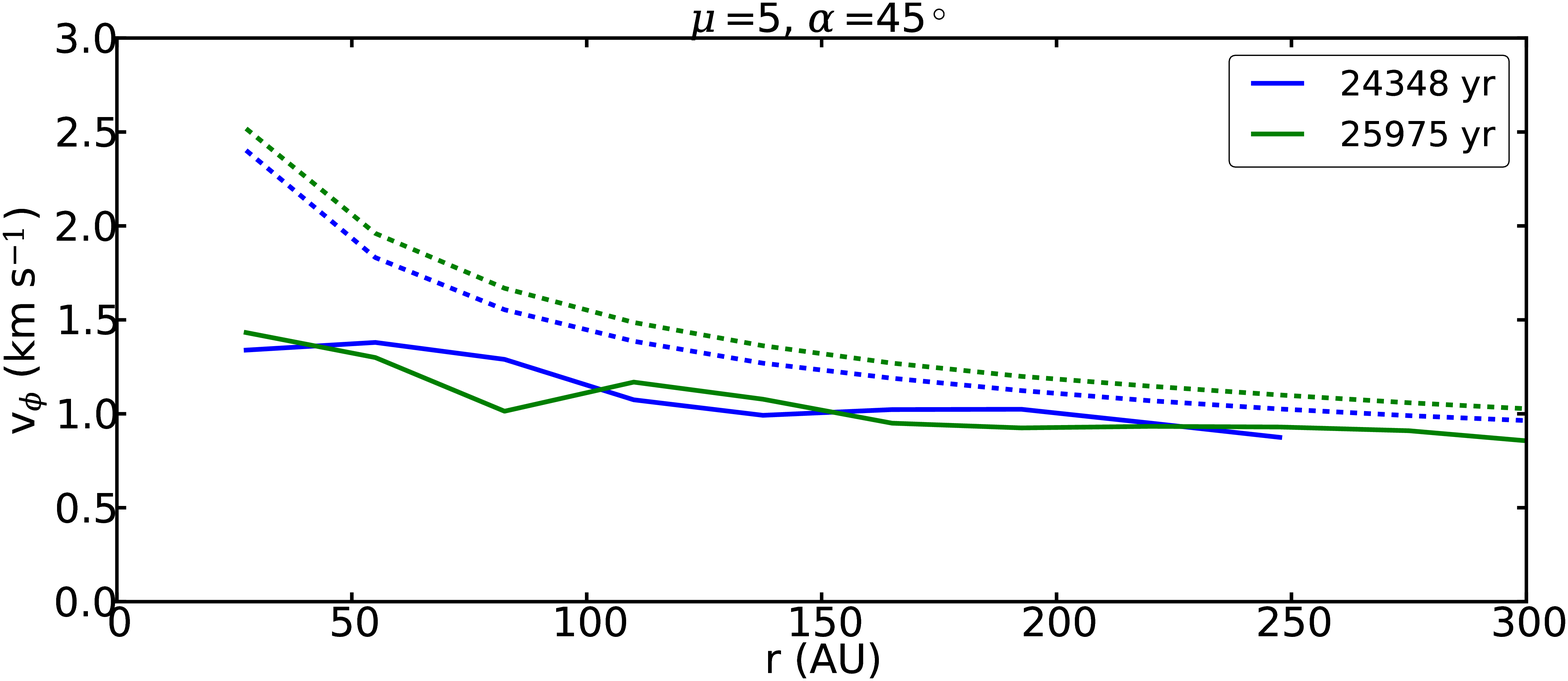}}
\subfigure[\label{img:vkep0316}]{\includegraphics[width = 0.47\textwidth]{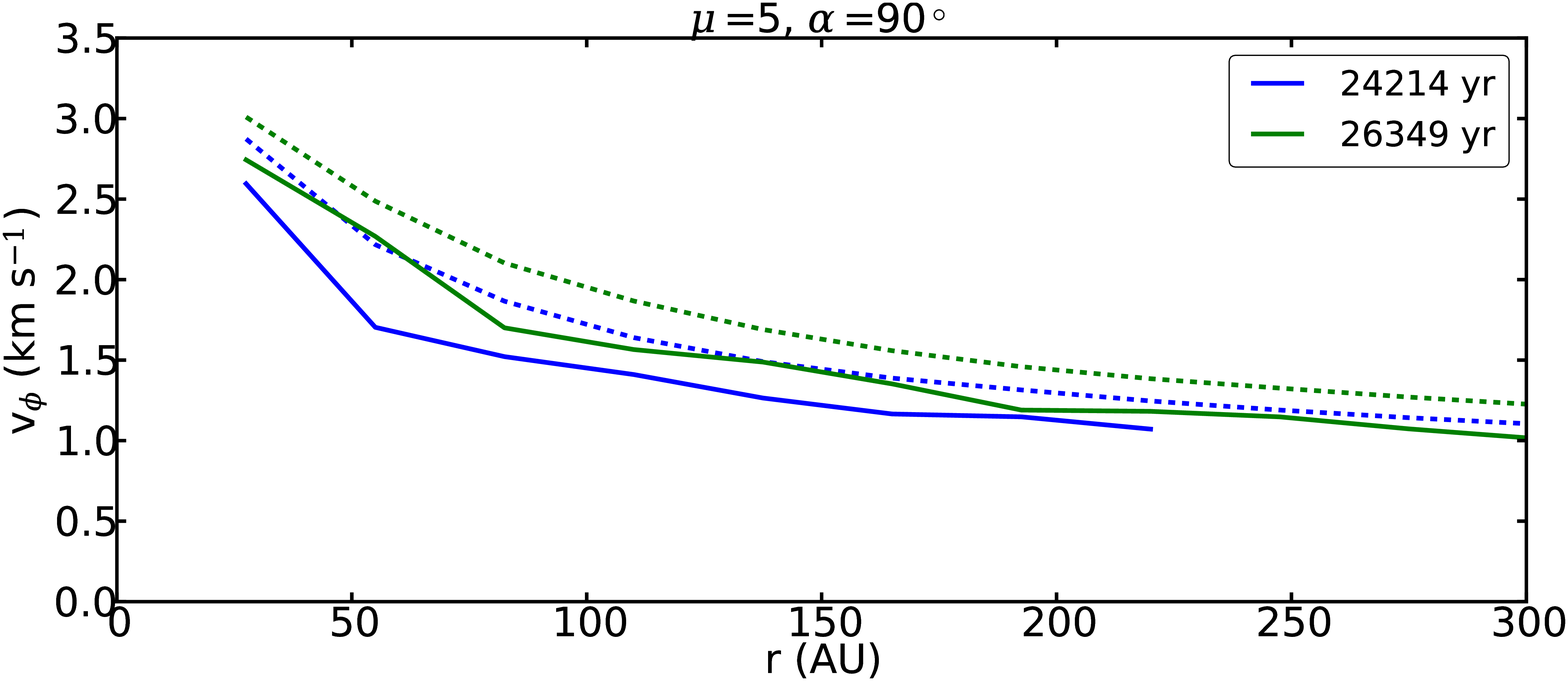}}
\subfigure[\label{img:vkep0516}]{\includegraphics[width = 0.47\textwidth]{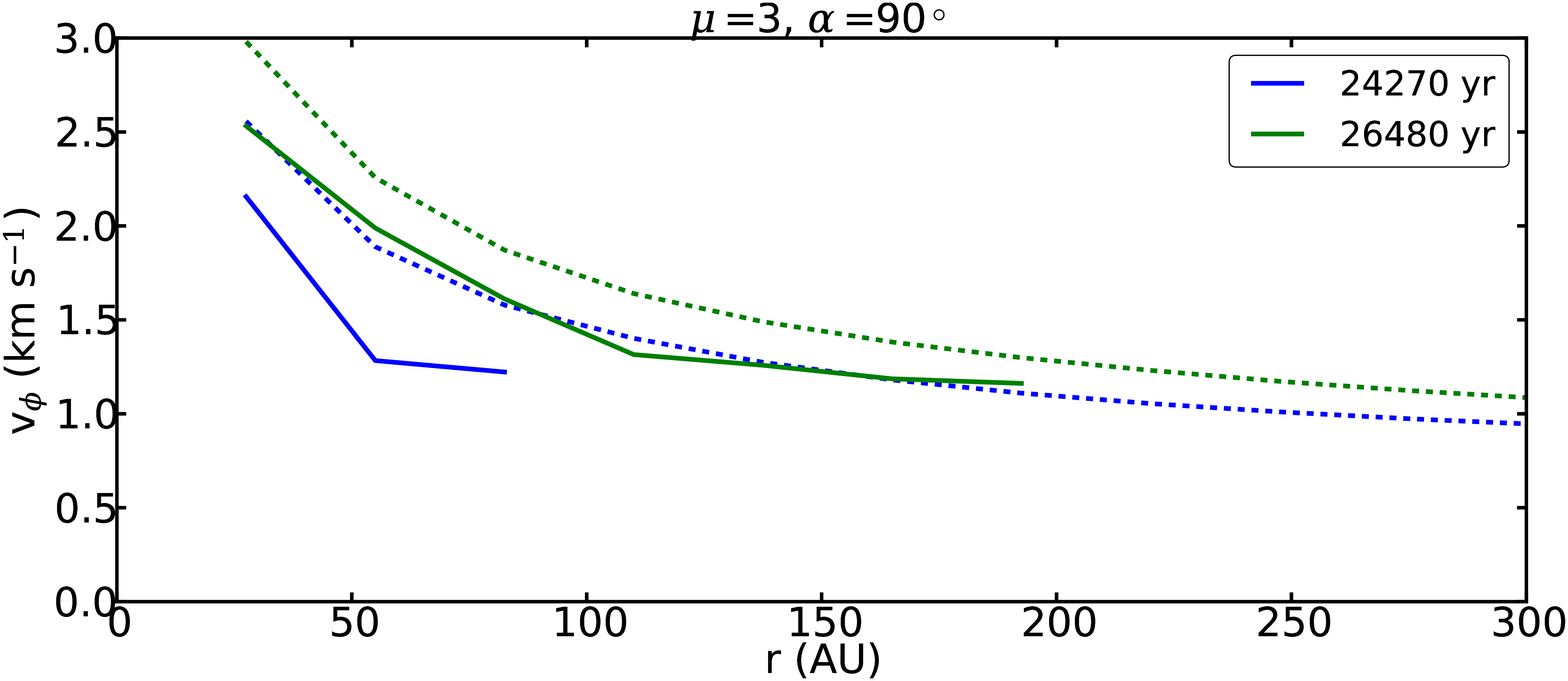}}
\caption{Radial profile of rotational velocity in the disk, for $\mu = 5, \alpha = 45 \text{ and } 90^{\circ}$ and $\mu = 3, \alpha = 90^{\circ}$. The straight lines are the rotational velocity for different time-steps; the dotted lines are the Keplerian velocity.}
\label{img:vkep}
\end{figure}

\begin{figure}
\subfigure[\label{img:pbVSek0308}]{\includegraphics[width = 0.47\textwidth]{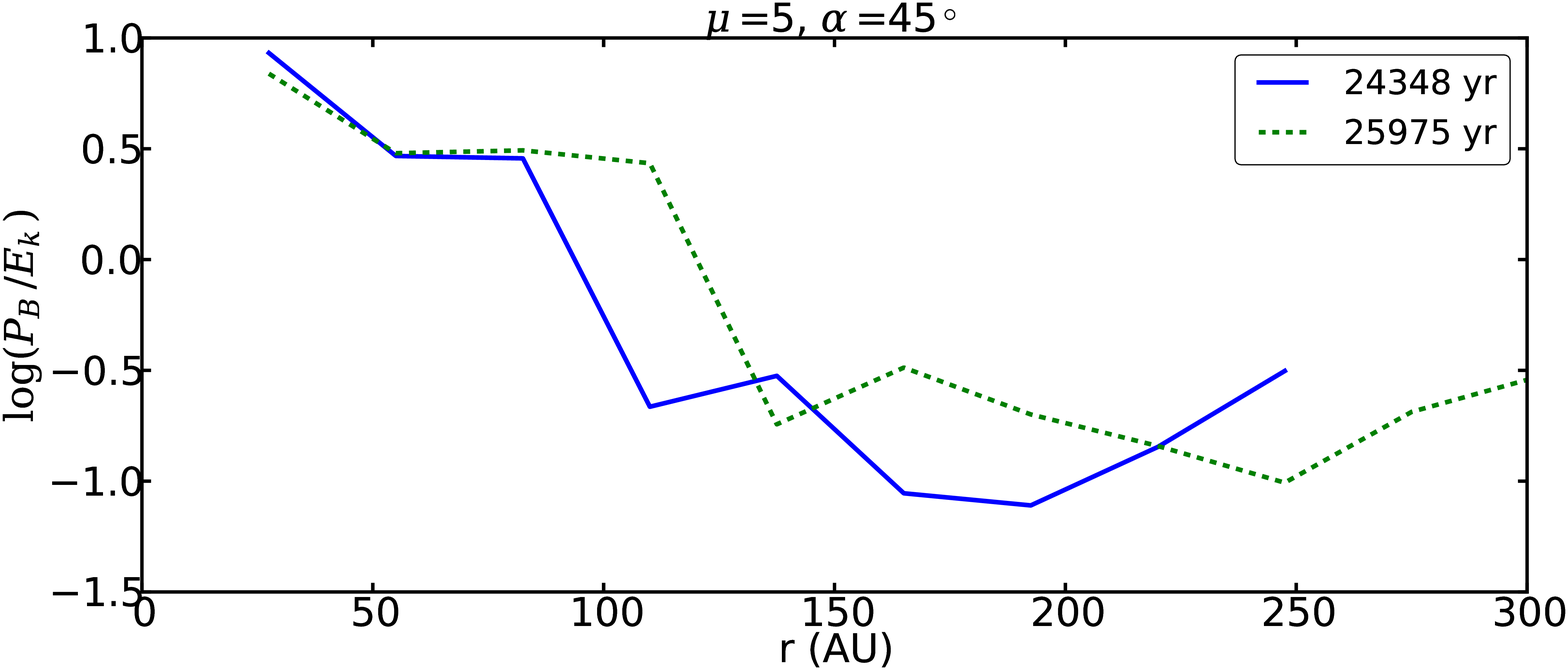}}
\subfigure[\label{img:pbVSek0316}]{\includegraphics[width = 0.47\textwidth]{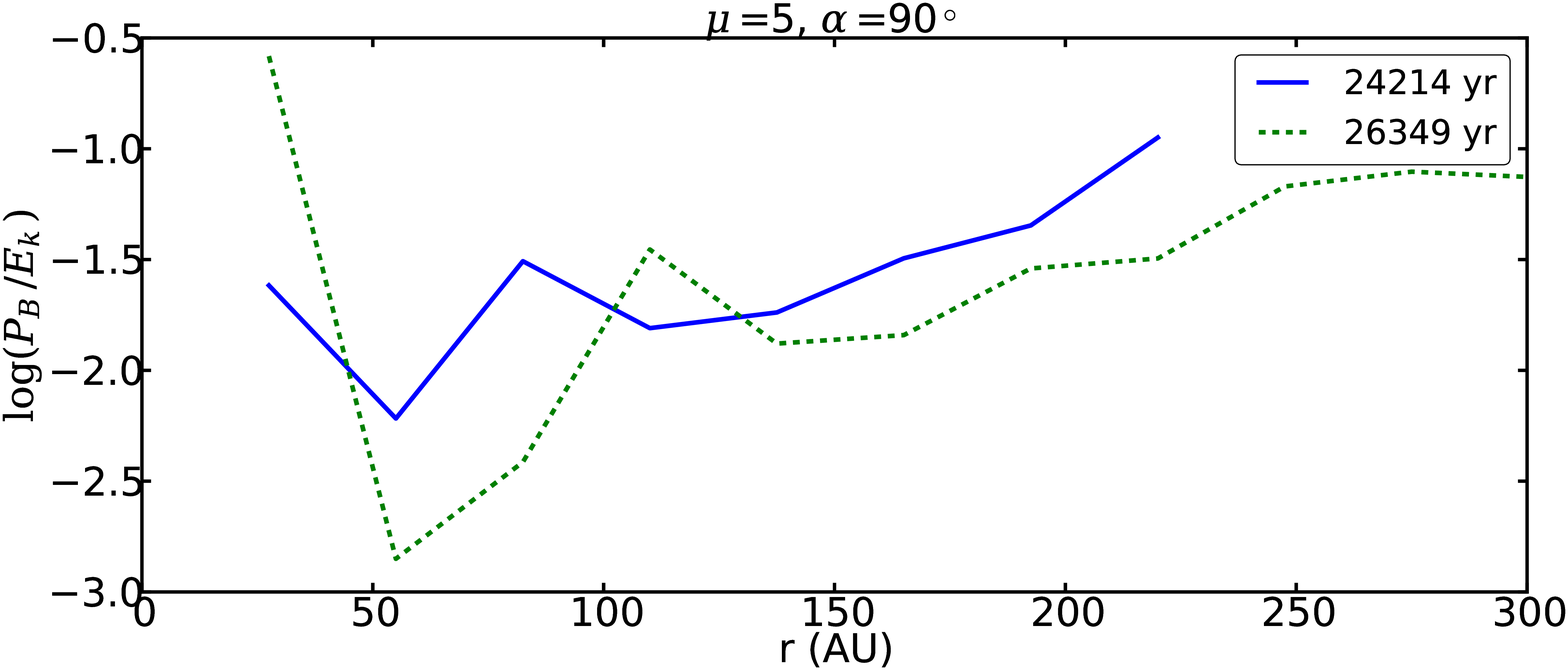}}
\subfigure[\label{img:pbVSek0516}]{\includegraphics[width = 0.47\textwidth]{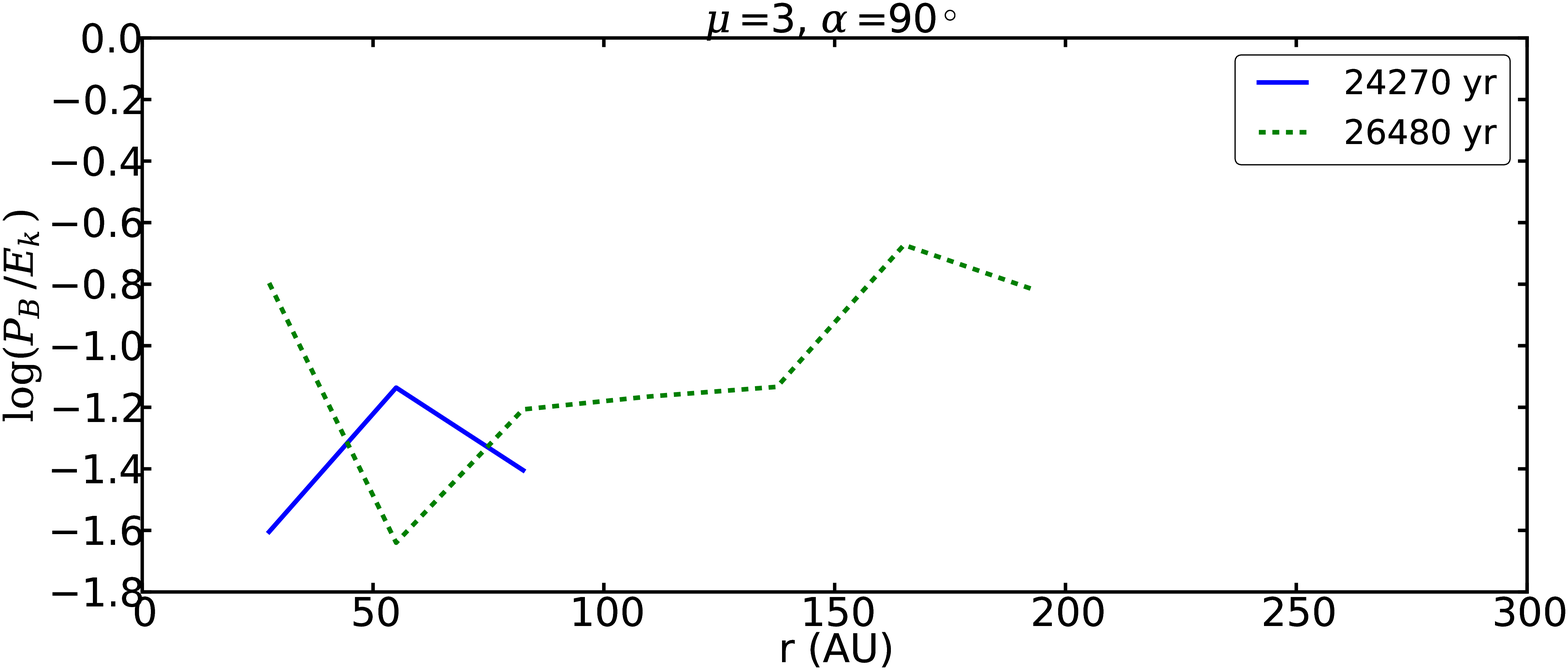}}
\caption{Ratio of magnetic pressure to rotational kinetic energy in the disk, for $\mu = 5, \alpha = 45 \text{ and } 90^{\circ}$ and $\mu = 3, \alpha = 90^{\circ}$.}
\label{img:pbVSek}
\end{figure}

To estimate the rotational velocity in the disk, we average $v_{\phi}(r,\phi,z)$ azimuthally and axially over the thickness of the disk. We can compare it to the Keplerian velocity $v_K(r) = \sqrt{GM(r)/r}$, with $M(r)$ the mass within a sphere of radius $r$. We find that the rotational velocity in the disk is nearly Keplerian, as shown in Fig. \ref{img:vkep} for $\mu = 5$, $\alpha = 45$ and $90^{\circ}$, and $\mu = 3$, $\alpha = 90^{\circ}$. In the perpendicular case, for both magnetizations, the rotation velocity is nonetheless slightly sub-Keplerian; for $\alpha = 45^{\circ}$, the disk has a flat rotation curve.

To understand the presence of sub-Keplerian rotation velocity profiles, we plot in Figure \ref{img:pbVSek} the ratio of the magnetic pressure to the rotational kinetic energy $\log_{10}(P_B/E_k) = \log_{10}(B^2/4\pi \rho v^2_{\phi})$ for the same magnetizations and angles. This shows that these velocities are sub-Keplerian because of the magnetic support in the disk: the more tilted the axis of rotation, the lower the ratio of the magnetic pressure to the rotational kinetic energy. In particular, in the perpendicular case, the vertical component of the magnetic field is smaller, whereas the density is higher since the disk is more massive.

\subsection{Disk shape} \label{subsec:diskshape}

\begin{figure}
\subfigure[\label{img:slice0316_1}]{\includegraphics[width = .5\textwidth]{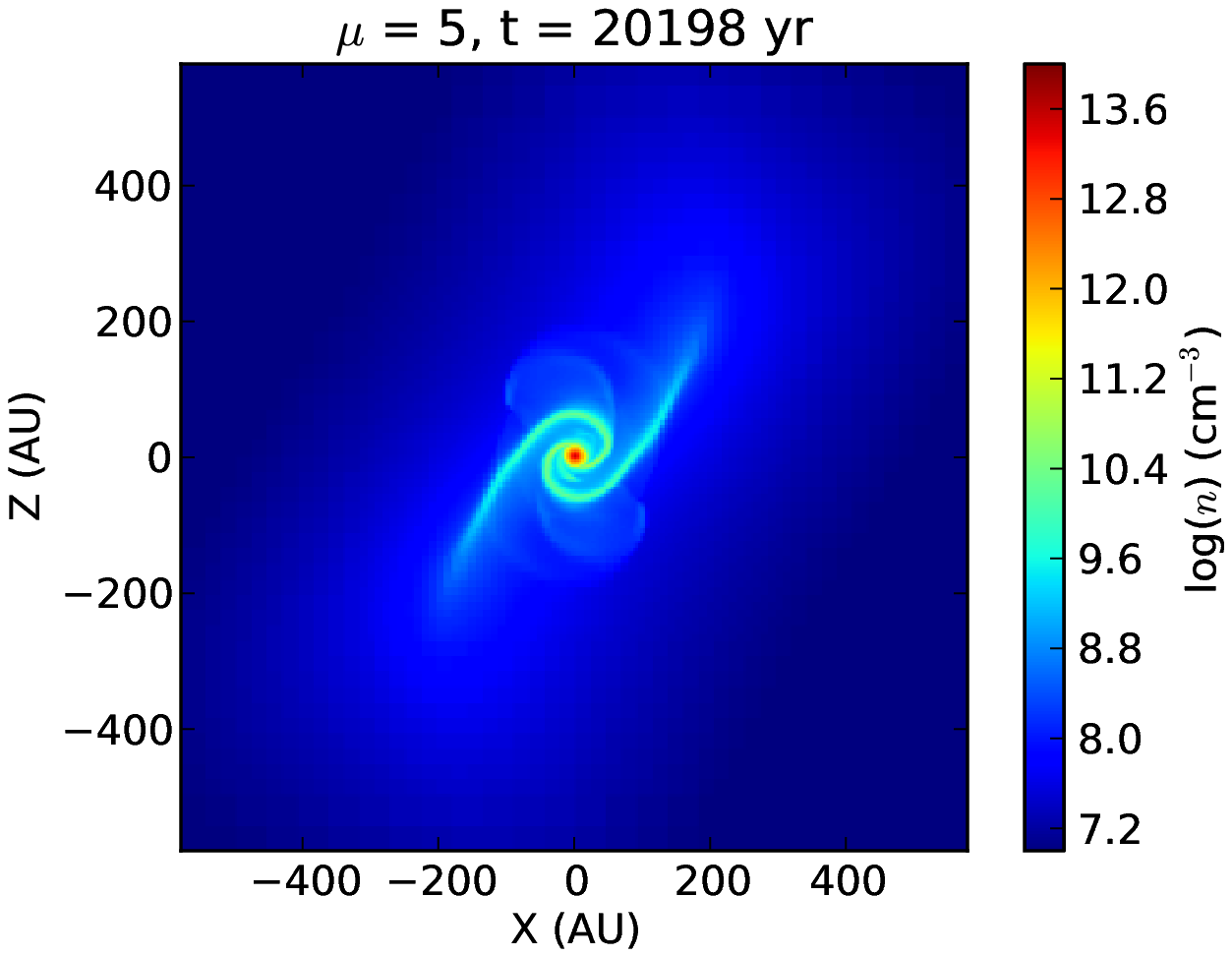}}
\subfigure[\label{img:slice0316_2}]{\includegraphics[width = .5\textwidth]{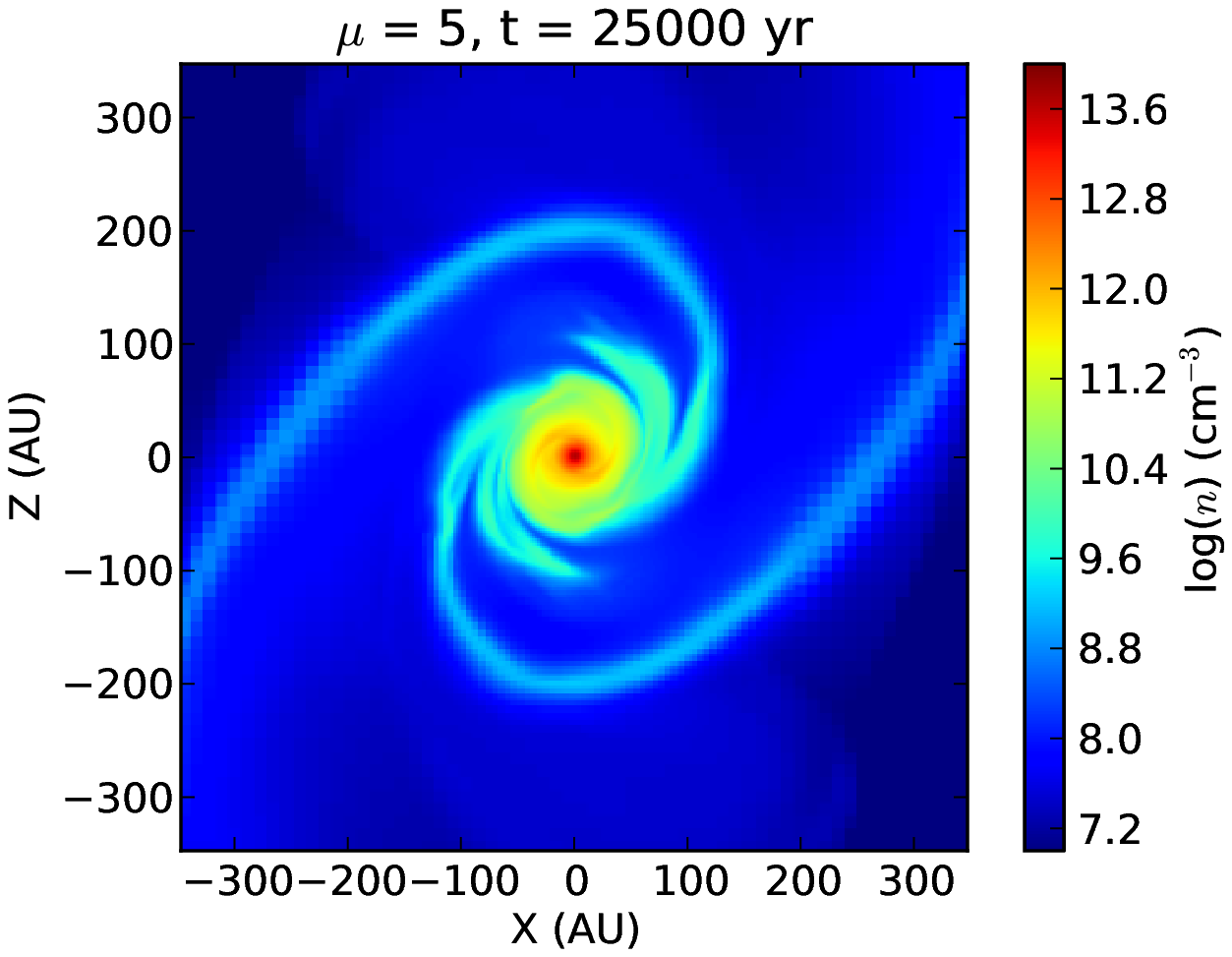}}
\caption{Density slice in the region of the disk for $\mu = 5$, $\alpha = 90^{\circ}$ at t = 20 000 and 25 000 yr.}
\label{img:slice0316}
\end{figure}

\begin{figure}
\subfigure[\label{img:slice030_1}]{\includegraphics[width = .5\textwidth]{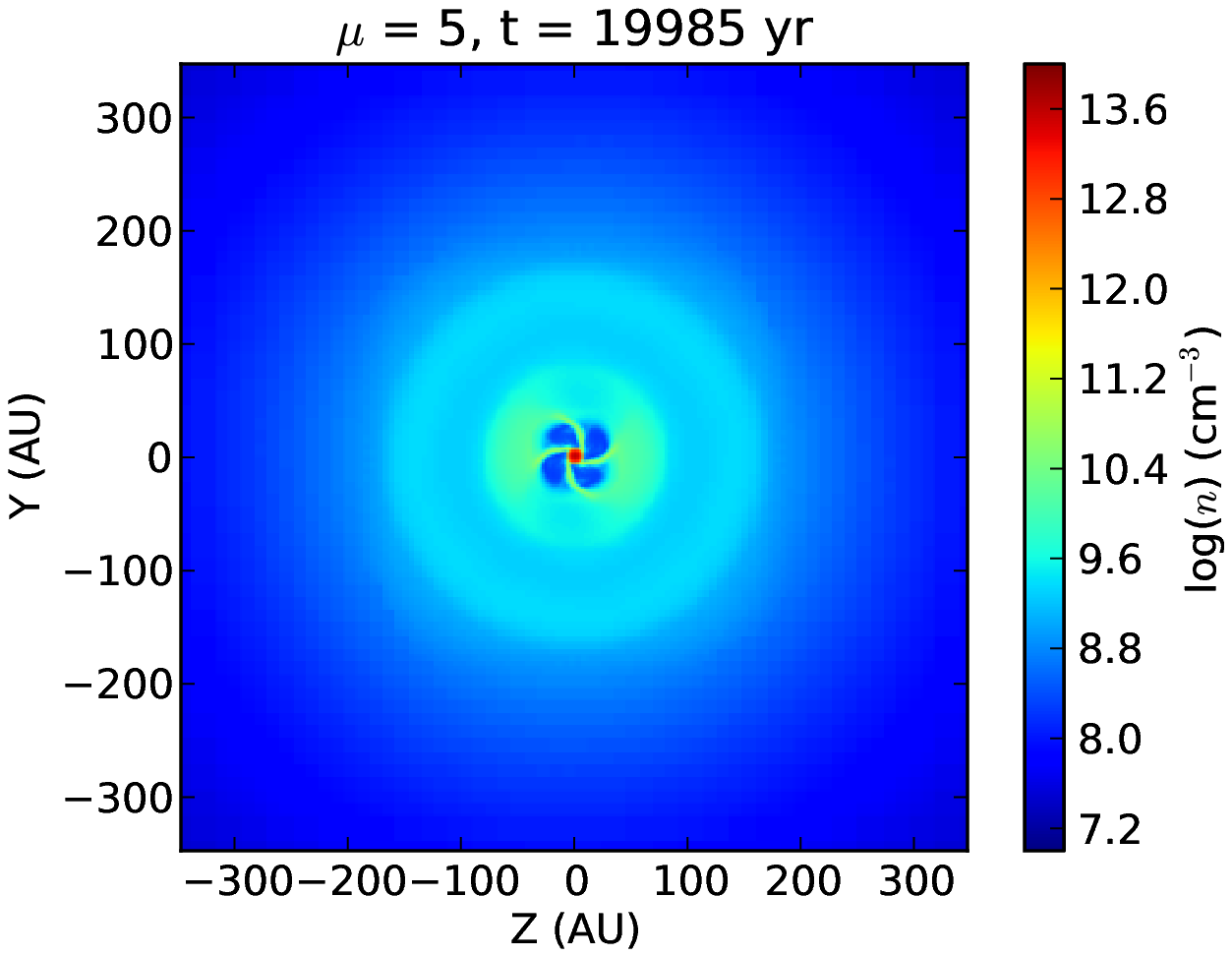}}
\subfigure[\label{img:slice030_2}]{\includegraphics[width = .5\textwidth]{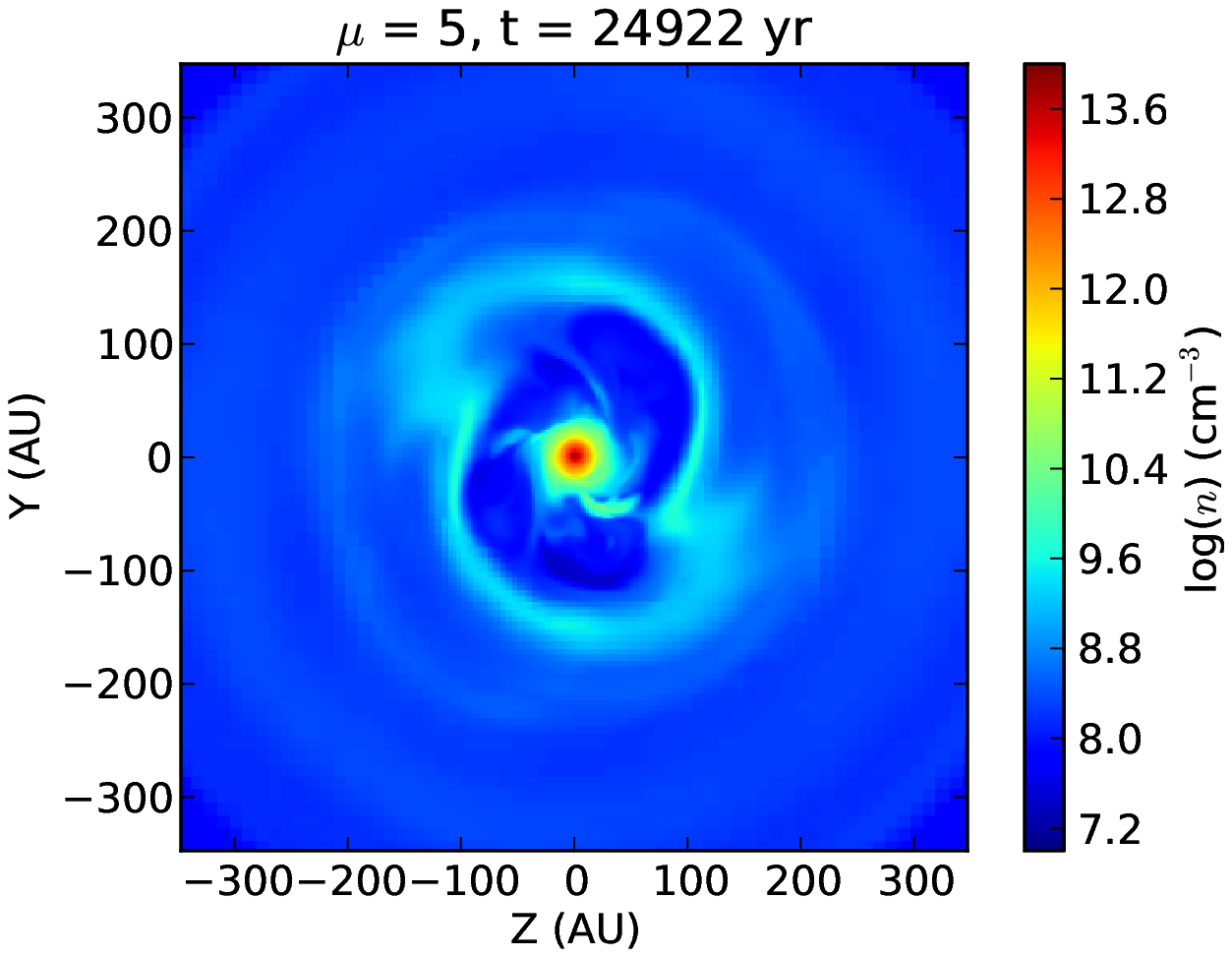}}
\caption{Density slice in the equatorial plane for $\mu = 5$, $\alpha = 0^{\circ}$ at t = 20 000 and 25 000 yr.}
\label{img:slice030}
\end{figure}

\begin{figure}
\subfigure[\label{img:slice070_1}]{\includegraphics[width = .5\textwidth]{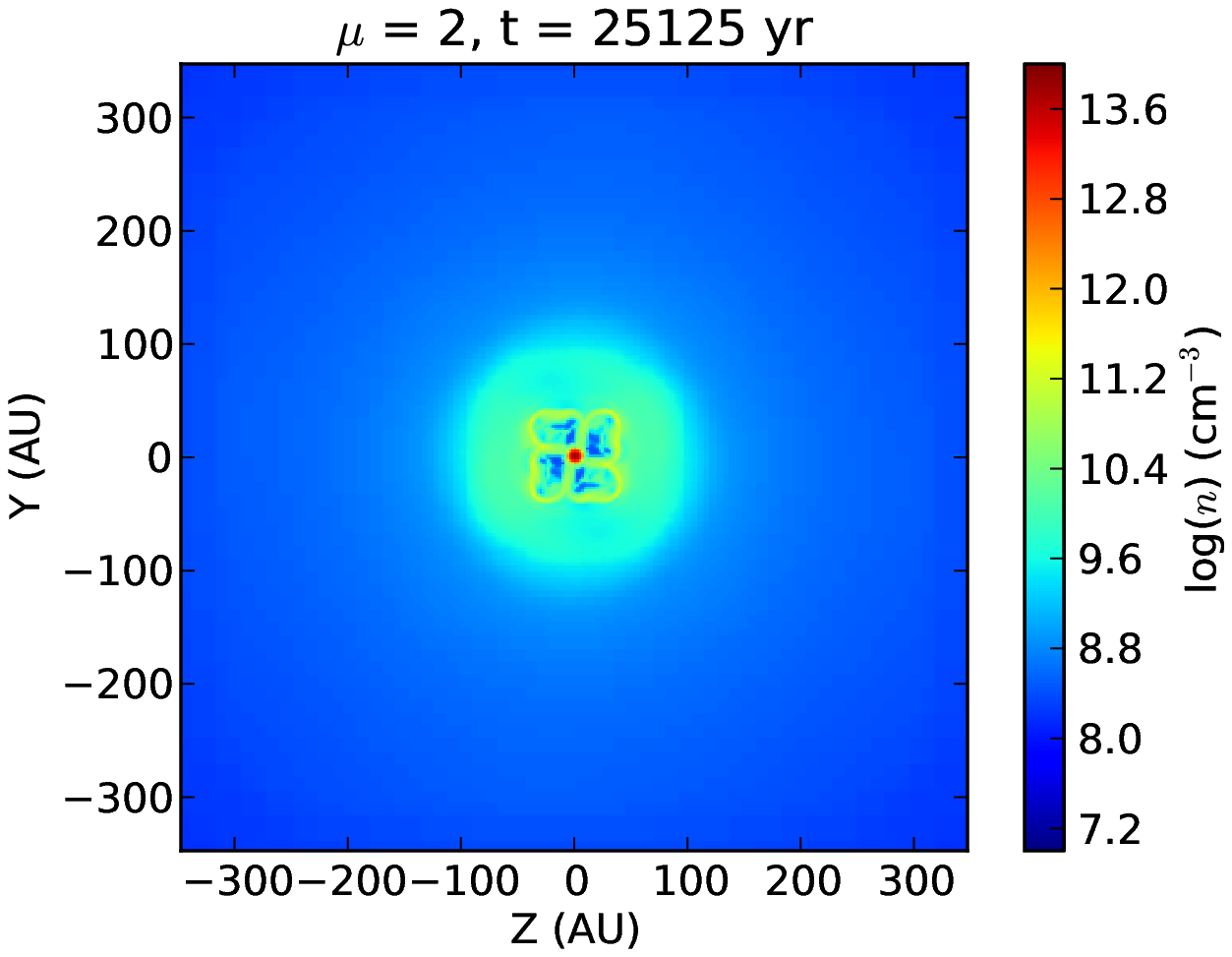}}
\subfigure[\label{img:slice070_2}]{\includegraphics[width = .5\textwidth]{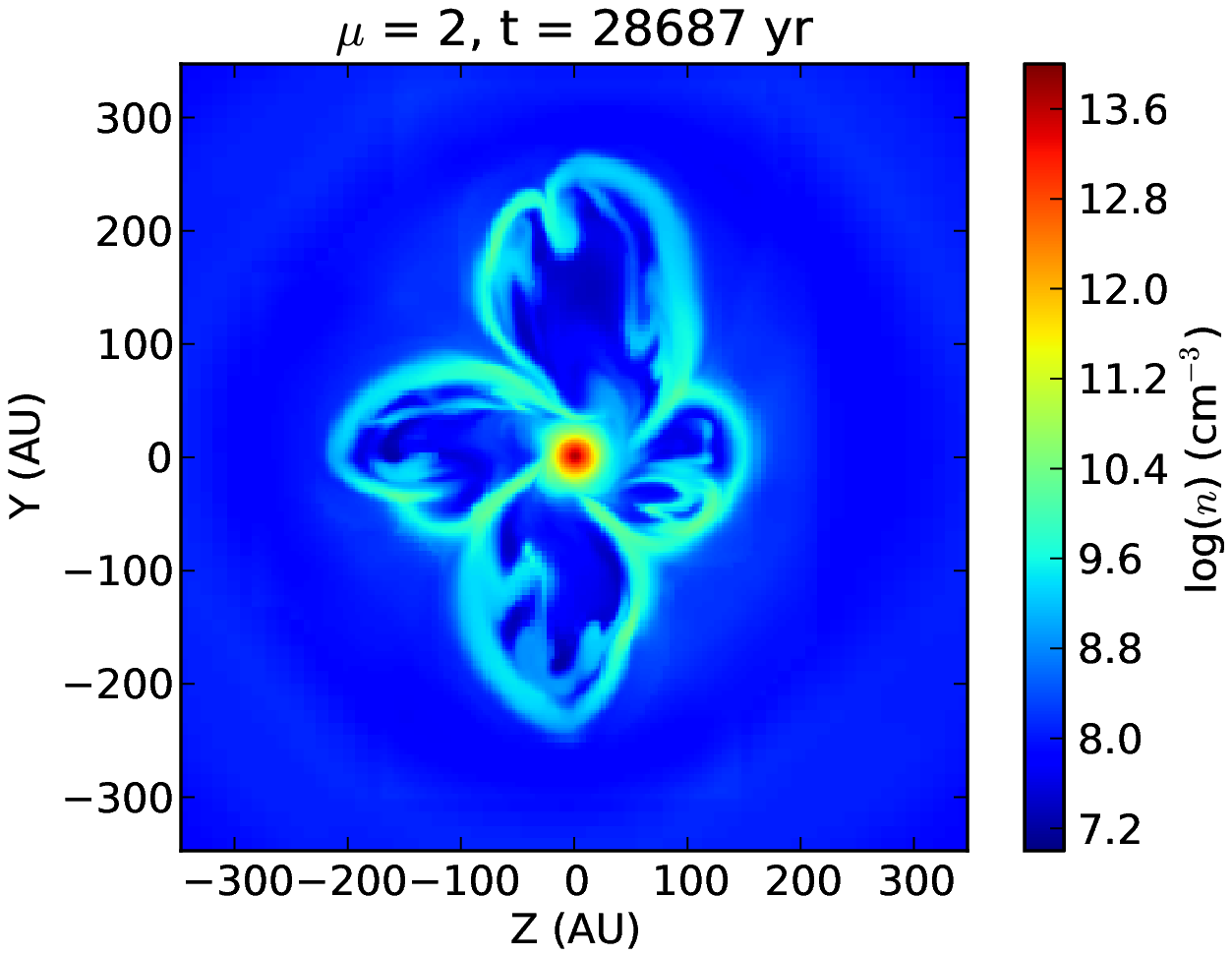}}
\caption{Same as Fig. \ref{img:slice030} for $\mu = 2$, $\alpha = 0^{\circ}$ at t = 25 000 and 29 000 yr.}
\label{img:slice070}
\end{figure}

\begin{figure}
\subfigure[\label{img:slice0716_1}]{\includegraphics[width = .5\textwidth]{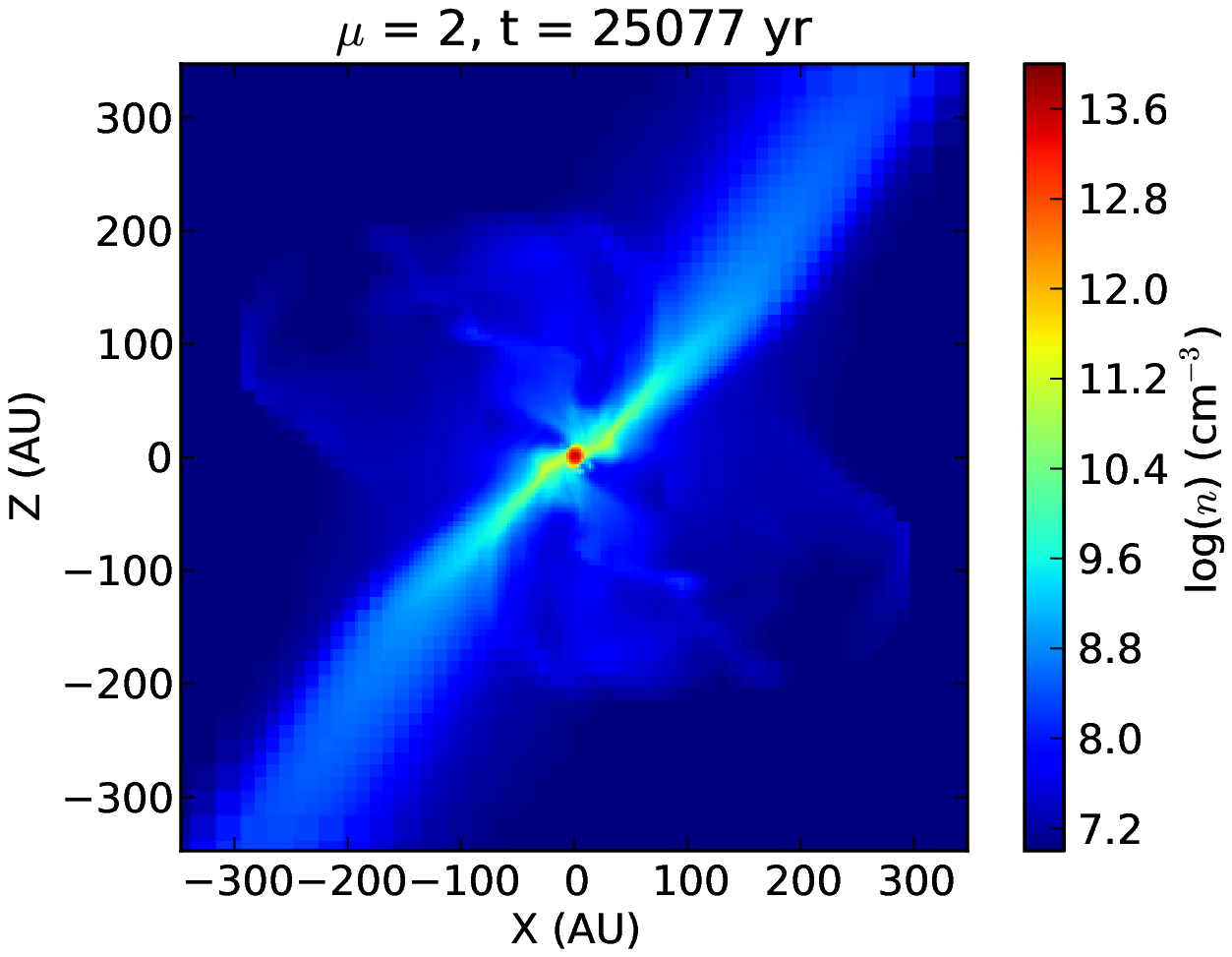}}
\subfigure[\label{img:slice0716_2}]{\includegraphics[width = .5\textwidth]{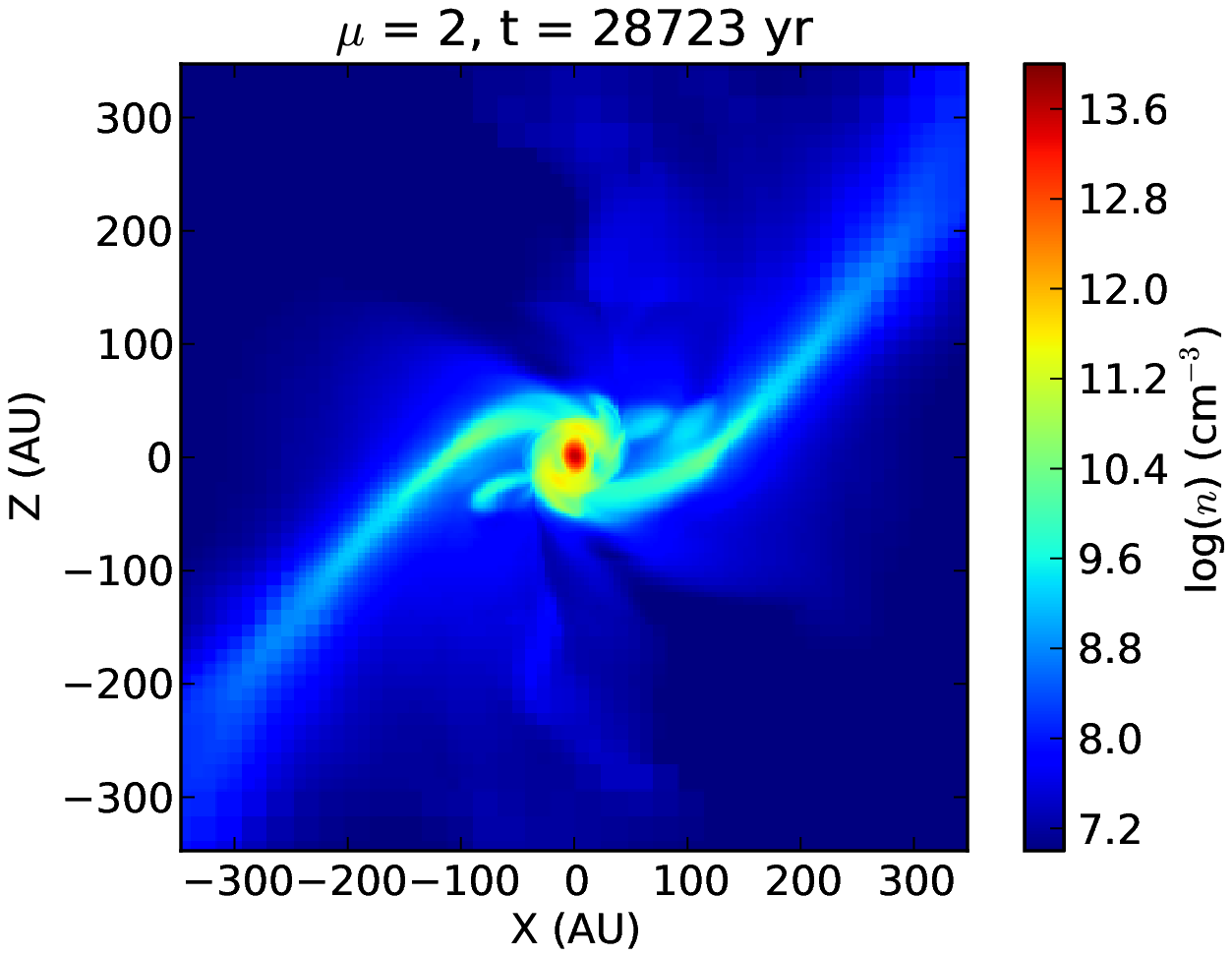}}
\caption{Same as Fig. \ref{img:slice030} for $\mu = 2$, $\alpha = 90^{\circ}$ at t = 25 000 and 29 000 yr.}
\label{img:slice0716}
\end{figure}

To estimate the radius and height of the disks, we use the first four criteria defined in section \ref{subsec:diskform} (\emph{cf.} Fig. \ref{img:shape0316}). However, as shown below, using only these criteria to infer the disk radius and height leads to an overestimate of their values, because of the large size of the spiral arms as evidently seen in Fig.~\ref{img:slice0316}. We thus consider them as upper limits for the disk radius and height. For a more realistic estimate, we constrain the disk to the central, denser rotating object and therefore consider the region of the disk with a density above $10^9$ cm$^{-3}$, which is our last criterion.

When the disk forms, its radius generally evolves with time, reaching values between 200 and 400~AU (for the minimum estimate of the radius, $R_{\rm min}$) from 500 to 800~AU (for the maximum estimate, $R_{\rm max}$). However, much smaller values are obtained when the braking is strong. The minimum height $H_{\rm min}$ is about from 20 to 40~AU, while the maximum height $H_{\rm max}$ reaches 140~AU. These values can be compared to analytical estimates of the characteristic height of a hydrostatic disk, $H_{\rm th} = \sqrt{c_s^2/4\pi G \rho}$, taking for $c_s$ and $\rho$ their mean values in the disk. The lower height estimates are in good agreement with the theoretical values. All results are summarized in Table \ref{tab:disk}. Finally, Figure \ref{img:slice0316} shows two density slices (at the beginning and the end of the simulation) in the perpendicular case for $\mu = 5$. A central well-shaped disk and two large spiral arms are clearly visible. We note that in our estimates, the maximum estimated radius corresponds to the maximum radius of the spiral arms. In contrast, the minimum radius accuratly describes the central denser object.

Figures \ref{img:slice030}, \ref{img:slice070}, and \ref{img:slice0716} show, in contrast to Fig.~\ref{img:slice0316}, density slices in the equatorial plane when no massive disk forms. In Fig.~\ref{img:slice030_1} and \ref{img:slice070_1}, the pseudo-disk can be clearly distinguished around the protostellar core. In Fig.~\ref{img:slice0716}, the two arms correspond to matter collapsing along the magnetic field lines.

\begin{table*}
\centering
\begin{tabular}{c c | c c c c c}
$\mu$ & $\alpha$ & $M_* (M_{\odot})$ & $M_{\rm disk} (M_{\odot})$ & $R_{\rm disk}$ (AU) & $H_{\rm disk}$ (AU) & $H_{\rm th}$ (AU) \\
\hline
\hline
17 & 0  & 0.43 & 0.15   & 250     & 30-140 & 33 \\
   & 45 & 0.43 & 0.25   & 250     & 30-70  & 13 \\
   & 90 & 0.66 & 0.25   & 400-800 & 20-130 & 33 \\
\hline
5  & 0  & 0.26 & 0.05   & 140     & 30-140 & 23 \\
   & 20 & 0.26 & 0.20   & 200-700 & 40-140 & 35 \\
   & 45 & 0.23 & 0.15   & 200-500 & 40-140 & 28 \\
   & 70 & 0.33 & 0.27   & 150-500 & 40-150 & 24 \\
   & 80 & 0.43 & 0.40   & 200-800 & 30-140 & 33 \\
   & 90 & 0.46 & 0.28   & 200-450 & 20-90  & 23 \\
\hline
3  & 0  & 0.19 & < 0.01 & < 50    & < 25   & 0 \\
   & 20 & 0.20 & < 0.01 & < 50    & < 10   & 12 \\
   & 45 & 0.29 & 0.11   & 200-800 & 30-120 & 31 \\
   & 90 & 0.37 & 0.10   & 200-800 & 30     & 17 \\
\hline
2  & 0  & 0.24 & 0      & 0       & 0      & 0 \\
   & 20 & 0.24 & 0.03   & 80      & 30-70  & 14 \\
   & 45 & 0.29 & 0.05   & 100     & 10     & 15 \\
   & 80 & 0.28 & < 0.01 & < 60    & < 10   & 6 \\
   & 90 & 0.25 & 0.02   & 50      & 10     & 5 \\
\hline
\hline
\end{tabular}
\caption{Disks characteristics (maximum mass of the protostellar core (which corresponds to $M(n > 10^{10}$ cm$^{-3}$)), maximum disk mass, disk radius ($R_{\rm min}-R_{\rm max}$), and disk height ($H_{\rm min}-H_{\rm max}$)).}
\label{tab:disk}
\end{table*}

\subsection{Discussions}

\subsubsection{Impact of the criteria}

\begin{figure}
\subfigure[\label{img:MDiskVel01}]{\includegraphics[width = .5\textwidth]{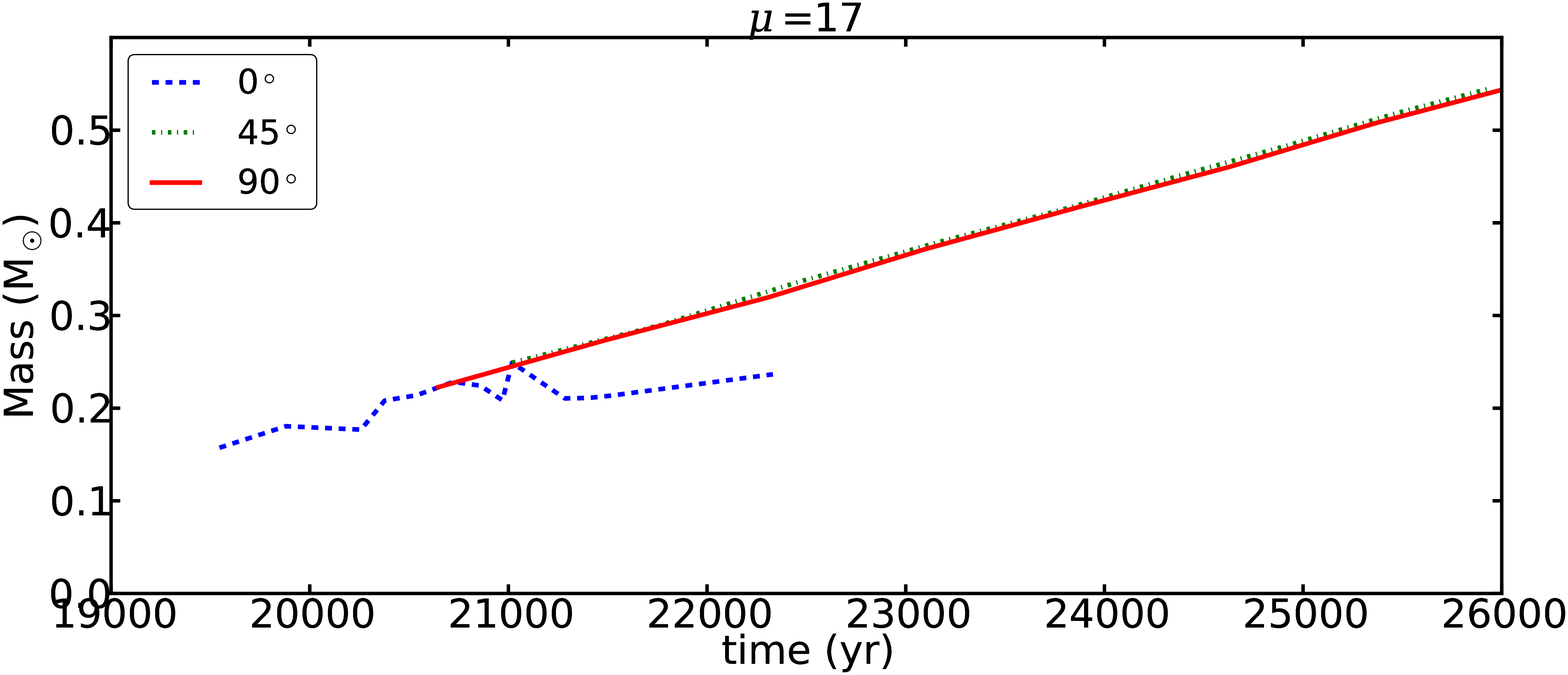}}
\subfigure[\label{img:MDiskVel03}]{\includegraphics[width = .5\textwidth]{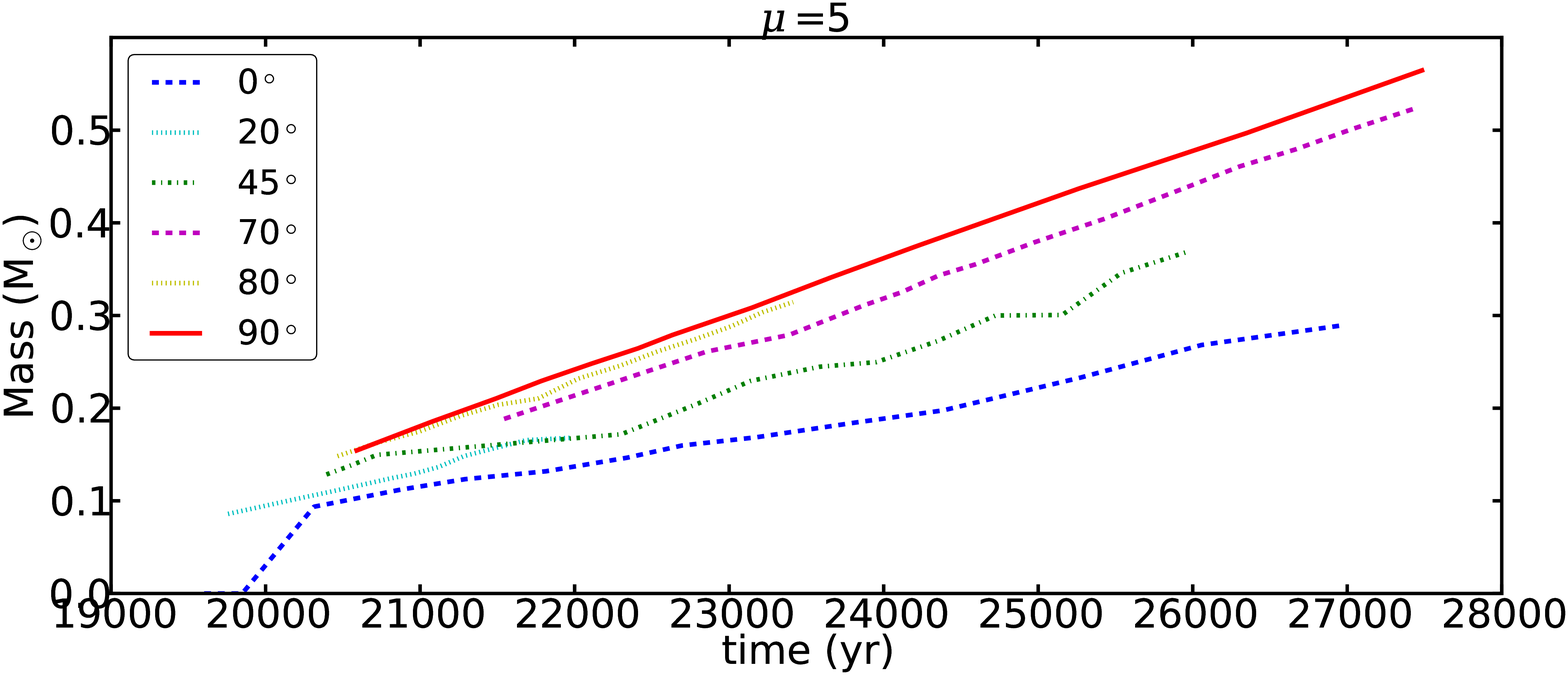}}
\subfigure[\label{img:MDiskVel05}]{\includegraphics[width = .5\textwidth]{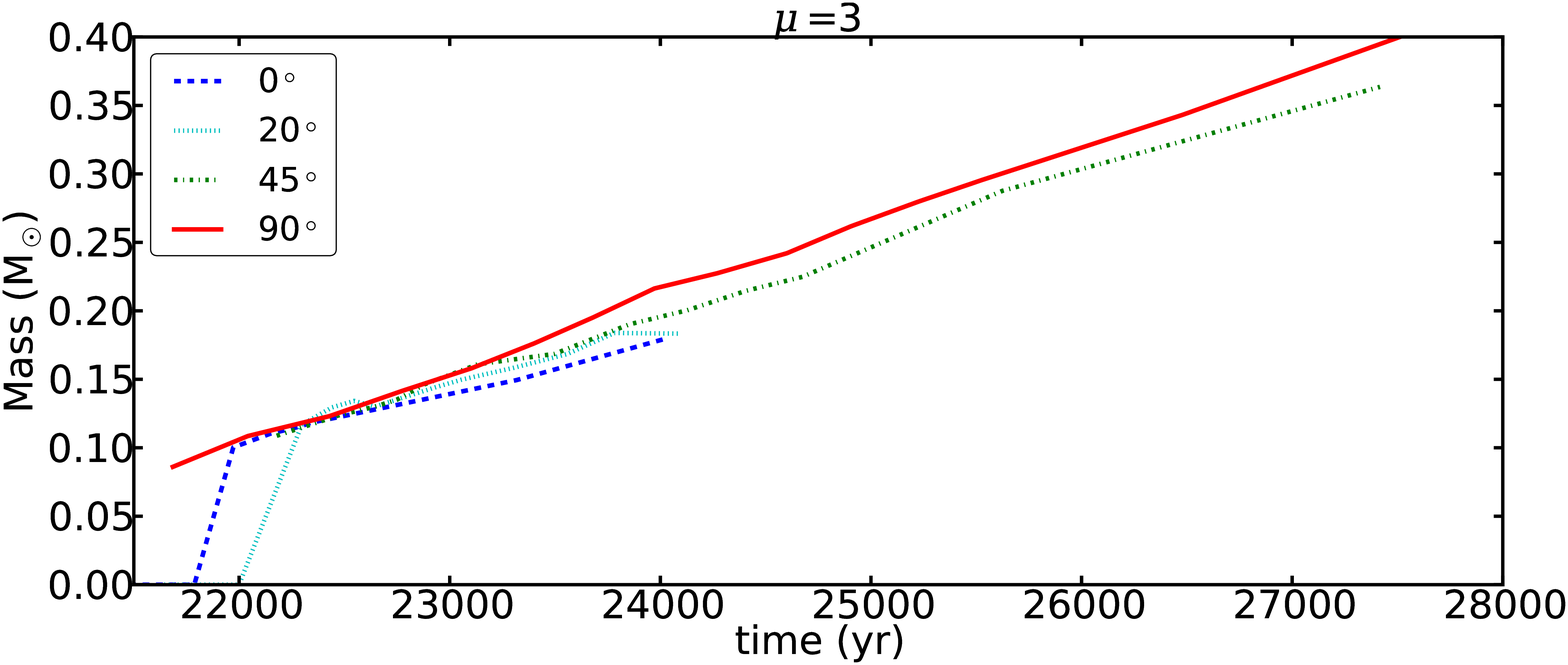}}
\subfigure[\label{img:MDiskVel07}]{\includegraphics[width = .5\textwidth]{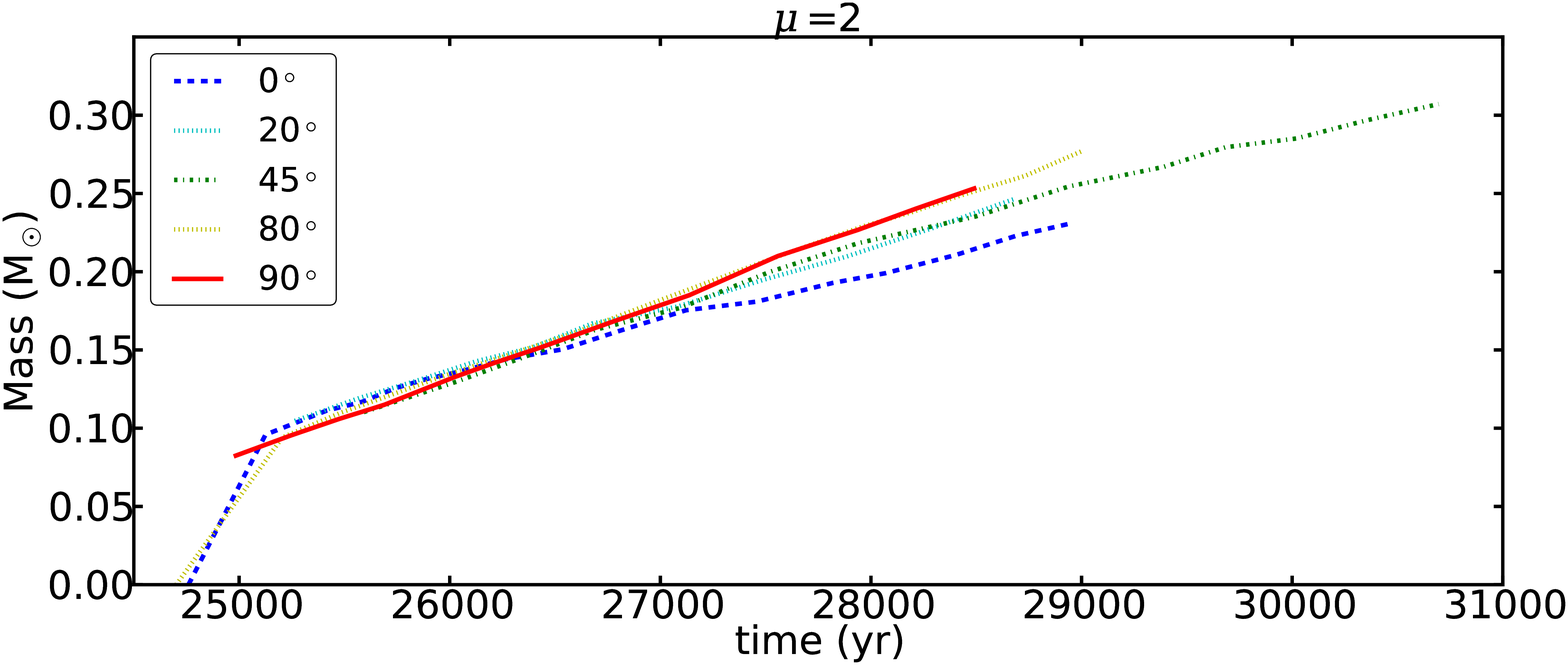}}
\caption{Mass of the ``disk'' evolution for $\mu = 17$ (Fig. \ref{img:MDiskVel01}), $\mu = 5$ (Fig. \ref{img:MDiskVel03}), $\mu = 3$ (Fig. \ref{img:MDiskVel05}), and $\mu = 2$ (Fig. \ref{img:MDiskVel07}), obtained with a very simple rotation criteria ($v_{\phi} > v_r$). The masses are significantly overestimated.}
\label{img:MDiskVel}
\end{figure}

Estimates of the disk mass generally depend on the working definition of the disk, and may lead to large overestimates. One example is to calculate the disk mass using a simple criterion based on a comparison between rotation and infall velocity (\emph{i.e.} $v_{\phi} > v_r$, which is the one used for example in \cite{Machida11a}). The panels in Fig. \ref{img:MDiskVel} display the mass of the ``disk'' found with this criterion, which range from 0.3 to 0.5 $M_{\odot}$ in all cases. They are more massive in the more tilted cases than in the aligned one, and for higher magnetizations, they are less massive, even if their formation is not prevented. Thus, while the trends are similar to those that we found previously, the mass of the disk can be greatly overestimated by using such a criterion.

\subsubsection{Comparison with observations}

Several studies have tried to infer disk masses from low resolution observations (with spatial resolutions of about 250~AU), without resolving the disk itself \citep{Enoch09, Enoch11}. For this purpose, they used a detailed emission model, coupled with an analytical model for the envelope, the cavity, and the disk. The envelope model is that of a rotating, collapsing sphere developed by \cite{Ulrich76}, with a cavity in which the density is set to zero to mimic the outflows. The disk density is given by a power-law dependence in radius and a Gaussian dependence in height (see \cite{Enoch09} for more details). Using these profiles, they ran a grid of models to find the parameters that most closely fit their observations, by computing detailed radiative transfer. Their most important parameters are the disk mass and radius. 

Following the same idea, we attempt to deduce the disk mass from our simulations using their method, but without the radiative transfer. At several time-steps, we compute column-density maps of our simulations, with an angle between the axis of rotation of the core and the line of sight of $15^{\circ}$ (which corresponds to the angle of the line of sight in their best-fit model). We consider the total mass of our simulation, $1 M_{\odot}$, to be the mass of the envelope, and adopt the same power-laws for the density profiles of the envelope and the disk, namely \citep{Ulrich76, Enoch09}
\begin{eqnarray}
\rho_{\rm env} & \propto & r^{-1.5} \\
\rho_{\rm disk} & \propto & r^{-1}e^{-(r/H(r))^2},
\end{eqnarray}
with $H(r) = r(H_0/R_{\rm disk})(r/R_{\rm disk})^{2/7}$ the height of the disk. We take an outflow opening angle of $20^{\circ}$ (which is the one that most closely fits their observations). As in \cite{Enoch09}, we infer a centrifugal radius $R_c$ from column-density profile corresponding to the radius where the slope of the column-density profile changes. The disk radius is another parameter that is varied; for simplicity, we only consider two different disk radii, $R_{\rm disk} = R_c$ and $R_c/2$. The disk vertical scale-height $H_0$ is equal to $0.2 R_{\rm disk}$. With these parameters, we run the models with $M_{\rm disk} = 0.01, \ 0.1, \ 0.2, \ 0.3, \ 0.4$, and $0.5 \ M_{\odot}$. The model grid, which is a grid of column-density maps, is compared to snapshots of our simulations by means of a mean squared error analysis to find the best-fit model.

\begin{figure}
\subfigure[\label{img:MDobs03}]{\includegraphics[width = .5\textwidth]{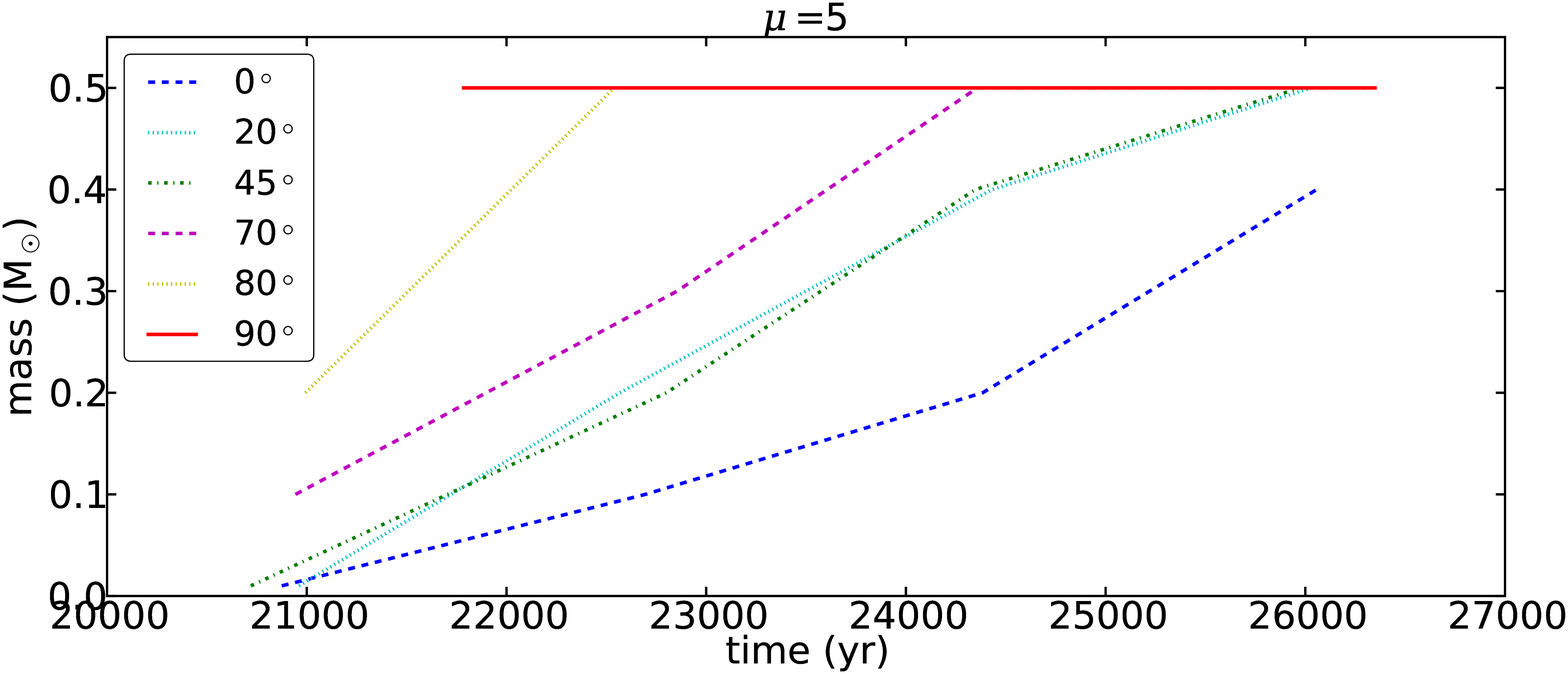}}
\subfigure[\label{img:MDobs05}]{\includegraphics[width = .5\textwidth]{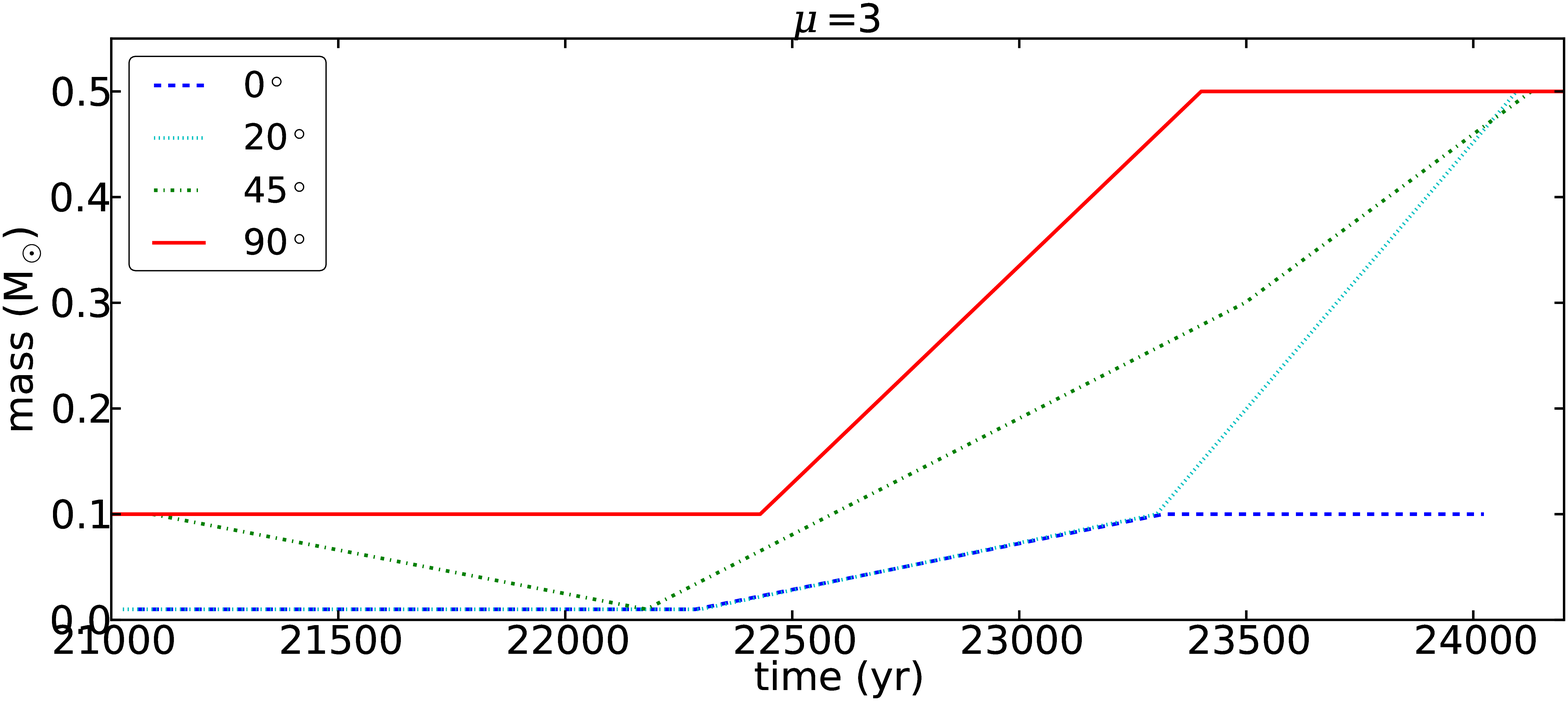}}
\subfigure[\label{img:MDobs07}]{\includegraphics[width = .5\textwidth]{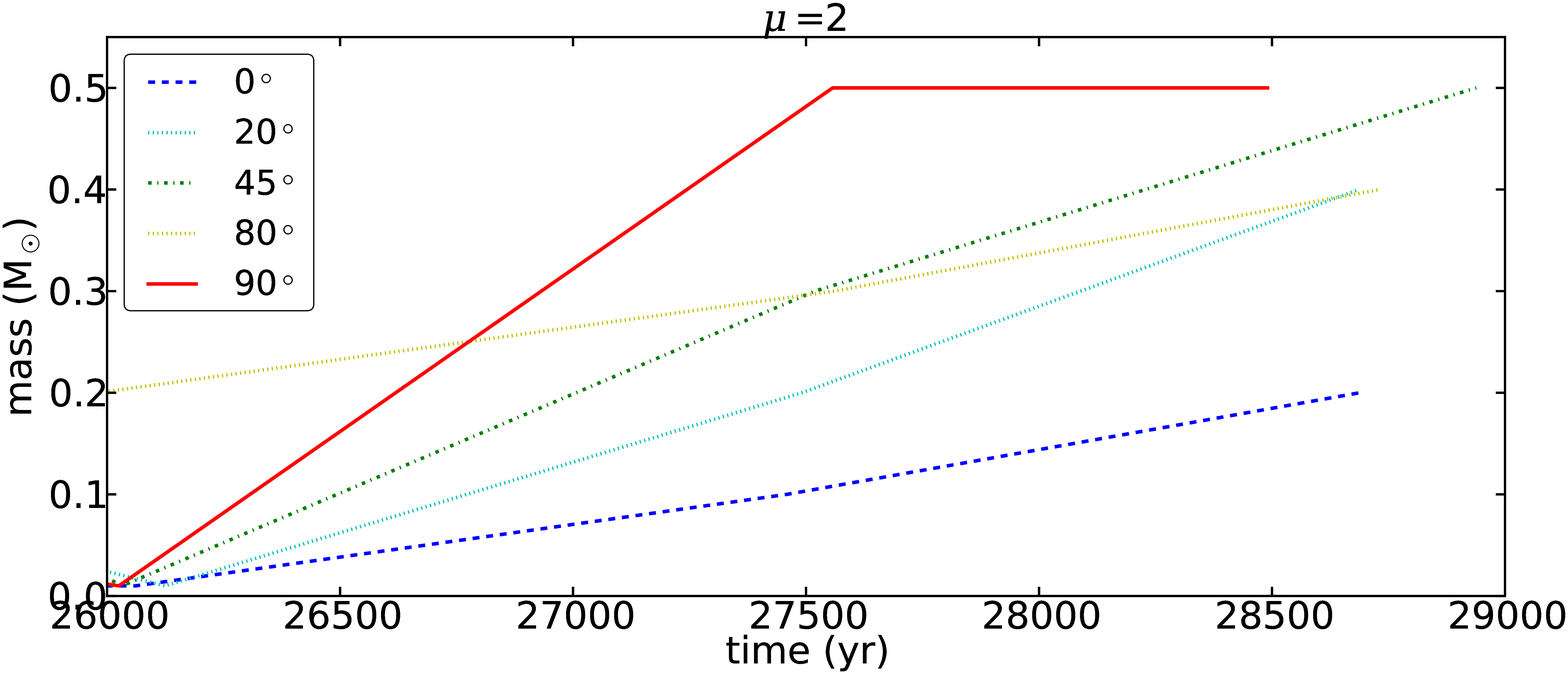}}
\caption{Observational estimate of the mass of the ``disk'' for $\mu = 5$ (Fig. \ref{img:MDobs03}), $\mu = 3$ (Fig. \ref{img:MDobs05}), and $\mu = 2$ (Fig. \ref{img:MDobs07}). These masses are inferred using a comparison between our simulations and analytical density profiles for the envelope, the outflows, and the disk (see text).}
\label{img:MDobs}
\end{figure}

Using this method, we find that the best fit to our simulations is characterized by a disk radius $R_{\rm disk} = R_c/2$ and masses between 0.4 and 0.5~$M_{\odot}$ for $\mu = 5$, masses between 0.1 and 0.5 $M_{\odot}$ for $\mu = 3$, and masses between 0.2 and 0.5 $M_{\odot}$ for $\mu = 2$ (see Fig.~\ref{img:MDobs}).

When massive disks actually form, which is not the case for the stronger magnetizations, this method provides results that are in relatively good agreement with the disk masses inferred from our simulations (with an overestimate of 30 to 40~\%). However, this comparison shows that in most cases density structures are detected that are mistaken for disks, particularly in either the aligned case or for the higher magnetizations when no disks form, leading to a large overestimate of disk masses. Those density structures, which also include the cavity of the outflows, actually correspond to a selection effect resulting from the use of a simple velocity criterion ($v_{\phi}>v_{r}$).

\begin{figure*}
  \includegraphics[width = 1.\textwidth]{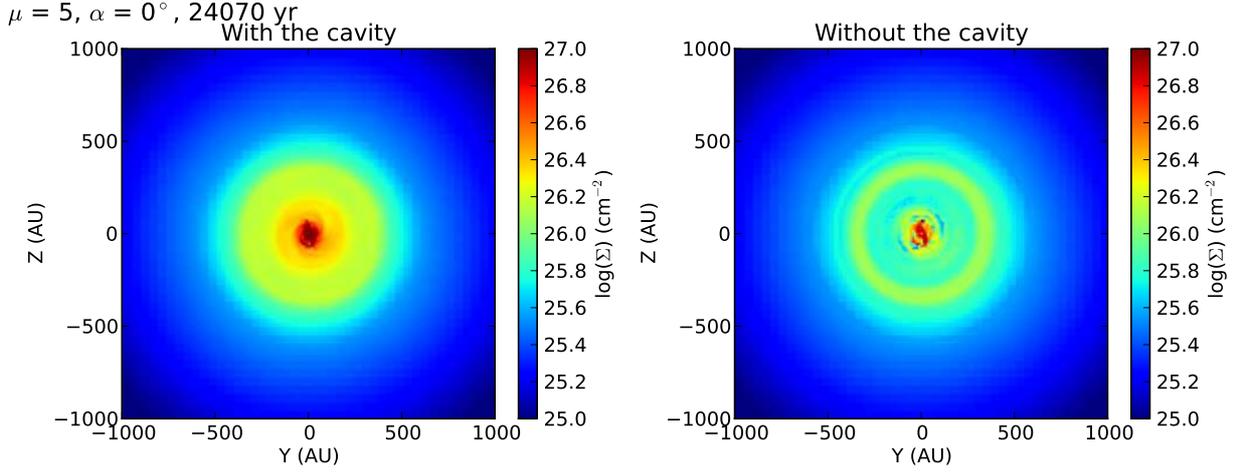}
  \caption{Column-density map for $\mu = 5, \alpha = 0^{\circ}$ at $t = 24000$~yr with and without the cavity of the outflow (left and right panel respectively).}
  \label{img:dcolCavity}
\end{figure*}

To verify this assumption, we remove the cavity of the outflows in the $\mu = 5, \alpha = 0^{\circ}$ case and repeat the analysis. Figure \ref{img:dcolCavity} shows column-density maps after taking into account all the gas (left panel) and removing the gas that belongs to the cavity (right panel). In the central region, within 500~AU, there is a discrepancy in the column-density of a factor three between those two maps: the cavity is a massive structure, which can be mistaken for a disk by projection effects. The best-fit model for a snapshot of this simulation without the cavity is $M_{\rm disk} = 0.01 M_{\odot}$, where it was $M_{\rm disk} = 0.5 M_{\odot}$ for the simulation with the cavity. Therefore, observed massive disks may actually be outflow cavities.

\subsubsection{Time formation of disks}

Our study shows that in very magnetized cases ($\mu \sim 1 - 3$) disks beginning to form at the earliest time of star formation (in Class 0 stage) will not eventually form or remain small. This may appear to contradict the ubiquity of disks at the Class I and later phases \citep[e.g.][]{Haisch01}. However, it is worth stressing that the magnetic braking represents an exchange of angular momentum between the inner and outer parts of the prestellar core. In particular, the envelope should then operate as a reservoir able to accept the excess of angular momentum present in the densest regions of the prestellar core. As accretion proceeds, the mass is the envelope diminishes and it is unlikely that magnetic braking remains efficient. We therefore speculate that disks will always form but that their formation time and size will depend strongly on the magnetization and the angle between the rotation axis and the magnetic field; later on, we may be able to reduce the formation time and increase the size.

Our conclusion -- that both the formation time and disk size depend strongly on the magnetization -- is qualitatively similar to that of \cite{Dapp10}, although the underlying reason is different.

\section{Conclusions}

We have presented an analytical analysis of a collapsing magnetic cloud that demonstrates that the magnetic field can remove angular momentum less efficiently when the rotation axis is perpendicular to the magnetic field than when they are both aligned.

We have then presented simulations of the collapse of prestellar dense cores with different magnetizations $\mu$ and in both aligned and various misaligned configurations. The orientation of the rotation axis with respect to the magnetic field, $\alpha$, has a strong effect on the formation of the adiabatic first core and the disk formation. In particular, we have performed a detailed analysis of the transport of the angular momentum in the simulations, and characterized the disks when they formed. Our main results are the following:
\begin{itemize}
\item Magnetic braking decreases with $\alpha$, but increases with $\mu$.
\item Misalignment has a strong impact on the outflows and can suppress them; consequently, the angular momentum transport by the outflows decreases with $\alpha$.
\item Angular momentum transport by gravity increases with $\alpha$, owing to the presence of the disks, particularly their asymmetric structures.
\item The mass in the disks increases with $\alpha$.
\item For increasing magnetic fields, the disk masses decrease, with a limiting case being that of $\mu = 2$, where disk formation is prevented.
\item disks have typical mass up to 0.3~$M_{\odot}$ and typical radii of from about 200 to 400~AU.
\item In general, magnetic braking is the most important mechanism for transporting angular momentum. It always dominates the transport of angular momentum by the flow and, except for low magnetization ($\mu \gtrsim 17$), also dominates the transport by means of gravitational torques.
\end{itemize}
We have shown that our conclusions depends on the criterion we use to define disks. We conclude that a simple rotation criteria is insufficient and leads to estimates of disk masses that are far too high.

We also analyzed our simulations following the method described in \cite{Enoch09}, demonstrating that low resolution observations can mistake density structures for disks.
\newline

{\small
\emph{Acknowledgements.} We thank the anonymous referee for his thorough reading of the manuscript and his helpful comments and suggestions, which helped to significantly improve the quality of this article. PH thanks Telemachos Mouschovias for enlightening discussions about magnetic braking.
}

\appendix

\section{Euler's equation and angular momentum transports} \label{appendix:euler}

The Euler's equation for a magnetized fluid can be written as
\begin{equation}
\rho \partial_t{\bf v} + \rho\left({\bf v}\cdot\nabla\right){\bf v} = - \nabla\left(P + \frac{B^2}{4\pi}\right) - \rho{\bf g} + \left(\frac{{\bf B}}{4\pi}\cdot\nabla\right){\bf B}, \label{eq:euler}
\end{equation}
with $\rho$ the density, ${\bf v}$ the velocity, $P$ the gas pressure, ${\bf B}$ the magnetic field, and ${\bf g}$ the gravitational acceleration where ${\bf g} = \nabla \Phi$ and $\Phi$ is the gravitational potential.

In cylindrical coordinates, the azimuthal component of the left hand side of \ref{eq:euler} can be written
\begin{equation}
\rho\partial_tv_{\phi} + \rho\left( v_r\partial_rv_{\phi} + \frac{v_rv_{\phi}}{r}+ \frac{v_{\phi}}{r}\partial_{\phi}v_{\phi} + v_z\partial_zv_{\phi} \right).
\end{equation}
Using the continuity equation
\begin{eqnarray}
\nabla\cdot(\rho{\bf v}) + \partial_t\rho & = & 0 \\
\frac{1}{r}\partial_r(r\rho v_r) + \frac{1}{r}\partial_{\phi}(\rho v_{\phi}) + \partial_z(\rho v_z) + \partial_t\rho & = & 0,
\end{eqnarray}
it can be written as
\begin{eqnarray}
\rho\partial_tv_{\phi} + v_{\phi}\partial_t\rho & + & \rho\frac{v_r}{r}\partial_r(rv_{\phi}) + \frac{v_{\phi}}{r}\partial_r(r\rho v_r) \nonumber \\
 & + & \rho\frac{v_{\phi}}{r}\partial_{\phi}(v_{\phi}) + \frac{v_{\phi}}{r}\partial_{\phi}(\rho v_{\phi}) \nonumber \\
 & + & \rho\partial_zv_{\phi} + v_{\phi}\partial_z(\rho v_z) \nonumber \\
= \partial_t(\rho v_{\phi}) & + & \frac{1}{r}\partial_r(r\rho v_rv_{\phi}) + \rho\frac{v_rv_{\phi}}{r} \nonumber \\
 & + & \frac{1}{r}\partial_{\phi}(\rho v_{\phi}v_{\phi}) + \partial_z(\rho v_zv_{\phi}) \nonumber \\
= \partial_t(\rho v_{\phi}) & + & \nabla\cdot(\rho v_{\phi}{\bf v}) + \rho\frac{v_rv_{\phi}}{r}
\end{eqnarray}

We can do the same for the magnetic tension component of the equation
\begin{equation}
\frac{1}{4\pi}\left(B_r\partial_rB_{\phi} + \frac{B_rB_{\phi}}{r} + \frac{B_{\phi}}{r}\partial_{\phi}B_{\phi} + B_z\partial_zB_{\phi} \right),
\end{equation}
using the solenoidal constraint ($\nabla \cdot {\bf B} = 0$); it comes
\begin{equation}
\frac{1}{4\pi}\left(\nabla\cdot(B_{\phi}{\bf B}) + \frac{B_rB_{\phi}}{r}\right).
\end{equation}

The density $\rho$ can be expressed as a function of the gravitational acceleration, using the Poisson's equation
\begin{equation}
\rho = \frac{1}{4\pi G}\nabla\cdot{\bf g}.
\end{equation}
The azimuthal component of the gravitational term of Euler's equation thus becomes
\begin{eqnarray}
\rho g_{\phi} & = & \frac{g_{\phi}}{4\pi G}\nabla\cdot{\bf g} \nonumber \\
 & = & \frac{g_{\phi}}{4\pi G}\left(\frac{1}{r}\partial_r(rg_r) + \frac{1}{r}\partial_{\phi}g_{\phi} + \partial_zg_z\right) \nonumber \\
 & = & \frac{1}{4\pi G}\left[\frac{g_rg_{\phi}}{r} + \partial_r(g_rg_{\phi}) + \frac{1}{r}\partial_{\phi}g_{\phi}g_{\phi} + \partial_zg_zg_{\phi}\right. \nonumber \\
 &   & - \left.\left(g_r\partial_rg_{\phi} + \frac{1}{r}g_{\phi}\partial_{\phi}g_{\phi} + g_z\partial_zg_{\phi}\right)\right],
\end{eqnarray}
and using Schwarz' theorem
\begin{eqnarray}
\rho g_{\phi} & = & \frac{1}{4\pi G}\left[\frac{g_rg_{\phi}}{r} + \frac{1}{r}\partial_r(rg_rg_{\phi}) + \frac{1}{r}\partial_{\phi}g_{\phi}g_{\phi} + \partial_zg_zg_{\phi}\right. \nonumber \\
 &   & - \left. \frac{1}{2r}\partial_{\phi}\left(g_r^2 + g_{\phi}^2 + g_z^2\right)\right] \nonumber \\
\rho g_{\phi} & = & \frac{1}{4\pi G}\left(\frac{g_rg_{\phi}}{r} + \nabla\cdot(g_{\phi}{\bf g}) - \frac{1}{2}\left({\bf e_{\phi}}\cdot\nabla\right){\bf g}^2\right),
\end{eqnarray}
we can identify a curvature term ($g_rg_{\phi}/r$), a tension term ($\nabla\cdot(g_{\phi}{\bf g})$), and a pressure term ($\left({\bf e_{\phi}}\cdot\nabla\right){\bf g}^2/8\pi G$).

Therefore, the azimuthal component of the Euler's equation can eventually be written as
\begin{eqnarray}
\partial_t(\rho v_{\phi}) & + & \nabla\cdot(\rho v_{\phi}{\bf v}) + \rho\frac{v_rv_{\phi}}{r} \nonumber \\
& & = - ({\bf e_{\phi}}\cdot\nabla)\left(P + \frac{B^2}{8\pi} - \frac{g^2}{8\pi G}\right) \nonumber \\
& & + \frac{1}{4\pi}\left(\nabla\cdot(B_{\phi}{\bf B}) + \frac{B_rB_{\phi}}{r}\right) \nonumber \\
& & - \frac{1}{4\pi G}\left(\nabla\cdot(g_{\phi}{\bf g}) + \frac{g_rg_{\phi}}{r}\right).
\end{eqnarray}
If we multiply this equation by $r$, it becomes
\begin{eqnarray}
\partial_t\left(\rho r v_{\phi}\right) & + & \nabla\cdot r\left[ \rho v_{\phi}{\bf v} + \left(P + \frac{B^2}{8\pi} - \frac{g^2}{8\pi G}\right){\bf e_{\phi}} \right. \nonumber \\
& - & \left. \frac{B_{\phi}}{4\pi}{\bf B} + \frac{g_{\phi}}{4\pi G}{\bf g} \right] = 0, \label{eq:angmom}
\end{eqnarray}
since $r$ is time-invariant. This equation expresses the angular momentum conservation; we can identify the magnetic and the gravitational torques $\nabla \cdot (r B_{\phi} {\bf B}/4 \pi)$ and $\nabla \cdot (r g_{\phi} {\bf g}/4 \pi G)$ with $i \in \{r, z\} $, which are responsible for the angular momentum transport by means of a magnetic field and gravitation, respectively.

\section{Convergence}

Additional sets of simulations were run to test the numerical convergence. These sets of simulations show that the numerical dissipation does not significantly change our results, which are qualitatively invariant.

In the first set of simulations, we change the Jeans refinement strategy to increase the spatial resolution. A cell was previously refined if its size exceeded one-tenth of a Jeans' length ($c_s (\pi/G \rho)^{1/2}$). We run simulations with 15 cells per Jeans' length (HR1) and another one with 20 cells per Jeans' length (HR2).

In the last set of simulations, we change the Courant number (from 0.8 to 0.4) to increase the temporal resolution.

All sets were run with $\mu = 5$ and 2, and $\alpha = 0, 45$, and 90$^{\circ}$ for increasing spatial resolution, and with $\mu = 5$, and $\alpha = 0, 45$, and 90$^{\circ}$ for increasing temporal resolution.

Our results remain qualitatively similar even if convergence is not achieved. These convergence runs show that the mass of the disk is slightly overestimated in our previous analysis.

\subsection{High spatial resolution simulations}

\begin{figure*}
\subfigure[\label{img:am03}]{\includegraphics[width = 1.\textwidth]{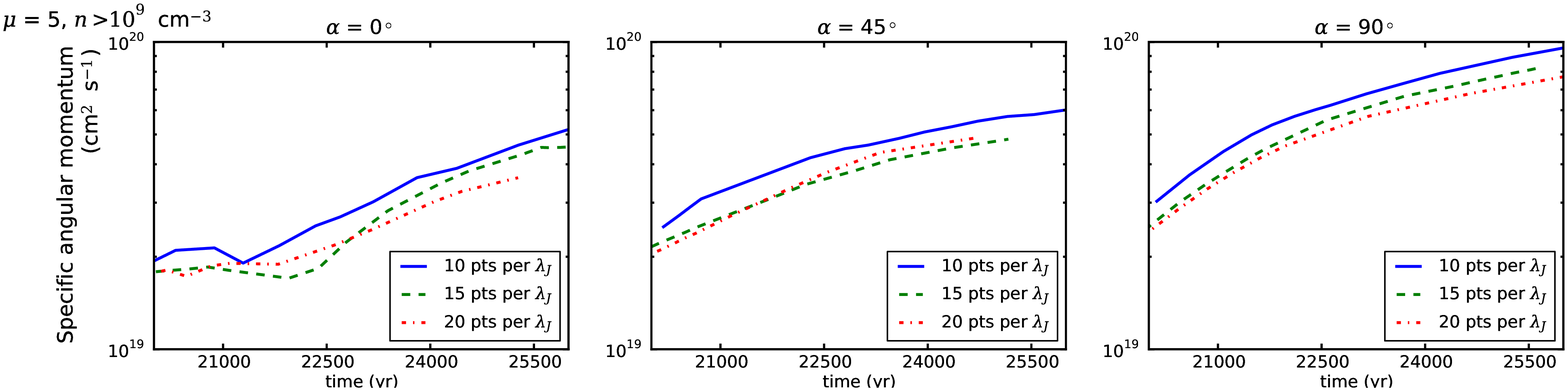}}
\subfigure[\label{img:am07}]{\includegraphics[width = 1.\textwidth]{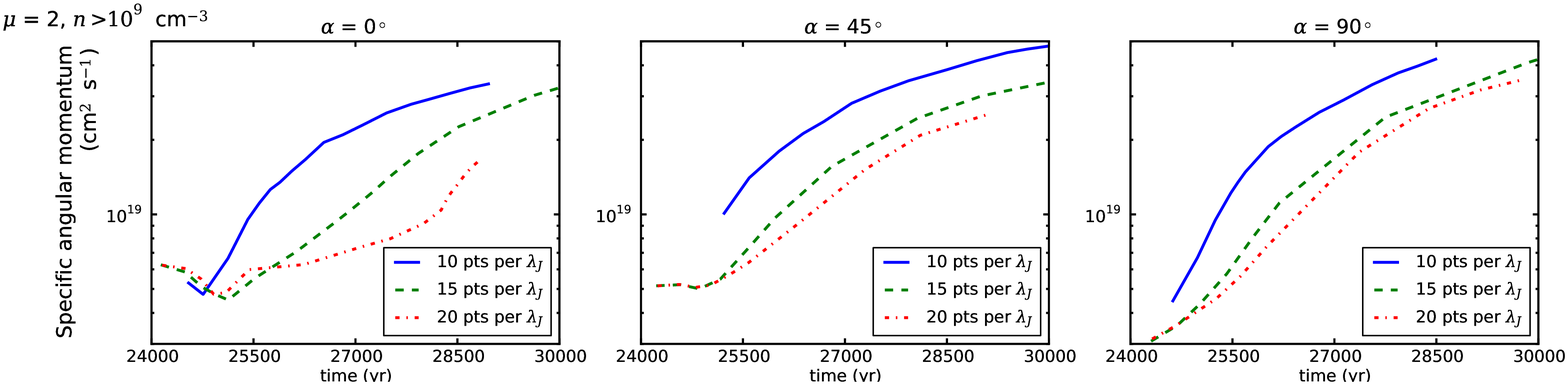}}
\caption{Specific angular momentum for $\mu = 5 \text{ (upper panels) and } \mu = 2 \text{(lower panels)}, n > 10^9$ cm$^{-3}$, for $\alpha = 0, 45$, and 90$^{\circ}$, with 10, 15, and 20 resolved points per Jeans' length.}
\label{img:amj}
\end{figure*}

\subsubsection{Angular momentum}

Figures \ref{img:amj} show the evolution of specific angular momentum for $n > 10^9$ cm$^{-3}$ for, respectively, $\mu = 5$ and $\mu = 2$. The left panel display its evolution for 10 points, the central panel for 15 points, and the right panel for 20 points resolved per Jeans' length. For all angles ($\alpha = 0, 45$ and 90$^{\circ}$), the specific angular momentum decreases with the resolution.

With increasing resolution, more momentum is transported by the magnetic field and outflows, and the steady state is reached at a later stage.

It is clear that numerical convergence has not yet been reached and that treating magnetic braking requires a very high spatial resolution \citep{Commercon10, Hennebelle11}.

Even though these results are quantitatively different from the simulations presented in this paper, they are qualitatively similar: magnetic braking still decreases significantly when the angle between the rotation axis and the magnetic field increases, and a comparable amount of angular momentum is carried away by magnetic braking and outflows.

\subsubsection{Disk mass}

\begin{figure*}
\subfigure[\label{img:massConv5}]{\includegraphics[width=1.\textwidth]{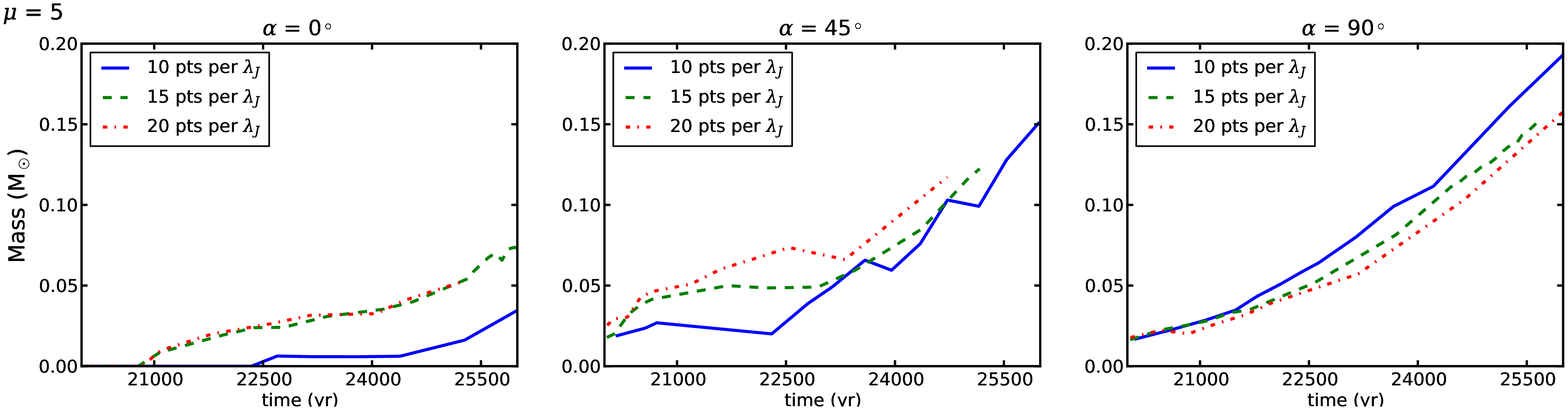}}
\subfigure[\label{img:massConv2}]{\includegraphics[width=1.\textwidth]{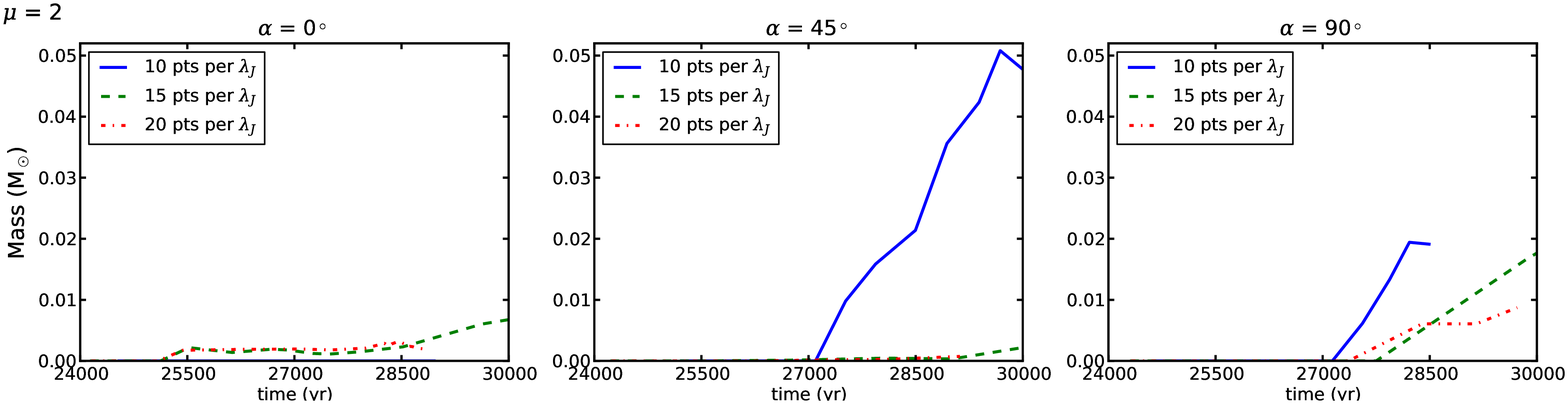}}
\caption{Mass of the disk as a function of time for $\mu = 5$ (upper panels) and $\mu = 2$ (lower panels), for $\alpha = 0, 45$, and 90$^{\circ}$, with 10, 15, and 20 resolved points per Jeans' length.}
\label{img:massConv}
\end{figure*}

The mass of the disk stays roughly the same or decreases with increasing resolution, as shown in Fig. \ref{img:massConv}. The mass of the disk can decrease with increasing resolution because of the more efficient transport of angular momentum in the higher resolution cases. For example for $\mu = 5, \alpha = 90^{\circ}$, the mass of the disk at t = 25 500~yr is 0.18 $M_{\odot}$ (LR) compared to 0.15 $M_{\odot}$ (HR1) and 0.13 $M_{\odot}$ (HR2).

However, our previous conclusions still hold: when magnetization is relatively strong ($\mu = 5$), disks form only when the rotation axis is misaligned with the magnetic field, and for lower $\mu$ (meaning stronger magnetization), magnetic braking acts so strongly it prevents disk formation.

\subsection{High temporal resolution simulations}

\subsubsection{Angular momentum}

\begin{figure*}
\includegraphics[width = 1.\textwidth]{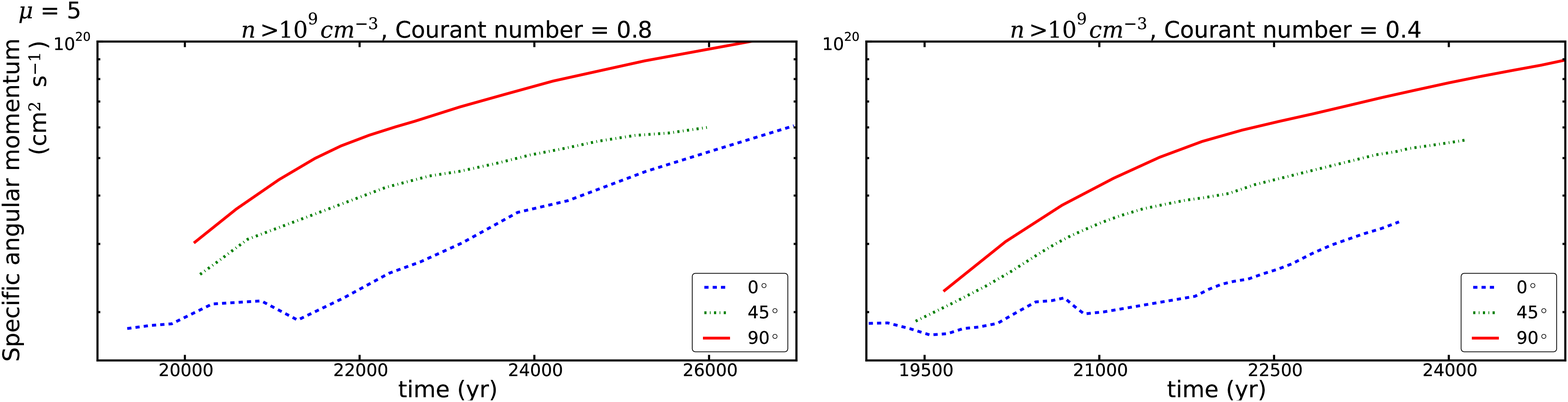}
\caption{Specific angular momentum for $\mu = 5, n > 10^9$ cm$^{-3}$, with a Courant number of 0.8 and 0.4.}
\label{img:amc}
\end{figure*}

Figure \ref{img:amc} displays the evolution of the specific angular momentum for $n > 10^9$ cm$^{-3}$, $\mu = 5$, for a Courant number of 0.8 (left panel) and 0.4 (right panel), which corresponds to a smaller time-step. As the resolution increases, more momentum is transported outward.

\subsubsection{Disk mass}

\begin{figure}
\includegraphics[width = .5\textwidth]{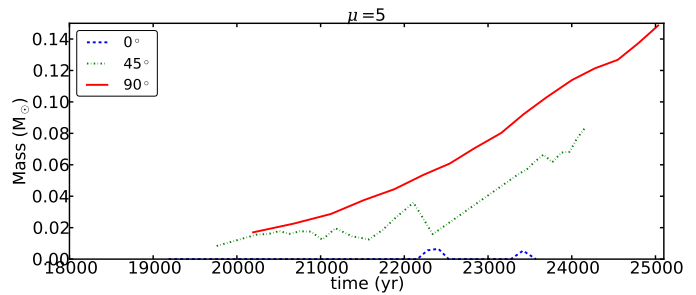}
\caption{Disk mass for $\mu = 5$ and a Courant number of 0.4.}
\label{img:massConvC}
\end{figure}

Figure \ref{img:massConvC} shows the evolution of the mass of the disk for Courant number of 0.4.

\bibliographystyle{aa}
\nocite{*}
\bibliography{bib_disk}

\end{document}